\documentclass[preprint,3p,times]{elsarticle}

\usepackage{amssymb}
\usepackage{amsmath}

\usepackage{xcolor}
\DeclareMathOperator{\E}{\mathbb{E}}
\newcommand{\Var}{\mathbb{V}\mathrm{ar}}
\newcommand{\Cov}{\mathbb{C}\mathrm{ov}}
\DeclareMathOperator{\Y}{\mathcal{Y}}

\DeclareMathOperator{\KL}{KL}

\newcommand{\reals}{\mathbb{R}}
\DeclareMathOperator{\Tr}{Tr}
\newcommand{\pdense}{p}
\usepackage{comment}
\newtheorem{task}{Task}

\newtheorem{theorem}{Theorem}

\newtheorem{remark}{Remark}

\newcommand{\BlackBox}{\rule{1.5ex}{1.5ex}}  % end of proof
\ifdefined\proof
    \renewenvironment{proof}{\par\noindent{\bf Proof\ }}{\hfill\BlackBox\\[2mm]}
\else
    \newenvironment{proof}{\par\noindent{\bf Proof\ }}{\hfill\BlackBox\\[2mm]}
\fi
\usepackage{graphicx}
\usepackage{caption}
\usepackage{subcaption}
\usepackage{float}
\usepackage{pifont}
\usepackage{tabularx}

\usepackage[linesnumbered,ruled,vlined]{algorithm2e}

\journal{Journal of Computational Physics}

\begin{document}

\begin{frontmatter}

\title{Sequential Bayesian parameter–state estimation in dynamical systems with noisy and incomplete observations via a variational framework}

\author[label1]{Liliang Wang}
\author[label2]{Alex Gorodetsky}
\affiliation[label1]{organization={Department of Aerospace Engineering, University of Michigan},
            addressline={2013 FXB, 1320 Beal Avenue},
            city={Ann Arbor},
            postcode={48109},
            state={MI},
            country={USA}}

\affiliation[label2]{organization={Department of Aerospace Engineering, University of Michigan},
            addressline={3025 FXB, 1320 Beal Avenue},
            city={Ann Arbor},
            postcode={48109},
            state={MI},
            country={USA}}

\begin{abstract}
Online joint estimation of a dynamical model’s unknown parameters and states with uncertainty quantification is crucial in many applications. For example, digital twins dynamically update their knowledge of model parameters and states to support prediction and decision-making. Reliability and computational speed are vital for DTs. Online parameter–state estimation ensures computational efficiency, while uncertainty quantification is essential for making reliable predictions and decisions. In parameter-state estimation, the joint distribution of the state and model parameters conditioned on the data, termed the \textit{joint posterior}, provides accurate uncertainty quantification. Because the joint posterior is generally intractable to compute, this paper presents an online variational inference framework to compute its approximation at each time step. The approximation is factorized into a marginal distribution over the model parameters and a state distribution conditioned on the parameters. This factorization enables recursive updates through a two-stage procedure: first, the parameter posterior is approximated via variational inference; second, the state distribution conditioned on the parameters is computed using Gaussian filtering based on the approximate parameter posterior. The algorithmic design is supported by a theorem establishing upper bounds on the joint posterior approximation error. Numerical experiments demonstrate that the proposed method (i) matches the performance of the joint particle filter in low-dimensional problems, accurately inferring both unobserved states and unknown parameters of dynamical and observation models; (ii) remains robust under noisy, partial observations and model discrepancies in a chaotic Lorenz'96 system; and (iii) scales effectively to a high-dimensional state-space system arising from the spatial discretization of a convection–diffusion equation, outperforming the joint ensemble Kalman filter in this setting.
\end{abstract}


\end{frontmatter}

\section{Introduction}
\label{sec:introduction}
In many data-driven techniques, there is a great need to continuously update knowledge of a system’s dynamics and states with proper uncertainty quantification in an online manner. A prominent example arises in the context of digital twins (DTs) of operating assets. A digital twin is a continuously updated virtual representation of a physical system, using data from its physical counterpart~\cite{XiuTartakovsky2025DigitalTwins}. One key consideration in DT applications is the limitation of computational resources~\cite{dt_gap}. In such settings, online inference methods are particularly well suited. Unlike offline approaches which process data in batches, online methods operate recursively: at each time step $k$, they update the inference results from the previous step $k-1$ by assimilating only the newly received data point $y_k$. This recursive structure removes the need to store and re-assimilate historical data, offering substantially higher computational efficiency compared with offline approaches. In DTs, models and data work together to enable monitoring, prediction and decision-making~\cite{dt_gap}.
However, real-world observational data are normally noisy and often incomplete. Therefore, DTs must not only learn model parameters but also estimate the states of the underlying system to effectively achieve monitoring and prediction. Moreover, there is often a mismatch between the inference model and the true system, either due to limited knowledge or the complexity of the real dynamics. The noisy, incomplete data together with the model misspecification introduce significant uncertainty. This makes uncertainty quantification essential for establishing trust in models and supporting risk-aware decision-making~\cite{pash2025predictive, dt_gap}. In parameter-state estimation, the \textit{joint posterior}, defined as the joint distribution of state and parameters conditioned on the data, provides accurate uncertainty quantification. It captures not only the individual uncertainties of the states and parameters but also their dependencies. However, the joint posterior is typically intractable to compute in practice. Therefore, we instead consider approximating it online. This task is formalized below.

\begin{figure}
     \centering
     \includegraphics[width=0.8\linewidth]{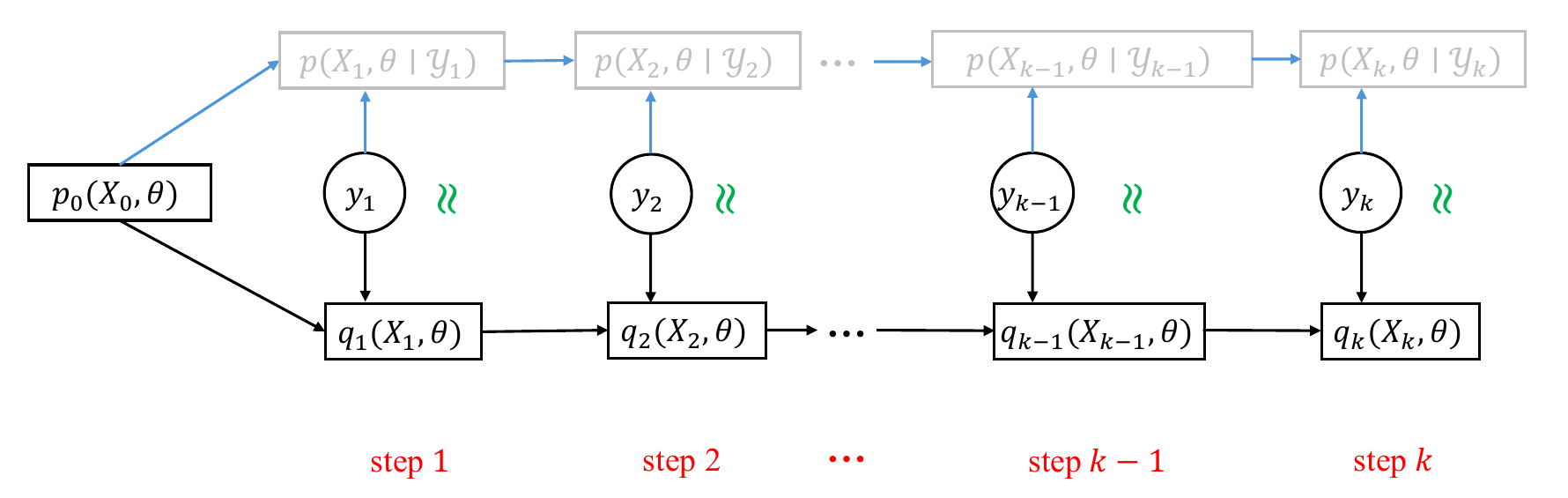}
     \caption{Visualization of the task considered in this paper. A prior distribution $p_0(X_0,\theta)$ that incorporates prior knowledge of the unknown initial state and model parameters is given. The objective is to sequentially approximate the joint posterior $p(X_k,\theta \mid \Y_k)$ by another distribution $q_k(X_k,\theta)$ as new data arrive. At each time step $k$, the approximation $q_k(X_k,\theta)$ is computed using only the newly received data $y_k$ and the previous approximation $q_{k-1}(X_{k-1},\theta)$, without reprocessing past data.}
    \label{fig:task}
\end{figure}

\begin{task}
\label{task}
Given sequentially arriving data points $y_1,y_2,\cdots$, the goal is to compute a distribution of the current state $X_k$ and unknown model parameters $\theta$, denoted by $q_k(X_k,\theta)$, that approximates the joint posterior $\pdense(X_k,\theta \mid \Y_k)$, defined as the distribution of $X_k$ and $\theta$ conditioned on all observations up to time $k$, $\Y_k=(y_1,y_2,\cdots,y_k)$. At each time step $k$, the task is to compute the current approximation $q_k(X_k,\theta)$ using only the previous approximation $q_{k-1}(X_{k-1},\theta)$ and the newly received data point $y_k$. This task is illustrated in Figure~\ref{fig:task}.
\end{task}

In recent years, some methods have been developed to address this task, namely, to approximate the joint posterior in an online manner. Among these methods, the most commonly used approaches are based on state augmentation. They augment the state vector with the unknown parameters and apply standard filtering methods to the augmented system. Examples of such methods include joint Gaussian filters (e.g., joint unscented Kalman filter (UKF)~\cite{joint_ukf2,joint_ukf3} and joint ensemble Kalman filter (EnKF)~\cite{joint_enkf}) and joint particle filters (PFs)~\cite{joint_pf_1,joint_pf_2}. Joint Gaussian filters approximate the joint posterior with a Gaussian distribution. Let $Y_k$ denote the observation at time step $k$, with the data $y_k$ as its realization. Joint Gaussian filters assume that the joint distribution $p(X_k, \theta, Y_k \mid \Y_{k-1})$ is Gaussian. Under this assumption, the joint posterior $p(X_k,\theta \mid \Y_k)$ remains Gaussian and can be updated recursively. This yields computationally efficient algorithms, but the Gaussian assumption is overly restrictive. It can introduce large approximation errors when the relation between $\theta$ and $X_k$ is highly nonlinear, a situation that can arise in both linear and nonlinear systems. Joint PFs, by contrast, avoid Gaussian assumptions by representing the joint posterior with an empirical distribution constructed from weighted particles, achieving significantly higher approximation accuracy for low-dimensional systems. However, the presence of static parameters in the augmented state amplifies particle weight degeneracy~\cite{joint_pf_issue}, a well-known issue of PFs. As a result, particle weights degenerate faster than in standard PFs, and the number of particles required to counteract this issue grows rapidly with the dimensions of the state and parameter spaces~\cite{Snyder2008Obstacles}. Introducing artificial dynamics for the parameters~\cite{joint_pf_1} mitigates this issue to some extent, but at the cost of introducing estimation bias that is difficult to quantify~\cite{joint_pf_issue}. Consequently, joint PFs may perform poorly even for moderately sized problems~\cite{fbovi}. 

To avoid the severe weight degeneracy resulting from state augmentation, another PF-based method for approximating the joint posterior online, SMC$^2$~\cite{Chopin2013SMC2}, was developed. SMC$^2$ employs a nested filtering structure to compute an approximation of the joint posterior. For each sample $\theta^{(i)}$ drawn from the previous approximate parameter posterior, an inner-loop particle filter estimates the conditional state posterior $p(X_k \mid \Y_k,\theta^{(i)})$. The conditional state posterior estimates are then used by an outer-loop filter to update the approximate parameter posterior. While effective, SMC$^2$ has a major drawback: its per-step computational cost grows with time, making it unsuitable for online tasks.

In contrast to PFs, variational inference (VI)~\cite{vi_def}, which approximates the posterior distribution by minimizing the Kullback-Leibler (KL) divergence of the approximate posterior from the true posterior, is widely regarded as scalable. Moreover, unlike Gaussian filters, VI is not restricted to a specific distribution family~\cite{online_variational_filtering}, making it a promising tool for accurate joint posterior approximation. Nevertheless, most of the existing VI methods for joint posterior approximation are designed for offline settings (e.g.,~\cite{IFT,Ryder2018BBVI_SDE,variational_state_parameter}). Only a limited number of VI-based approaches have been proposed for online joint posterior approximation~\cite{online_rl, online_VI_linear}, and these methods apply only to linear systems. Furthermore, these methods typically approximate the joint posterior as the product of marginal distributions over parameters and states. Such factorization neglects the intrinsic dependence between model parameters and states and often results in inaccurate joint posterior approximations.

It is noteworthy that another class of online parameter–state estimation methods focuses on computing only the approximate marginal posteriors of the states and parameters by running two filters—a state filter and a parameter filter. Examples include the United Filter~\cite{UnitedFilter} and classical dual Gaussian filters such as the dual UKF~\cite{dual_ukf_1} and dual EnKF~\cite{dual_enkf}. However, since the outputs of these methods do not capture the dependency between parameter and state uncertainties, they often fail to provide accurate uncertainty quantification. 

Beyond the aforementioned limitations of the existing posterior approximation-based methods for online parameter-state estimation, few of them provide theoretical guarantees on their approximation accuracy. However, theoretical foundations are essential for establishing trustworthiness, since numerical experiments cannot capture the full range of practical scenarios.

In summary, there is a clear need for an online approach that (i) provides accurate joint posterior approximations across diverse problem settings and scales, and (ii) is supported by theoretical guarantees of effectiveness. In this paper, we introduce a flexible and scalable variational inference framework to address the task of computing an approximate joint posterior online using noisy and incomplete data. We refer to this framework as \textit{factorization-based online variational inference} (FBOVI). FBOVI factorizes the approximate joint posterior $q_k(X_k,\theta)$ into a parameter marginal $\nu_k(\theta)$ and a conditional state distribution $\rho_k(X_k \mid \theta)$. Guided by a theorem establishing approximation error bounds, the framework proceeds in two stages: Stage~1 computes $\nu_k(\theta)$ via VI, while Stage~2 computes $\rho_k(X_k \mid \theta)$ using a filtering method based on $\nu_k(\theta)$. The family of $\nu_k(\theta)$ can be chosen flexibly, while $\rho_k(X_k \mid \theta)$ is represented as a Gaussian with mean and covariance depending on $\theta$, enabling the use of Gaussian filtering in Stage~2. Although this paper primarily focuses on Gaussian representations for $\rho_k(X_k \mid \theta)$ and using Gaussian filtering in Stage 2, our framework also supports non-Gaussian representations and filtering methods not in the family of Gaussian filters.

This work is an extension and expansion of our previous study~\cite{fbovi}, which also factorizes $q_k(X_k,\theta)$ and employs a two-stage procedure. The present work, however, introduces several key advances: (i) the framework supports a wide range of Gaussian filtering methods, including the EnKF, thus enhancing scalability; (ii) it provides a simplified procedure for linear systems by integrating the Kalman filter; (iii) it establishes theoretical results on approximation accuracy; and (iv) it validates the method with extensive numerical experiments, including cases with partially unknown observation models, model misspecification, and a high-dimensional system described by a partial derivative equation (PDE), whereas ~\cite{fbovi} focuses only on low-dimensional problems with correct model forms. 

The main contributions of this paper are summarized as follows:
\begin{enumerate}
\item We propose a flexible variational inference framework for online joint posterior approximation. By avoiding restrictive assumptions on the structure of the joint posterior, the framework enables higher approximation accuracy. A variety of Gaussian filtering methods (including EnKF) can be integrated into the proposed framework, which allows practitioners to balance inference accuracy and computational efficiency.
\item We establish a theoretical result (Theorem~\ref{thm:error_bound}) that provides an upper bound on the error between the true joint posterior and the approximate joint posterior. This result demonstrates the effectiveness of the factorization-based strategy and informs our algorithmic design. Building on this, we further derive Theorem~\ref{thm:error_bound_linear}, which gives estimable error bounds of our method for linear systems.
\item Numerical experiments show that the proposed method (i) achieves estimation and prediction performance comparable to joint PF in low-dimensional systems, successfully inferring unobserved states and parameters in partially unknown dynamical and observation models, (ii) exhibits robustness when noisy and incomplete data, model misspecification, and chaotic dynamics occur simultaneously, and (iii) scales well to a high-dimensional system arising from the spatial discretization of a convection--diffusion equation.
\end{enumerate}

The contributions of this work can support the advancement of emerging data-driven techniques that rely on online parameter–state estimation. For example, the proposed framework can assist DTs in improving monitoring and predictive performance, and providing reliable uncertainty quantification and theoretical guarantees that contribute to achieving trustworthiness.

The remainder of this paper is organized as follows. Section~\ref{sec:background} introduces the notation, formulates the online parameter-state estimation problem, and reviews the Kullback-Leibler divergence and the general framework of Gaussian filtering. Section~\ref{sec:methodology} presents the proposed factorization-based online variational inference framework. Section~\ref{sec:computational_complexity} analyzes the computational complexities of our framework. In Section~\ref{sec:experiments}, we evaluate the proposed method through numerical experiments and compare its performance with several online joint posterior approximation methods.  Finally, conclusions are given in Section~\ref{sec:conclusion}. 

\section{Background}
\label{sec:background}
In this section, we formally present the notation and problem formulation, followed by an introduction to Kullback-Leibler divergence. Lastly, we introduce the fundamental concepts of Gaussian filtering.

\subsection{Notation}
Let $\mathbb{R}$ denote the set of real numbers and $\mathbb{Z}_{\geq 0}$ the set of non-negative integers. For a square matrix $M$, let $\lvert M \rvert$ denote its determinant. Let the triple $(\Omega,\mathcal{F},\mathbb{P})$ denote a probability space, where $\mathbb{P}$ is a probability measure with density $\pdense$. The notation $p(X = x \mid Y = y)$ denotes the value of the probability density function at $x$, conditioned on the event that the random variable $Y$ takes the value $y$. For simplicity, we use $p(x \mid y)$ to represent $p(X = x \mid Y = y)$ throughout this paper. Let $\E[X]$, $\Var[X]$, and $\Cov[X, Y]$ denote the expectation and variance of the random variable $X$, and the covariance between the random variables $X$ and $Y$, respectively. Similarly, let $\E[X \mid Z]$, $\Var[X \mid Z]$, and $\Cov[X, Y \mid Z]$ denote, respectively, the expectation and variance of $X$, and the covariance between $X$ and $Y$, conditioned on $Z$.

We use $\mathcal{N}(\mu,R)$ to denote a Gaussian distribution with mean $\mu$ and covariance $R$. The value of the probability density function of this distribution at $x$ is written as $\pdense_\mathcal{N}(x;\mu,R)$, that is,
\[
\pdense_\mathcal{N}(x;\mu,R) := \frac{1}{(2\pi)^{d_x/2}\lvert R \rvert^{1/2}} 
\exp\!\left(-\tfrac{1}{2}(x-\mu)^\top R^{-1}(x-\mu)\right),
\]
where $d_x$ is the dimension of $x$. 

\subsection{Problem formulation}
\label{sec:problem}
Consider the following discrete-time system with partially observable states:
\begin{equation}\label{sys:general}
\begin{split}
X_{k+1} &= \Phi(X_k;\theta)+W_k, \quad W_k \overset{\mathrm{i.i.d.}}{\sim} \mathcal{N}(0,\Sigma(\theta)), \\
Y_k &= h(X_k;\theta)+V_k, \quad V_k \overset{\mathrm{i.i.d.}}{\sim} \mathcal{N}(0,\Gamma(\theta)),
\end{split}
\end{equation}
where $k \in \mathbb{Z}_{\geq 0}$ denotes a time index; and $X_k \in \reals^n$, $Y_k \in \reals^r$, $W_k \in \reals^n$, and $V_k \in \reals^r$ represent the state, observation, process noise, and measurement noise at time step $k$, respectively. Here $\theta$ represents the vector of unknown system parameters. The functions $\Phi(X_k;\theta)$ and $h(X_k;\theta)$ are parameterized by the parameters $\theta$. The process noise covariance $\Sigma(\theta)$ and the measurement noise covariance $\Gamma(\theta)$ are determined by $\theta$. The quantities $X_k$, $Y_k$, $W_k$, $V_k$, and $\theta$ are random variables.

Let $y_k$ denote the observed value of $Y_k$ and let $\Y_k = (y_1,\ldots,y_k)$ represent the set of all observations obtained up to time $k$. The \textit{joint posterior} at time step $k$ is defined as the joint distribution of state $X_k$ and parameters $\theta$ conditioned on $\Y_k$, represented by the probability density $\pdense(X_k,\theta \mid \Y_k)$. The goal is to compute the joint posterior $\pdense(X_k,\theta \mid \Y_k)$ at each time step $k$, using only the previous joint posterior $\pdense(X_{k-1},\theta \mid \Y_{k-1})$ and the newly received data point $y_k$, without reprocessing all historical data $\Y_{k-1}$. 

\begin{figure}[htbp]
    \centering
    \includegraphics[width=0.6\linewidth]{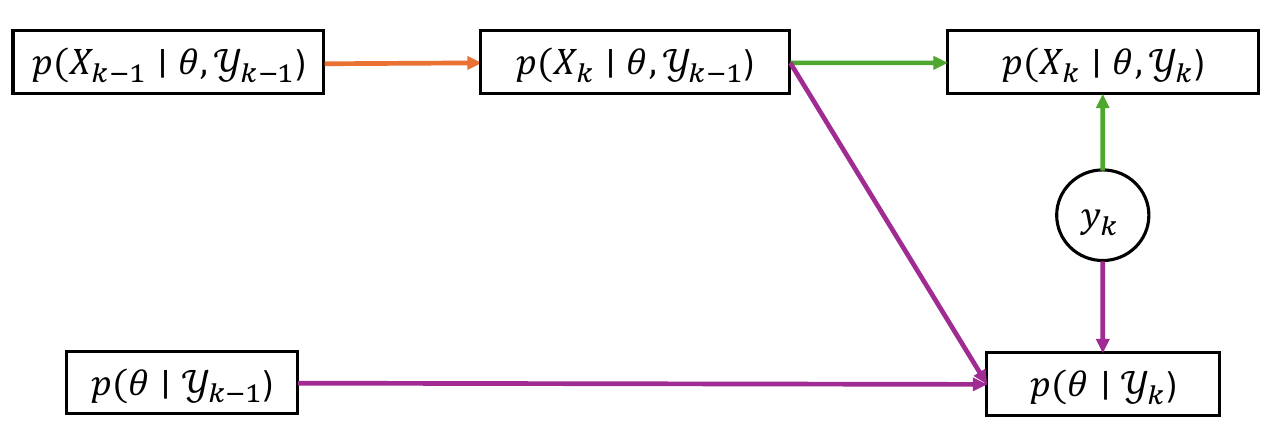}
    \caption{Illustration of the prediction–update procedure for determining the conditional filtering distribution $p(X_k \mid \theta, \Y_k)$ and the parameter posterior $p(\theta \mid \Y_k)$. The predictive distribution $p(X_k \mid \theta, \Y_{k-1})$ is obtained by propagating the previous conditional filtering distribution $p(X_{k-1} \mid \theta, \Y_{k-1})$ through the dynamics. The data $y_k$ is then assimilated to update the predictive distribution, yielding $p(X_k \mid \theta, \Y_k)$. The predictive distribution $p(X_k \mid \theta, \Y_{k-1})$ and the previous parameter posterior $\pdense(\theta \mid \Y_{k-1})$ are used for determining the current parameter posterior $p(\theta \mid \Y_k)$.}
    \label{fig:pred_update}
\end{figure}

Theoretically, the expression of the current joint posterior $\pdense(X_k,\theta \mid \Y_k)$ in terms of the previous joint posterior $\pdense(X_{k-1},\theta \mid \Y_{k-1})$ and the newly received data $y_k$ can be derived through a prediction–update procedure. Specifically, at each time step $k$, the joint posterior satisfies
\begin{equation*}
\pdense(X_k,\theta \mid \Y_k) = \pdense(X_k \mid \theta, \Y_k)\,\pdense(\theta \mid \Y_k),
\end{equation*}
where $\pdense(X_k \mid \theta, \Y_k)$ denotes the filtering distribution conditioned on the parameters $\theta$, and $\pdense(\theta \mid \Y_k)$ represents the posterior of $\theta$. During the prediction step, the predictive distribution $p(X_k \mid \theta, \Y_{k-1})$ is obtained by propagating the previous conditional filtering distribution $p(X_{k-1} \mid \theta, \Y_{k-1})$ through the system dynamics of~\eqref{sys:general}. During the update step, the newly observed data $y_k$ is assimilated to update both the predictive distribution $p(X_k \mid \theta, \Y_{k-1})$ and the previous parameter posterior $\pdense(\theta \mid \Y_{k-1})$, yielding the conditional filtering distribution $\pdense(X_k \mid \theta, \Y_k)$ and the updated parameter posterior $\pdense(\theta \mid \Y_k)$~\cite{fbovi}. This prediction–update mechanism is illustrated in Figure~\ref{fig:pred_update}. The subsequent subsections provide a detailed description of each step.

\subsubsection{Prediction step}
\label{sec:pred_ana}
The predictive distribution $p(X_k \mid \theta, \Y_{k-1})$ is obtained by propagating the previous conditional filtering distribution forward through the dynamics:
\begin{align}
p(X_k \mid \theta, \Y_{k-1})
&= \int \pdense(X_k, X_{k-1} \mid \theta, \Y_{k-1})\,dX_{k-1} \notag\\
&= \int \pdense_{\mathcal{N}}\!\left(X_k; \Phi(X_{k-1};\theta), \Sigma(\theta)\right)\pdense(X_{k-1} \mid \theta, \Y_{k-1})\,dX_{k-1}.
\end{align}
Let $\mathcal{T}$ denote the operator that maps the previous conditional filtering distribution $\pdense(X_{k-1} \mid \theta, \Y_{k-1})$ to the predictive distribution $\pdense(X_k \mid \theta, \Y_{k-1})$. We refer to $\mathcal{T}$ as the \textit{prediction operator}. For example, the distribution $\pdense(X_k \mid \theta, \Y_{k-1})$ can be represented as $\pdense(X_k \mid \theta, \Y_{k-1}) = \mathcal{T}p(X_{k-1} \mid \theta, \Y_{k-1})$.

\subsubsection{Update step}
Given the predictive distribution $\pdense(X_k \mid \theta, \Y_{k-1})$, we now derive the posterior distributions $\pdense(X_k \mid \theta, \Y_k)$ and $\pdense(\theta \mid \Y_k)$.  By Bayes’ rule, the conditional filtering distribution is
\begin{equation}
\pdense(X_k \mid \theta, \Y_k)
= \frac{1}{Z_{X,k}(\theta)}\,\pdense_{\mathcal{N}}\!\left(y_k; h(X_k;\theta), \Gamma(\theta)\right)
\pdense(X_k \mid \theta, \Y_{k-1}),
\end{equation}
where
\[
Z_{X,k}(\theta)
= \int \pdense_{\mathcal{N}}\!\left(y_k; h(X_k;\theta), \Gamma(\theta)\right)
\pdense(X_k \mid \theta, \Y_{k-1})\,dX_k
\]
is a normalization constant for a given $\theta$.  
Let $\mathcal{F}_k^X$ denote the operator mapping $\pdense(X_{k-1} \mid \theta, \Y_{k-1})$ to $\pdense(X_k \mid \theta, \Y_k)$. Similarly, by Bayes’ rule, the parameter posterior can be written as
\begin{equation}
\pdense(\theta \mid \Y_k)
= \frac{1}{Z_k}\,\pdense(\theta \mid \Y_{k-1})\,p(y_k \mid \theta, \Y_{k-1})
= \frac{1}{Z_k}\,\pdense(\theta \mid \Y_{k-1})
\int \pdense_{\mathcal{N}}\!\left(y_k; h(X_k;\theta), \Gamma(\theta)\right)
\pdense(X_k \mid \theta, \Y_{k-1})\,dX_k,
\end{equation}
where
\[
Z_k
= \int \pdense(\theta \mid \Y_{k-1})
\left (\int \pdense_{\mathcal{N}}\!\left(y_k; h(X_k;\theta), \Gamma(\theta)\right)
\pdense(X_k \mid \theta, \Y_{k-1})\,dX_k\right )d\theta
\]
is a normalizing constant.  
Let $\mathcal{F}_k$ denote the operator that maps the pair $\big(\pdense(X_{k-1} \mid \theta, \Y_{k-1}),\, \pdense(\theta \mid \Y_{k-1})\big)$ to the updated parameter posterior $\pdense(\theta \mid \Y_k)$.

\subsection{Kullback-Leibler divergence}
\label{sec:metrics_dist}
This section introduces the Kullback–Leibler (KL) divergence, which is employed in our framework described in detail in Section~\ref{sec:methodology}. Let $\mu_1$ and $\mu_2$ be two probability measures with corresponding probability density functions $f_1(x)$ and $f_2(x)$. The KL divergence of $\mu_1$ relative to $\mu_2$ is defined as:
\begin{equation}
\KL(\mu_1 \lVert \mu_2) := \int f_1(x)\log \frac{f_1(x)}{f_2(x)}dx.
\end{equation}

\subsection{Gaussian filtering}
\label{sec:gf_intro}
Gaussian filtering will be employed in the proposed framework. In this section, we present the fundamental concept of Gaussian filtering. 
Gaussian filtering assumes that the joint distribution of the current state $X_k$ and the current observation $Y_k$, conditioned on past observations $\Y_{k-1}$, namely, $\pdense \left(X_k,Y_k \mid \Y_{k-1}\right)$, is Gaussian. Under this assumption, the posterior $p(X_k \mid \Y_k)$ is also Gaussian and can be written as
\begin{equation*}
p(X_k \mid \Y_k) = p_{\mathcal{N}}(X_k; m_k, C_k),
\end{equation*}
with mean $m_k$ and covariance $C_k$ given by
\begin{align}
m_k &= m_k^- + U_k S_k^{-1}(y_k - \mu_k), \qquad 
C_k = C_k^- - U_k S_k^{-1} U_k^\top,
\label{eq:post_mean_cov}
\end{align}
where $m_k^-$, $C_k^-$, $\mu_k$, $U_k$, and $S_k$ are defined as
\begin{equation}\label{eq:GF_eqs}
\begin{split}
m_k^- &:= \E[X_k \mid \Y_{k-1}], \quad 
C_k^- := \Var[X_k \mid \Y_{k-1}], \quad 
\mu_k := \E[Y_k \mid \Y_{k-1}], \\
U_k &:= \Cov[X_k,Y_k \mid \Y_{k-1}], \quad 
S_k := \Var[Y_k \mid \Y_{k-1}].
\end{split}
\end{equation}
Various Gaussian filtering methods differ in how they evaluate the quantities $m_k^-$, $C_k^-$, $\mu_k$, $U_k$, and $S_k$ \cite{bayesian_filter}. Commonly used Gaussian filters include the Kalman filter (KF), extended Kalman filter (ExKF), unscented Kalman filter (UKF), Gauss–Hermite Kalman filter (GHKF), and ensemble Kalman filter (EnKF).

\section{Methodology}
\label{sec:methodology}
In this section, we introduce our proposed framework, FBOVI, which aims to obtain the joint posterior $p(X_k,\theta \mid \Y_k)$ in an online fashion.

As noted in Section~\ref{sec:introduction}, computing the joint posterior $\pdense(X_k,\theta \mid \Y_k)$ is typically intractable. We therefore approximate $\pdense(X_k,\theta \mid \Y_k)$ at each time step $k$ with another distribution $q_k(X_k,\theta)$, referred to as the \emph{approximate joint posterior}. The approximation is computed recursively: $q_k(X_k,\theta)$ is obtained using only the newly received data $y_k$ and the previous approximation $q_{k-1}(X_{k-1},\theta)$. We construct $q_k$ as the product of a conditional distribution of $X_k$ given $\theta$ and a marginal distribution over $\theta$:
\begin{equation*}
q_k(X_k,\theta) := \rho_k(X_k \mid \theta)\,\nu_k(\theta).
\end{equation*}
In principle, one could approximate the joint posterior $\pdense(X_k,\theta \mid \Y_k)=\pdense(X_k \mid \theta,\Y_k)\,\pdense(\theta \mid \Y_k)$ by directly approximating $\pdense(X_k \mid \theta,\Y_k)$ with $\rho_k(X_k \mid \theta)$ and $\pdense(\theta \mid \Y_k)$ with $\nu_k(\theta)$. However, this direct approximation requires the previous conditional filtering distribution $\pdense(X_{k-1}\mid \theta,\Y_{k-1})$ and the previous parameter posterior $\pdense(\theta \mid \Y_{k-1})$, both of which are unavailable. Therefore, such a direct approximation is infeasible.

\begin{figure}[htbp]
    \centering
    \includegraphics[width=0.8\linewidth]{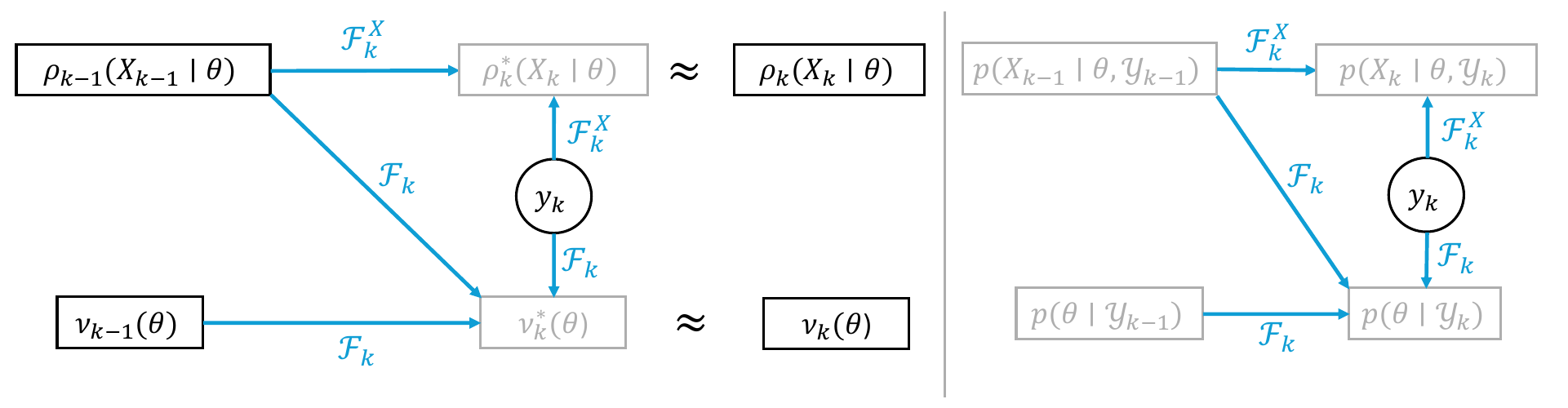}
    \caption{Core idea of the proposed framework: the online conditional filtering distribution $\rho_k^*(X_k \mid \theta)$ is approximated by $\rho_k(X_k \mid \theta)$, and the online parameter posterior $\nu_k^*(\theta)$ is approximated by $\nu_k(\theta)$. Their product, $\rho_k(X_k \mid \theta)\nu_k(\theta)$, forms the approximate joint posterior $q_k(X_k,\theta)$. The blue arrows represent the maps $\mathcal{F}_k^X$ and $\mathcal{F}_k$. For comparison, the transformations of the exact solutions are shown on the right. The distributions shown in gray are those that are intractable to compute.}
    \label{fig:strategy}
\end{figure}

Our solution to this issue is to introduce ``online targets". We let the distribution $\rho_k(X_k \mid \theta)$ approximate the \emph{online conditional filtering distribution} $\rho_k^*(X_k \mid \theta)$ and let $\nu_k(\theta)$ approximate the \emph{online parameter posterior} $\nu^*_k(\theta)$. The online conditional filtering distribution $\rho_k^*(X_k \mid \theta)$ is the resulting $\pdense(X_k \mid \theta, \Y_k)$ obtained by replacing the previous conditional filtering distribution $\pdense(X_{k-1} \mid \theta, \Y_{k-1})$ with $\rho_{k-1}(X_{k-1} \mid \theta)$. The online parameter posterior $\nu_k^*(\theta)$ represents the resulting $\pdense(\theta \mid \Y_k)$ when the distributions $\pdense(X_{k-1} \mid \theta, \Y_{k-1})$ and $\pdense(\theta \mid \Y_{k-1})$ are replaced by $\rho_{k-1}(X_{k-1} \mid \theta)$ and $\nu_{k-1}(\theta)$, respectively. Formally, $\rho^*_k$ and $\nu_k^*$ are defined as
\begin{equation*}
\rho_k^* := \mathcal{F}_k^X \rho_{k-1}, \qquad 
\nu_k^* := \mathcal{F}_k(\rho_{k-1},\nu_{k-1}),
\end{equation*}
where $\mathcal{F}_k^X$ and $\mathcal{F}_k$ are the operators introduced in Section~\ref{sec:problem}. Generally, the distributions $\rho_k^*(X_k \mid \theta)$ and $\nu_k^*(\theta)$ cannot be computed analytically. The core idea of our framework is therefore to use $\rho_k(X_k \mid \theta)$ to approximate $\rho_k^*(X_k \mid \theta)$, and $\nu_k(\theta)$ to approximate $\nu_k^*(\theta)$. This procedure is illustrated in Figure~\ref{fig:strategy}.

As \( \rho_k^*(X_k \mid \theta) \) and \( \nu_k^*(\theta) \) normally differ from \(p(X_k \mid \theta, \Y_k)\) and \(p(\theta \mid \Y_k)\), respectively, one may ask whether approximating these online targets is sufficient to effectively approximate the true joint posterior $\pdense(X_k,\theta \mid \Y_k)$. The following theorem, which establishes an upper bound on the error between the true and approximate joint posteriors, demonstrate the effectiveness of our strategy and offers principled guide for algorithmic design:
\begin{theorem}
\label{thm:error_bound}
Assume $\inf_{\theta} \big \lvert \Gamma(\theta) \big \rvert = \tilde{C} > 0$. 
The total variation and Hellinger distances between $p(X_k,\theta \mid \Y_k)$ and $q_k(X_k,\theta)$ satisfy:
\begin{align}
d(p(X_k,\theta \mid \Y_k), q_k(X_k,\theta)) &\leq \sum_{j=1}^{k-1}C(\Y_k,\tilde{C},j,k)\sqrt{\E_{\theta \sim \nu_j(\cdot)}\left[\KL\left(\rho_j(\cdot \mid \theta) \lVert \rho_j^*(\cdot \mid \theta) \right) \right] +\KL\left(\nu_j \lVert \nu_j^*\right)} \nonumber \\
&\qquad  + \frac{1}{\sqrt{2}} \sqrt{\E_{\theta \sim \nu_k(\cdot)}\left[\KL\left(\rho_k(\cdot \mid \theta) \lVert \rho_k^*(\cdot \mid \theta) \right) \right] +\KL\left(\nu_k \lVert \nu_k^*\right)},
\label{ieq:learning_error_bound_tv}
\end{align}
where $C(\Y_k,\tilde{C},j,k)=\frac{(2\pi)^{-\frac{r(k-j)}{2}}\tilde{C}^{-\frac{k-j}{2}}}{\sqrt{2}p(\Y_{j+1:k}\mid\Y_j)}$ when $d$ is the total variation distance, and $C(\Y_k,\tilde{C},j,k)=\frac{2^{(k-j)}(2\pi)^{-\frac{r(k-j)}{4}}\tilde{C}^{-\frac{k-j}{4}}}{\sqrt{2p(\Y_{j+1:k}\mid\Y_j)}}$ when $d$ is the Hellinger distance. Here the notation $\Y_{j+1:k}$ denotes the data received from time step $j+1$ through $k$.
\end{theorem}
The total variation (TV) and Hellinger distances are two commonly used metrics for measuring the difference between two probability distributions. Their definitions are provided in~\ref{app:metrics_dist}. 

\begin{proof}
See~\ref{app:proof_thm}.
\end{proof}

The terms $C(\Y_k,\tilde{C},j,k)$ depend only on the observational data and model quantities and are thus constant with respect to the choices for $\rho_j$ and $\nu_j$ for $j \in [1,k]$. The controllable terms in the bounds are the divergences
\[
\E_{\theta \sim \nu_j(\cdot)}\!\left[\KL\!\left(\rho_j(\cdot \mid \theta)\,\big\|\, \rho_j^*(\cdot \mid \theta)\right)\right]
\quad \text{and} \quad
\KL\!\left(\nu_j \,\big\|\, \nu_j^*\right), \qquad j=1,\dots,k,
\]
which measure how well the approximate conditional filtering distribution $\rho_j(X_j \mid \theta)$ and the approximate parameter posterior $\nu_j(\theta)$ track their online targets. Hence, to reduce the error between the true and approximate joint posteriors, we adopt the following two-stage strategy at each time step $k$:
\begin{enumerate}
    \item \textbf{Minimize} \( \KL(\nu_k  \| \nu_k^*) \) to obtain \( \nu_k(\theta) \).
    \item \textbf{Minimize} \( \E_{\theta \sim \nu_k(\cdot)}[\KL(\rho_k(\cdot \mid \theta) \| \rho_k^*(\cdot\mid \theta))] \) to obtain \( \rho_k(X_k \mid \theta) \), using the computed \( \nu_k(\theta) \).
\end{enumerate}
We allow \( \nu_k(\theta) \) to be chosen from an arbitrary variational family, while \( \rho_k(X_k \mid \theta) \) is modeled as a conditional Gaussian:
\[
\rho_k(X_k \mid \theta) = \pdense_{\mathcal{N}}(X_k; m_k(\theta), C_k(\theta)).
\]
We now describe how each of these two stages is performed in detail.

\subsection{Stage 1: computation of $\nu_k(\theta)$}
\label{sec:step1}
To compute $\nu_k(\theta)$ by minimizing $\KL(\nu_k \Vert \nu_k^*)$, we first derive the expression of $\nu_k^*(\theta)$. 
Let $\rho_k^{*-}(X_k \mid \theta)$ denote the corresponding predictive distribution $p(X_k \mid \theta, \Y_{k-1})$ when the previous conditional filtering distribution $\pdense(X_{k-1} \mid \theta, \Y_{k-1})$ is replaced with $\rho_{k-1}(X_{k-1} \mid \theta)$, i.e., $\rho_k^{*-}:=\mathcal{T}\rho_{k-1}$. We can express $\rho_k^{*-}(X_k \mid \theta)$ as 
\begin{equation*} \rho_k^{*-}(X_k \mid \theta) := \int \pdense_{\mathcal{N}}(X_k; \Phi(X_{k-1};\theta), \Sigma(\theta))\rho_{k-1}(X_{k-1} \mid \theta)dX_{k-1}. \end{equation*} 
We refer to $\rho_k^{*-}(X_k \mid \theta)$ as the \emph{online conditional predictive distribution}. Throughout this paper, the superscript ‘$-$’ indicates quantities associated with predictive distributions.

By Bayes' rule, the online parameter posterior $\nu_k^*(\theta)$ is given by:
\begin{equation}
\nu_k^*(\theta) = \frac{1}{\hat{Z}_{k}} \nu_{k-1}(\theta) \int \pdense_\mathcal{N}(y_k; h(X_k;\theta), \Gamma(\theta))\, \rho_k^{*-}(X_k \mid \theta)\, dX_k,
\label{eq:nu_star}
\end{equation}
where \( \hat{Z}_{k} \) is a normalizing constant.
Given Equation~\eqref{eq:nu_star}, the KL divergence $\KL(\nu_k  \| \nu_k^*)$ can be written as
\begin{equation}
\KL(\nu_k  \| \nu_k^*) = \log \hat{Z}_{k} - \left( \E_{\theta \sim \nu_k(\cdot)} \left[\log \int \pdense_\mathcal{N}(y_k; h(X_k;\theta), \Gamma(\theta))\, \rho_k^{*-}(X_k \mid \theta)\, dX_k \right]- \KL(\nu_k \| \nu_{k-1})\right).
\label{eq:kl_step1}
\end{equation}
Since the first term on the right-hand side of Equation~\eqref{eq:kl_step1}, $\log \hat{Z}_{k}$, is constant, VI minimizes the KL divergence $\KL(\nu_k \| \nu_k^*)$ by maximizing the negative of the remaining terms on the right-hand side of Equation~\eqref{eq:kl_step1}, which is defined as the evidence lower bound (ELBO) $\mathcal{L}_k(\nu_k)$:
\begin{equation}
\mathcal{L}_k(\nu_k) := \E_{\theta \sim \nu_k(\cdot)}\!\left[\log I(\theta)\right] - \KL\!\left(\nu_k\,\|\, \nu_{k-1}\right),
\label{eq:elbo_step1}
\end{equation}
where 
\begin{equation*}
I(\theta) := \int \pdense_\mathcal{N}(y_k; h(X_k;\theta), \Gamma(\theta))\, \rho_k^{*-}(X_k \mid \theta)\, dX_k.
\end{equation*}
In the following subsections, we describe how the integral $I(\theta)$ can be computed for both linear and nonlinear systems.

\subsubsection{Linear case}
\begin{figure}
     \centering
     \includegraphics[width=0.6\linewidth]{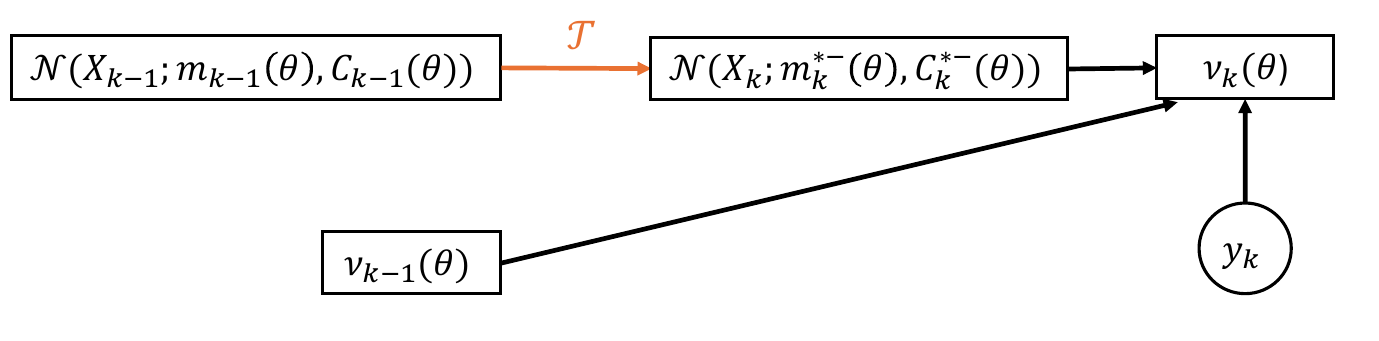}
     \caption{Computation of the distribution $\nu_k(\theta)$ for linear systems. The predictive distribution $\rho_k^{*-}(X_k \mid \theta)$ can be computed in the linear case. The prediction operator $\mathcal{T}$ is introduced in Section~\ref{sec:pred_ana}. It represents the operator that maps the previous conditional filtering distribution $\pdense(X_{k-1} \mid \theta, \Y_{k-1})$ to the predictive distribution $\pdense(X_k \mid \theta, \Y_{k-1})$. }
    \label{fig:stage1_linear}
\end{figure}

If the dynamical and observation models of system~\eqref{sys:general} are linear in the state $X_k$ and have the following forms
\begin{equation}\label{sys:linear}
\begin{split}
    X_{k+1} &= A(\theta)X_k + W_k, \qquad W_k \overset{\mathrm{i.i.d.}}{\sim} \mathcal{N}(0,\Sigma(\theta)), \\
    Y_k &= H(\theta)X_k + V_k, \qquad V_k \overset{\mathrm{i.i.d.}}{\sim} \mathcal{N}(0,\Gamma(\theta)),
\end{split}
\end{equation}
where $A(\theta) \in \mathbb{R}^{n \times n}$ and $H(\theta) \in \mathbb{R}^{r \times n}$ are matrices determined by the parameters $\theta$, the online conditional predictive distribution $\rho^{*-}_k$ is Gaussian and given by
\begin{equation*}
\rho^{*-}_k(X_k \mid \theta) =  \pdense_{\mathcal{N}}\left(X_k; m_k^{*-}(\theta), C_k^{*-}(\theta)\right),
\end{equation*}
where
\begin{equation}
m_k^{*-}(\theta) = A(\theta)m_{k-1}(\theta), \qquad C_k^{*-}(\theta)=A(\theta)C_{k-1}(\theta)A(\theta)^T+\Sigma(\theta).
\label{eq:mean_cov_pred_linear}
\end{equation}
Thus the integral $I(\theta)$ has a closed-form expression:
\begin{align}
&I(\theta) = \pdense_{\mathcal{N}}\left(y_k;H(\theta)m_k^{*-}(\theta), H(\theta)C_k^{*-}(\theta)H(\theta)^T+\Gamma(\theta)\right).
\end{align}
The procedure of computing the distribution $\nu_k(\theta)$ for the linear case is illustrated in Figure~\ref{fig:stage1_linear}. 

\subsubsection{Nonlinear case}
When the dynamical model is nonlinear in the state variable $X_k$, closed-form expressions of the online conditional predictive distribution $\rho_k^{*-}(X_k \mid \theta)$ are generally unavailable, rendering the computation of the ELBO $\mathcal{L}_k(\nu_k)$ intractable. In such cases, we approximate $\rho_k^{*-}(X_k \mid \theta)$ with a Gaussian distribution:
\begin{equation}
\rho_k^{*-}(X_k \mid \theta) \approx \pdense_{\mathcal{N}}(X_k; m_k^-(\theta), C_k^-(\theta)),
\label{eq:predictive_nonlinear}
\end{equation}
where the mean $m_k^-(\theta)$ and the covariance $m_k^-(\theta)$ are defined as
\begin{align}
m_k^-(\theta) &:= \int \Phi(X_{k-1};\theta)\pdense_{\mathcal{N}}(X_{k-1}; m_{k-1}(\theta), C_{k-1}(\theta))dX_{k-1},
\label{eq:int_pred_mean} \\
C_k^-(\theta) &:= \int \left(\Phi(X_{k-1};\theta)-m_k^-(\theta)\right)\left(\Phi(X_{k-1};\theta)-m_k^-(\theta)\right)^T\pdense_{\mathcal{N}}(X_{k-1}; m_{k-1}(\theta), C_{k-1}(\theta))dX_{k-1}+\Sigma(\theta).
\label{eq:int_pred_cov}
\end{align}
The integrals in Equations~\eqref{eq:int_pred_mean} and~\eqref{eq:int_pred_cov} can be evaluated using the unscented transform or Gaussian quadrature~\cite{bayesian_filter}.

\begin{figure}
     \centering
     \includegraphics[width=0.7\linewidth]{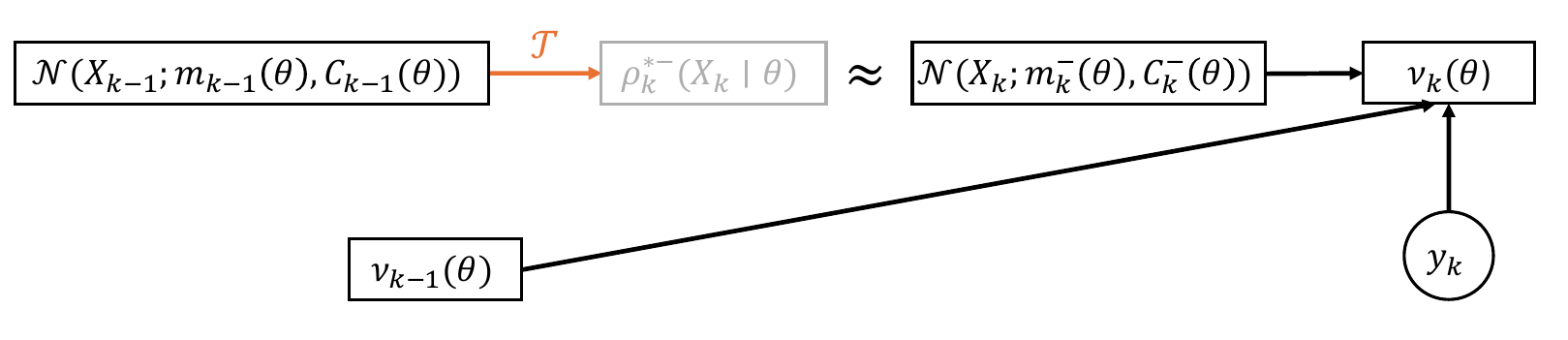}
     \caption{Computation of the distribution $\nu_k(\theta)$ for nonlinear systems. The predictive distribution $\rho_k^{*-}(X_k \mid \theta)$ is intractable to compute and thus approximated by a Gaussian distribution $\mathcal{N}(X_k;m_k^-(\theta),C_k^-(\theta))$ first.}
    \label{fig:stage1_nonlinear}
\end{figure}

With the Gaussian approximation $\pdense_{\mathcal{N}}\left(X_k;m_k^-(\theta), C_k^-(\theta)\right)$ for $\rho_{k}^{*-}(X_{k} \mid \theta)$, we approximate the integral $I(\theta)$ by
\begin{equation}
I(\theta) \approx \int \pdense_{\mathcal{N}}(y_k;h(X_k;\theta),\Gamma(\theta))\pdense_{\mathcal{N}}\left(X_k;m_k^-(\theta), C_k^-(\theta)\right)dX_k.
\label{eq:int_approx}
\end{equation}
The right-hand side of Equation~\eqref{eq:int_approx} can be evaluated using the unscented transform, Gaussian quadrature, or the approximation
\[
\int \pdense_{\mathcal{N}}(y_k;h(X_k;\theta),\Gamma(\theta))
\pdense_{\mathcal{N}}\left(X_k;m_k^-(\theta), C_k^-(\theta)\right)dX_k
\approx
\pdense_{\mathcal{N}}(y_k;\mu_k(\theta), S_k(\theta)),
\]
where $\mu_k(\theta)$ and $S_k(\theta)$ are defined in Equation~\eqref{eq:GF_eqs} and can be computed by Gaussian filtering methods for a given $\theta$. In addition, when the observation model is linear in the state $X_k$:
\begin{equation}
Y_k = H(\theta)X_k + V_k, \qquad V_k \overset{\mathrm{i.i.d.}}{\sim} \mathcal{N}(0,\Gamma(\theta)),
\label{eq:obs_linear}
\end{equation}
the right hand side of Equation~\eqref{eq:int_approx} has a closed-form expression:
\begin{align}
\int \pdense_{\mathcal{N}}(y_k;h(X_k;\theta),\Gamma(\theta))\rho_k^{*-}(X_k \mid \theta)dX_k &= \pdense_{\mathcal{N}}\left(y_k;H(\theta)m_k^-(\theta), H(\theta)C_k^-(\theta)H(\theta)^T+\Gamma(\theta)\right).
\end{align}
The procedure of computing the distribution $\nu_k(\theta)$ for the nonlinear case is illustrated in Figure~\ref{fig:stage1_nonlinear}.
The choice of approximating the distribution $\rho^{*-}_k(X_k \mid \theta)$ with a Gaussian distribution is motivated by the overall computational efficiency of the proposed method. If we don't choose to approximate the distribution $\rho^{*-}_k(X_k \mid \theta)$ with a Gaussian, we can still estimate the integral $I(\theta)$ using the Monte Carlo method:
\begin{equation}
I(\theta) \approx \frac{1}{N}\sum_{i=1}^N\pdense_{\mathcal{N}}\left(y_k;h(X_k^{(i)};\theta),\Gamma(\theta)\right),
\end{equation}
where each $X_k^{(i)}$ is a sample from the distribution $\rho_k^{*-}(X_k \mid \theta)$, which is constructed by
\begin{equation*}
X_k^{(i)} = \Phi(X_{k-1}^{(i)};\theta) + \eta^{(i)}, \qquad X_{k-1}^{(i)} \sim \rho_{k-1}(X_{k-1} \mid \theta), \quad \eta^{(i)} \sim \Sigma(\theta).
\end{equation*}
When the integral $I(\theta)$ can be evaluated, we can compute the distribution $\nu_k(\theta)$ by maximizing the ELBO $\mathcal{L}_k(\nu_k)$. Stage 1 ends with $\nu_k(\theta)$ obtained and we proceed to Stage 2.

\subsection{Stage 2: computation of \( \rho_k(X_k \mid \theta) \)}
\label{sec:step2}
With $\nu_k(\theta)$ computed, we focus on computing the conditional distribution \( \rho_k(X_k \mid \theta) \) that minimizes
\[
\E_{\theta \sim \nu_k(\cdot)}\left[ \KL\!\left(\rho_k(\cdot\mid \theta) \,\|\, \rho_k^*(\cdot \mid \theta) \right) \right].
\]
In the following two subsections, we describe how \( \rho_k(X_k \mid \theta) \) is computed for both linear and nonlinear systems, given the marginal distribution $\nu_k(\theta)$.

\subsubsection{Linear case}
When the dynamical and observation models are linear and described by Equation~\eqref{sys:linear}, the online conditional filtering distribution \( \rho_k^*(X_k \mid \theta) \) is Gaussian:
\[
\rho_k^*(X_k \mid \theta) = \mathcal{N}\left(X_k; m_k^*(\theta), C_k^*(\theta)\right),
\]
with the mean \( m_k^*(\theta) \) and covariance \( C_k^*(\theta) \) given by the Kalman filter:
\begin{align}
m_k^*(\theta) =& m_k^{*-}(\theta)+C_k^{*-}(\theta)H(\theta)^T\left(H(\theta)C_k^{*-}(\theta)H(\theta)^T+\Gamma(\theta)\right)^{-1}\left(y_k-H(\theta)m_k^{*-}(\theta)\right), 
\label{eq:kf_mean_post} \\
C_k^*(\theta) =& C_k^{*-}(\theta) - C_k^{*-}(\theta)H(\theta)^T\left(H(\theta)C_k^{*-}(\theta)H(\theta)^T+\Gamma(\theta)\right)^{-1}H(\theta)C_k^{*-}(\theta),
\label{eq:kf_var_post}
\end{align}
where $m_k^{*-}(\theta)$ and $C_k^{*-}(\theta)$ are defined in Equation~\eqref{eq:mean_cov_pred_linear}.

\begin{figure}
     \centering
     \includegraphics[width=0.7\linewidth]{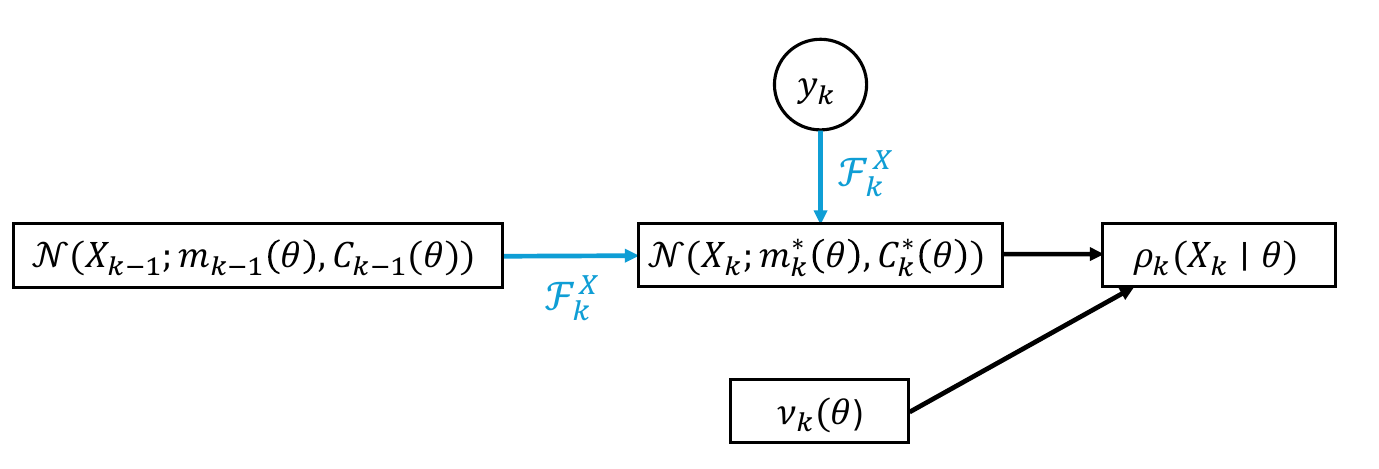}
     \caption{Procedure of computing the distribution $\rho_k(X_k \mid \theta)$ for linear systems. The online conditional filtering distribution $\rho_k^*(X_k \mid \theta)$ is a Gaussian distribution $\mathcal{N}(X_k;m_k^*(\theta),C_k^*(\theta))$ given by the Kalman filter. We compute the distribution $\rho_k(X_k \mid \theta) = \mathcal{N}(X_k ; m_k(\theta), C_k(\theta))$ using the target mean function $m_k^*(\theta)$, the target covariance function $C_k^*(\theta)$, and the distribution $\nu_k(\theta)$ which is inferred at Stage 1.}
    \label{fig:stage2_linear}
\end{figure}

While setting $m_k(\theta) = m_k^*(\theta)$ and $C_k(\theta) = C_k^*(\theta)$ makes the KL divergence $\KL(\rho_k(\cdot \mid \theta) \|\ \rho_k^*(\cdot \mid \theta))$ equal to $0$, this is computationally infeasible in practice. As shown in Equations~\eqref{eq:kf_mean_post} and~\eqref{eq:kf_var_post}, the complexities of \( m_k^*(\theta) \) and \( C_k^*(\theta) \) increase over time due to recursive dependence on past distributions. To maintain constant computational and storage costs over time, we represent $m_k(\theta)$ and $C_k(\theta)$ using neural networks\footnote{Throughout the paper, when we state that a neural network is used to represent the covariance function $C_k(\theta)$, we in practice use a neural network to represent a lower triangular matrix $c_k(\theta)$. Specifically, this neural network outputs a vector containing all entries in the lower triangular part of $c_k(\theta)$. This vector is then reshaped into the lower triangular matrix $c_k(\theta)$, with the entries in the strictly upper triangular part set to zero. The covariance matrix $C_k(\theta)$ is then represented as
$C_k(\theta)=c_k(\theta)c_k(\theta)^T$. This construction ensures that the resulting matrix $C_k(\theta)$ is positive semidefinite.} whose architectures are fixed across time steps $k$. This design ensures that the per-step inference cost does not grow with time, while leveraging the universal approximation capability of neural networks to reduce error. The procedure of computing the distribution $\rho_k(X_k \mid \theta)$ for the linear case is illustrated in Figure~\ref{fig:stage2_linear}.

The following theorem demonstrates the joint posterior approximation accuracy of our method, establishing an upper bound on the error between the true joint posterior $p(X_k,\theta \mid \Y_k)$ and the approximate joint posterior $q_k(X_k,\theta)$ obtained by our method, for linear system~\eqref{sys:linear} when the conditional prior distribution $\pdense(X_0 \mid \theta)$ is Gaussian: $\pdense(X_0 \mid \theta) = \mathcal{N}(X_0;m_0(\theta), C_0(\theta))$. 

\begin{theorem}
\label{thm:error_bound_linear}
Assume $\inf_{\theta} \big \lvert \Gamma(\theta) \big \rvert = \tilde{C} > 0$. Suppose the ELBO $\mathcal{L}_i(\nu_i)$ satisfies $\mathcal{L}_i(\nu_i) \geq \epsilon_i$ for all $i \in [1,k]$.
The total variation and Hellinger distances between $p(X_k,\theta \mid \Y_k)$ and $q_k(X_k,\theta)$ satisfy:
\begin{align}
&d(p(X_k,\theta \mid \Y_k), q_k(X_k,\theta)) \nonumber \\
&\quad \leq \sum_{j=1}^{k-1}C^\prime(\Y_k,\tilde{C},j,k, m_{j:k-1}, C_{j:k-1}, \nu_{j:k-1})\sqrt{\E_{\theta \sim \nu_j(\cdot)}\left[ \Psi_j(C_j(\theta), m_j(\theta), C_{j-1}(\theta), m_{j-1}(\theta), \theta) \right] -\frac{r}{2}\log(2\pi)-\frac{1}{2}\log \tilde{C} - \epsilon_j} \nonumber \\
&\qquad \qquad  + \frac{1}{\sqrt{2}} \sqrt{\E_{\theta \sim \nu_k(\cdot)}\left[\Psi_k(C_k(\theta), m_k(\theta), C_{k-1}(\theta), m_{k-1}(\theta), \theta) \right] -\frac{r}{2}\log(2\pi)-\frac{1}{2}\log \tilde{C} - \epsilon_k},
\label{eq:bound_linear}
\end{align}
where $m_{j:k-1}$, $C_{j:k-1}$, and $\nu_{j:k-1}$ denote the sequences $\{m_i\}_{i=j}^{k-1}$, $\{C_i\}_{i=j}^{k-1}$, and $\{\nu_i\}_{i=j}^{k-1}$, respectively. For any integer $j \in [1,k]$, $\Psi_j$ is defined as:
\begin{align*}
&\Psi_j(C_j(\theta), m_j(\theta), C_{j-1}(\theta), m_{j-1}(\theta), \theta) \\
& \qquad := \frac{1}{2}\log \big \lvert F_c(m_{j-1}(\theta), C_{j-1}(\theta), \theta) \big \rvert - \frac{1}{2} \log \big \lvert C_j(\theta) \big \rvert - \frac{n}{2} + \frac{1}{2}\Tr\left(\left(F_c(m_{j-1}(\theta), C_{j-1}(\theta), \theta)\right)^{-1}C_j(\theta)\right) \\
&\qquad \qquad + \frac{1}{2}\left(m_j(\theta)-F_m(m_{j-1}(\theta), C_{j-1}(\theta),\theta)\right)^T\left(F_c(m_{j-1}(\theta), C_{j-1}(\theta), \theta)\right)^{-1}\left(m_j(\theta)-F_m(m_{j-1}(\theta), C_{j-1}(\theta),\theta)\right),
\end{align*}
with $F_m$ and $F_c$ defined as
\begin{align*}
F_m(m_{j-1}(\theta), C_{j-1}(\theta), \theta) :=& m_j^{*-}(\theta)+C_j^{*-}(\theta)H(\theta)^T\left(H(\theta)C_j^{*-}(\theta)H(\theta)^T+\Gamma(\theta)\right)^{-1}\left(y_j-H(\theta)m_j^{*-}(\theta)\right), \nonumber \\
F_c(m_{j-1}(\theta), C_{j-1}(\theta), \theta) :=& C_j^{*-}(\theta) - C_j^{*-}(\theta)H(\theta)^T\left(H(\theta)C_j^{*-}(\theta)H(\theta)^T+\Gamma(\theta)\right)^{-1}H(\theta)C_j^{*-}(\theta),
\end{align*}
where
\begin{equation*}
m_j^{*-}(\theta):= A(\theta)m_{j-1}(\theta), \quad C_j^{*-}(\theta) := A(\theta)C_{j-1}(\theta)A(\theta)^T+\Sigma(\theta).
\end{equation*}
When $d$ is the total variation distance, the term $C^\prime(\Y_k,\tilde{C},j,k, m_{j:k-1}, C_{j:k-1}, \nu_{j:k-1})$ in  Equation~\eqref{eq:bound_linear} is defined as 
\[C^\prime(\Y_k,\tilde{C},j,k, m_{j:k-1}, C_{j:k-1}, \nu_{j:k-1})=\frac{(2\pi)^{-\frac{r(k-j)}{2}}\tilde{C}^{-\frac{k-j}{2}}}{\sqrt{2}\prod_{i=j+1}^kZ_i(m_{i-1}, C_{i-1},\nu_{i-1})}.\]
When $d$ is the Hellinger distance, the term $C^\prime(\Y_k,\tilde{C},j,k, m_{j:k-1}, C_{j:k-1}, \nu_{j:k-1})$ is:
\[C^\prime(\Y_k,\tilde{C},j,k, m_{j:k-1}, C_{j:k-1}, \nu_{j:k-1})=\frac{2^{(k-j)}(2\pi)^{-\frac{r(k-j)}{4}}\tilde{C}^{-\frac{k-j}{4}}}{\sqrt{2\prod_{i=j+1}^kZ_i(m_{i-1}, C_{i-1},\nu_{i-1})}}.\] Here $Z_i(m_{i-1}, C_{i-1},\nu_{i-1})$ is defined as $Z_i(m_{i-1}, C_{i-1},\nu_{i-1}):=\E_{\theta \sim \nu_{i-1}(\cdot)}[\pdense_{\mathcal{N}}(y_i;H(\theta)m_i^{*-}(\theta), H(\theta)C_i^{*-}(\theta)H(\theta)^T+\Gamma(\theta))]$.
\end{theorem}

\begin{proof}
See~\ref{app:proof_thm_linear}.
\end{proof}

\begin{remark}
The upper bound provided by Theorem~\ref{thm:error_bound_linear} on the learning error $d(p(X_k,\theta) \mid \Y_k), q_k(X_k, \theta))$ is determined by the outputs of FBOVI: $m_i(\theta), C_i(\theta)$ and $\nu_i(\theta)$ for $i=1,2,\cdots, k$. In practice, this upper bound can be estimated using the outputs of FBOVI by Monte Carlo method:
\begin{align*}
&\E_{\theta \sim \nu_j(\cdot)}\left[ \Psi_j(C_j(\theta), m_j(\theta), C_{j-1}(\theta), m_{j-1}(\theta), \theta) \right] \approx \frac{1}{N}\sum_{i=1}^N \Psi_j(C_j(\theta^{(i)}), m_j(\theta^{(i)}), C_{j-1}(\theta^{(i)}), m_{j-1}(\theta^{(i)}), \theta^{(i)}), \quad \theta^{(i)} \overset{\mathrm{i.i.d.}}{\sim} \nu_j(\cdot), \\
&\qquad \qquad Z_j(m_{j-1}, C_{j-1}, \nu_{j-1}) \approx \frac{1}{N^\prime}\sum_{i=1}^{N^\prime} \pdense_{\mathcal{N}}(y_j;H(\theta^{(i)})m_j^{*-}(\theta^{(i)}), H(\theta^{(i)})C_j^{*-}(\theta^{(i)})H(\theta^{(i)})^T+\Gamma(\theta^{(i)})), \quad \theta^{(i)} \overset{\mathrm{i.i.d.}}{\sim} \nu_{j-1}(\cdot).
\end{align*}

\end{remark}

\subsubsection{Nonlinear case}
\begin{figure}
     \centering
     \includegraphics[width=0.9\linewidth]{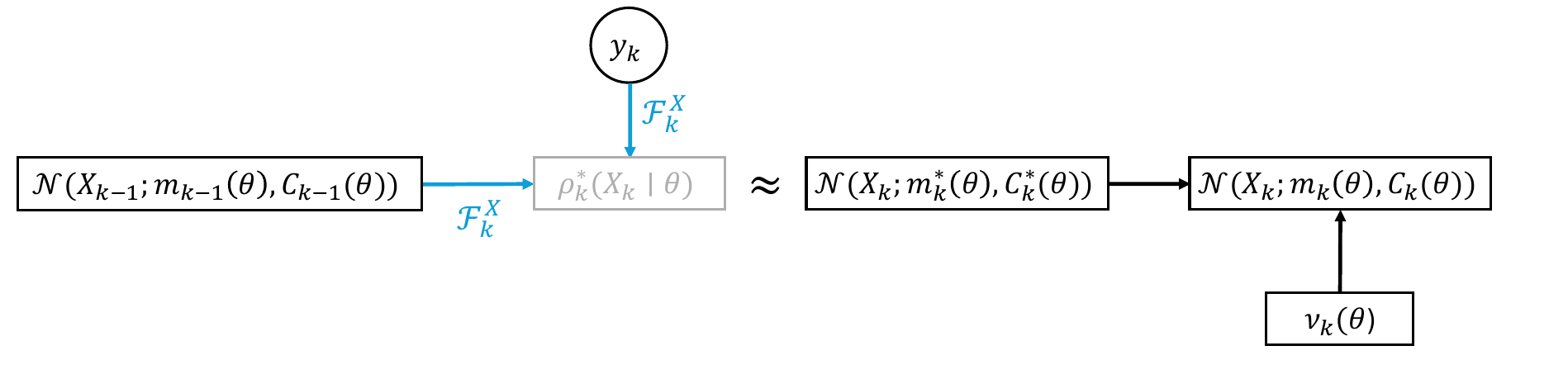}
     \caption{Procedure of computing the distribution $\rho_k(X_k \mid \theta)$ for nonlinear systems. The online conditional filtering distribution $\rho_k^*(X_k \mid \theta)$ is infeasible to compute and approximated by a Gaussian distribution $\mathcal{N}(X_k;m_k^*(\theta),C_k^*(\theta))$. We compute the distribution $\rho_k(X_k \mid \theta)$ using the target mean function $m_k^*(\theta)$, the target covariance function $C_k^*(\theta)$, and the distribution $\nu_k(\theta)$ which is computed at Stage 1.}
    \label{fig:stage2_nonlinear}
\end{figure}

\begin{figure}[t]
     \centering
     \includegraphics[width=0.7\linewidth]{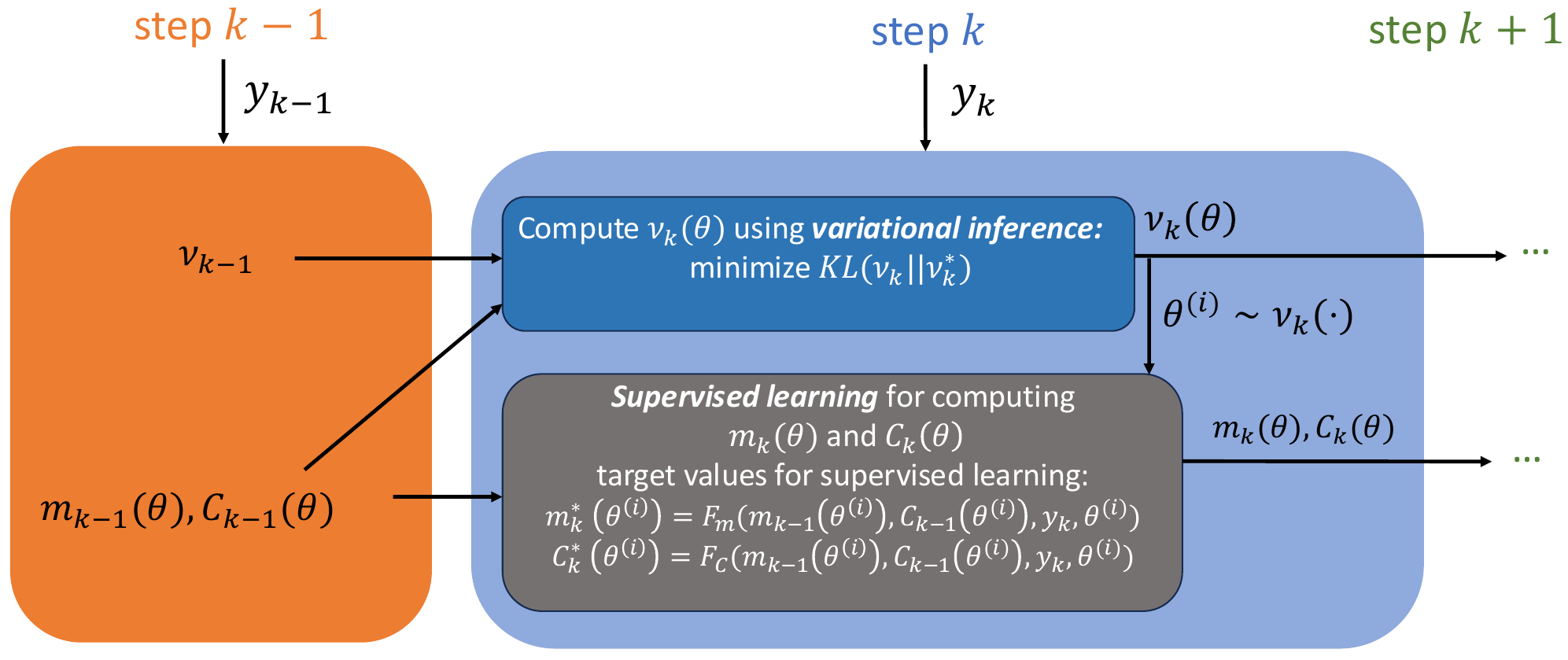}
     \caption{Scheme of FBOVI. Functions $F_m$ and $F_C$ are given by KF or Gaussian filtering methods for nonlinear systems.}
    \label{fig:procedure_onlineVI_2}
\end{figure}

When the dynamical model or the observation model is nonlinear, we first use Gaussian filtering to approximate \( \rho_k^*(X_k \mid \theta) \) by a Gaussian distribution:
\[
\rho_k^*(X_k \mid \theta) \approx \mathcal{N}\left(X_k; m_k^*(\theta), C_k^*(\theta)\right),
\]
where \( m_k^*(\theta) \) and \( C_k^*(\theta) \) are obtained via Gaussian filtering based on $m_{k-1}(\theta)$ and $C_{k-1}(\theta)$.
Again, to prevent the per-step computational cost from increasing over time, we represent \( m_k(\theta) \) and \( C_k(\theta) \) using neural networks with time-invariant architectures and train them to approximate \( m_k^*(\theta) \) and \( C_k^*(\theta) \).
The procedure of computing the distribution $\rho_k(X_k \mid \theta)$ for the nonlinear case is illustrated in Figure~\ref{fig:stage2_nonlinear}.
A schematic overview of the complete procedure is presented in Figure~\ref{fig:procedure_onlineVI_2}.

\subsubsection*{Training Strategy for $m_k$ and $C_k$}
In principle, the neural networks representing $m_k$ and $C_k$ should be trained by directly minimizing the expected KL divergence:
\begin{align}
&\E_{\theta \sim \nu_k(\cdot)}\left[\KL\left(\rho_k(\cdot  \mid \theta) \,\|\, \rho_k^*(\cdot\mid \theta)\right)\right] \approx \frac{1}{N} \sum_{i=1}^N \KL\left(\rho_k(\cdot\mid \theta^{(i)}) \,\|\, \rho_k^*(\cdot \mid \theta^{(i)})\right) \nonumber \\
&\quad  \approx \frac{1}{2N} \sum_{i=1}^N\log \big \lvert C_k^*(\theta^{(i)}) \big \rvert - \log \big \lvert C_k(\theta^{(i)}) \big \rvert - n + \Tr\left(\left(C_k^*(\theta^{(i)})\right)^{-1}C_k(\theta^{(i)})\right) + \delta(\theta^{(i)})^T\left(C_k^*(\theta^{(i)})\right)^{-1}\delta(\theta^{(i)}),
\label{eq:kl_linear_step2}
\end{align}
with \( \theta^{(i)} \sim \nu_k(\cdot) \) and $\delta(\theta^{(i)}) = m_k(\theta^{(i)})-m_k^*(\theta^{(i)})$. 

From Equation~\eqref{eq:kl_linear_step2}, we can see that evaluating the expected KL divergence $\E_{\theta \sim \nu_k(\cdot)}\left[\KL\left(\rho_k(\cdot \mid \theta) \,\|\, \rho_k^*(\cdot \mid \theta)\right)\right]$ requires evaluating the determinants of the covariances $C_k^*(\theta^{(i)})$ and $C_k(\theta^{(i)})$, as well as the inverse of the covariance $C_k^*(\theta^{(i)})$, for each sample $\theta^{(i)}$. This can be computationally expensive, especially for high state dimension. When the requirement for computational efficiency is high, the following two surrogate losses can be employed for the training of neural networks used to represent $m_k(\theta)$ and $C_k(\theta)$:
\begin{align}
\frac{1}{N} \sum_{i=1}^N \big  \lVert m_k(\theta^{(i)}) - m_k^*(\theta^{(i)})\big \rVert_2^2, \quad
\frac{1}{N} \sum_{i=1}^N \big  \lVert C_k(\theta^{(i)}) - C_k^*(\theta^{(i)})\big \rVert_F^2, \quad \theta^{(i)} \sim \nu_k(\cdot).
\end{align}

\subsubsection{Incorporation of ensemble Kalman filter}
In high-dimensional nonlinear systems, many Gaussian filtering methods, including the unscented Kalman filter (UKF) and the Gauss–Hermite Kalman filter (GHKF), encounter significant scalability challenges. For instance, with an $n$-dimensional state space, evaluating each quantity in Equation~\eqref{eq:GF_eqs} using the UKF requires $2n+1$ function evaluations, and, more critically, the UKF’s covariance estimates become increasingly unstable as the dimensionality grows. Similarly, the GHKF requires $l^n$ function evaluations for computing each quantity in Equation~\eqref{eq:GF_eqs} when employing an $l$-th-order Hermite polynomial, causing the computational cost to rise drastically with $n$. Consequently, directly incorporating these Gaussian filtering methods into Stage~2 of our framework would also constrain the overall scalability of the proposed approach. To overcome this, the proposed framework supports the use of the EnKF, which is known for its scalability. 

A distinctive aspect of our framework is that only the approximate parameter posterior \( \nu_k(\theta) \), and the function-based representations \( m_k(\theta) \) and \( C_k(\theta) \), are propagated to the next time step. State particles are not passed forward. As a result, the standard EnKF is not directly applicable. Instead, we propose an \textit{every-step resampling} procedure. For a given sample \( \theta^{(i)} \), we first draw \( N_{e,F} \) state particles from the previous approximate conditional posterior \( \rho_{k-1}(X_{k-1} \mid \theta^{(i)}) \):
\begin{equation*}
X_{k-1}^{(j)} \sim \mathcal{N}\left(m_{k-1}(\theta^{(i)}), C_{k-1}(\theta^{(i)})\right), \quad j = 1,\dots,N_{e,F}.
\end{equation*}
These samples \( \{X_{k-1}^{(j)}\}_{j=1}^{N_{e,F}} \) are then propagated using the standard EnKF update to generate an ensemble \( \{X_k^{(j)}\}_{j=1}^{N_{e,F}} \), which approximates the online conditional filtering distribution \( q_k^*(X_k \mid \theta^{(i)}) \). Finally, the sample mean and covariance of \( \{X_k^{(j)}\}_{j=1}^{N_{e,F}} \) are computed and assigned to \( m_k^*(\theta^{(i)}) \) and \( C_k^*(\theta^{(i)}) \), respectively. The full procedure is outlined in Algorithm~\ref{alg:enkf_update}.

\begin{algorithm}[t]
\caption{Single-step EnKF-based computation for $m_k^*(\theta)$ and $C_k^*(\theta)$}
\label{alg:enkf_update}
\KwIn{
    \\
    $\nu_k(\theta)$: approximate parameter posterior at time $k$ \\
    $m_{k-1}(\theta)$, $C_{k-1}(\theta)$: mean and covariance functions of the previous conditional posterior $\rho_{k-1}(X_{k-1} \mid \theta)$\\
    $y_k$: data received at time $k$ \\
    $\Phi(\cdot\,;\theta)$: state transition function \\
    $h(\cdot;\theta)$: deterministic component of observation model\\
    $\Sigma(\theta)$, $\Gamma(\theta)$: process noise covariance and measurement noise covariance \\
    $N_{e,F}$: number of ensemble members \\
    $M$: number of samples $\theta^{(i)} \sim \nu_k(\theta)$
}
\KwOut{
    Training data $\{ \theta^{(i)}, m_k^*(\theta^{(i)}), C_k^*(\theta^{(i)}) \}_{i=1}^{M}$
}

\For{$i = 1$ \KwTo $M$}{
    Sample $\theta^{(i)} \sim \nu_k(\theta)$ \\
    
    \For{$j = 1$ \KwTo $N_{e,F}$}{
        Sample $X_{k-1}^{(j)} \sim \mathcal{N}(m_{k-1}(\theta^{(i)}), C_{k-1}(\theta^{(i)}))$ {\color{red} $\leftarrow$ Resampling happens here}\\
        Sample $\xi_k^{(j)} \sim \mathcal{N}(0, \Sigma(\theta^{(i)}))$ \\
        $X_k^{(j)-} \leftarrow \Phi(X_{k-1}^{(j)};\theta^{(i)}) + \xi_k^{(j)}$
    }

    $m_k^- \leftarrow \frac{1}{N_{e,F}\sum_{j=1}^{N_{e,F}}}X_k^{(j)-}$ 

    \For{$j = 1$ \KwTo $N_{e,F}$}{
        Sample $\eta_k^{(j)} \sim \mathcal{N}(0, \Gamma(\theta^{(i)}))$ \\
        $y_k^{(j)} \leftarrow y_k - \eta_k^{(j)}$
    }

    $\mu_y \leftarrow \frac{1}{N_{e,F}} \sum_{j=1}^{N_{e,F}}(h(X_k^{(j)-};\theta^{(i)})  + \eta_k^{(j)})$ \\
    $S \leftarrow \frac{1}{N_{e,F}-1} \sum_{j=1}^{N_{e,F}}(h(X_k^{(j)-};\theta^{(i)}) + \eta_k^{(j)} - \mu_y)(h(X_k^{(j)-};\theta^{(i)}) + \eta_k^{(j)} - \mu_y)^T$ \\
    $U \leftarrow \frac{1}{N_{e,F}-1} \sum_{j=1}^{N_{e,F}} (X_k^{(j)-} - m_k^-)(h(X_k^{(j)-};\theta^{(i)}) + \eta_k^{(j)} - \mu_y)^T$

    \For{$j = 1$ \KwTo $N_{e,F}$}{
        $X_k^{(j)} \leftarrow X_k^{(j)-} + US^{-1} (y_k^{(j)} - h(X_k^{(j)-};\theta^{(i)}))$
    }

    $m_k^*(\theta^{(i)}) \leftarrow \frac{1}{N_{e,F}}\sum_{j=1}^{N_{e,F}} X_k^{(j)}$ \\
    $C_k^*(\theta^{(i)}) \leftarrow \frac{1}{N_{e,F}-1} \sum_{j=1}^{N_{e,F}} (X_k^{(j)} - m_k^*(\theta^{(i)})) (X_k^{(j)} - m_k^*(\theta^{(i)}))^T$
    
}

\Return $\{ \theta^{(i)}, m_k^*(\theta^{(i)}), C_k^*(\theta^{(i)}) \}_{i=1}^{M}$
\end{algorithm}

Once the pairs \( \{ \theta^{(i)}, m_k^*(\theta^{(i)}), C_k^*(\theta^{(i)}) \}_{i=1}^{M} \) are computed, they are used as training data to fit the neural networks representing \( m_k(\theta) \) and \( C_k(\theta) \). The trained neural networks are then passed to the next time step \( k+1 \). 

The framework introduced in this section is compatible with arbitrary variational distribution families for the approximate parameter posterior $\nu_k(\theta)$ and can incorporate a variety of Gaussian filtering methods.~\ref{sec:guidance} provides guidance on selecting the variational family for $\nu_k(\theta)$ and the Gaussian filter to meet different practical requirements. We note that the proposed framework can also accommodate non-Gaussian representations of the conditional distribution $\rho_k(X_k \mid \theta)$ and filtering methods beyond the class of Gaussian filters can be incorporated. Details on computing $\nu_k(\theta)$ and $\rho_k(X_k \mid \theta)$ in this setting are provided in~\ref{app:non_Gaussian}.

\section{Computational complexity}
\label{sec:computational_complexity}
In this section, we assess the computational efficiency of FBOVI and three other algorithms that also compute the joint posterior online: joint PF, joint UKF, and joint EnKF. We assume that the process noise covariance $\Sigma$ and measurement noise covariance $\Gamma$ are constant and known. Recall that the dimensions of the state $X_k$ and the observation $Y_k$ are denoted by $n$ and $r$, respectively. Let $d_{\theta}$ denote the dimension of $\theta$. Let $\bar{C}_{\Phi}$ and $\bar{C}_h$ denote the computational complexities of evaluating the dynamical model $\Phi(\cdot;\theta)$ and the observation model $h(\cdot;\theta)$ for a given parameter vector $\theta$, respectively. In addition, let $C_{\Phi}$ and $C_h$ denote the computational complexities of computing $\Phi(X_k;\theta)$ and $h(X_k;\theta)$, respectively, for a given state $X_k$, assuming that the corresponding dynamical and observation models, $\Phi(\cdot;\theta)$ and $h(\cdot;\theta)$, have already been evaluated. In the linear case, we have $C_{\Phi}=\mathcal{O}(n^2)$ and $C_h=\mathcal{O}(nr)$.

We analyze the computational complexity of FBOVI with KF incorporated in the linear case, and with UKF or EnKF incorporated in the nonlinear case. For the EnKF-incorporated version of FBOVI, let $N_{e,F}$ denote the ensemble size used by EnKF. We consider the setting in which the approximate parameter posterior $\nu_k(\theta)$ is represented by a Gaussian distribution. In FBOVI, optimization is performed using Adam, and gradients are computed using reverse-mode automatic differentiation. Let $N_1$ and $N_2$ denote the total numbers of optimization steps in Stage 1 and Stage 2 of FBOVI, respectively. Let $N_s$ denote the number of samples from $\nu_k(\theta)$ which are used to approximate the term $\E_{\theta\sim \nu_k(\cdot)}[\log I(\theta)]$ in the ELBO $\mathcal{L}_k(\nu_k)$ by Monte Carlo for a single optimization step. Specifically, the term $\E_{\theta\sim \nu_k(\cdot)}[\log I(\theta)]$ is computed by
\begin{equation}
\E_{\theta \sim \nu_k(\cdot)}[\log I(\theta)]
\approx
\frac{1}{N_s}\sum_{i=1}^{N_s}\log I(\theta^{(i)}),
\qquad
\theta^{(i)} \overset{\mathrm{i.i.d.}}{\sim} \nu_k(\cdot).
\label{eq:approx_E_logI}
\end{equation}
Let $\bar{C}_I$ denote the computational complexity of computing $\log I(\theta)$ for a given $\theta$, where $I(\theta)$ is evaluated as in Equation~\eqref{eq:int_approx}, with $m_k^-(\theta)$ and $C_k^-(\theta)$ already evaluated. For linear observation models, we have
\[
\bar{C}_I=\mathcal{O}(\bar{C}_h+r^3+nr^2+n^2r).
\]
For nonlinear observation models, $\bar{C}_I$ normally depends on $\bar{C}_h$, $C_h$, and the method used to evaluate the integral in Equation~\eqref{eq:int_approx}. Recall that, in practice, instead of directly representing the covariance function $C_k(\theta)$ with a neural network, we use a neural network to represent a lower triangule matrix $c_k(\theta)$. The covariance matrix $C_k(\theta)$ is then constructed as $C_k(\theta)=c_k(\theta)c_k(\theta)^T$. Let $d_m$ and $d_c$ represent the numbers of trainable parameters of the neural networks representing $m_k(\theta)$ and $c_k(\theta)$, respectively. Let $C_m$ and $C_c$ denote the computational complexities of one forward evaluation of the neural networks representing $m_k(\theta)$ and $c_k(\theta)$, respectively. For example, if the neural network representing $m_k(\theta)$ has $l$ fully connected layers, with each of the first $l-1$ layers followed by a ReLU activation function, and the $i$-th layer has size $(z_{i-1},z_i)$, then we have
\[
C_m=\mathcal{O}\bigg(d_{\theta}z_1+\sum_{i=2}^{l-1}z_{i-1}z_i+nz_{l-1}\bigg).
\]
Let $M$ denote the size of the dataset used to train the neural networks representing $m_k(\theta)$ and $c_k(\theta)$, i.e., the training dataset is given by
$\{\theta^{(i)},m_k^*(\theta^{(i)}),C_k^*(\theta^{(i)})\}_{i=1}^M$. Let $B$ denote the size of the minibatch used to compute the loss at each optimization step when training these neural networks. 

For the computational complexity analysis of joint PF and joint EnKF, we consider the version of joint PF based on the bootstrap PF with $N_p$ particles, and we let $N_e$ denote the number of ensemble members used by the joint EnKF. Table~\ref{tab:computational_complexity} reports the per-time-step computational complexities of FBOVI, the joint PF, the joint UKF, and the joint EnKF.

\begin{table}[H]
\centering
\caption{Computational complexities of FBOVI, the joint PF, the joint UKF, and the joint EnKF for a single time step. For FBOVI, we report the computational complexities under three choices of the incorporated Gaussian filter: the KF, the UKF, and the EnKF.}
\label{tab:computational_complexity}
\begin{tabular}{p{2cm} p{12cm}}
\hline
Algorithm & Computational Complexity \\
\hline
\rule{0pt}{13pt}
FBOVI(KF) & $\mathcal{O}\big(N_1d_{\theta}^3 + (N_1N_s+M)(d_{\theta}^2+\bar{C}_{\Phi}+\bar{C}_h+C_m+C_c+n^3+n^2r+nr^2+r^3)+N_2B(C_m+C_c+n^3)+N_2(d_m+d_c)\big)$  \\
FBOVI(UKF) & $\mathcal{O}\big(N_1d_{\theta}^3+ (N_1N_s+M)(d_{\theta}^2+\bar{C}_{\Phi}+C_m+C_c+nC_{\Phi}+n^3)+N_1N_s\bar{C}_I+M(\bar{C}_h+nC_h+n^2r+nr^2+r^3)+N_2B(C_m+C_c+n^3)+N_2(d_m+d_c)\big)$ \\
FBOVI(EnKF) & $\mathcal{O}\big(N_1d_{\theta}^3+(N_1N_s+M)(d_{\theta}^2+\bar{C}_{\Phi}+C_m+C_c+N_{e,F}(C_{\Phi}+n^2))+N_1N_s\bar{C}_I+M(\bar{C}_h+nr^2+r^3)+MN_{e,F}(C_h+nr+r^2)+N_2B(C_m+C_c+n^3)+N_2(d_m+d_c)\big)$  \\
Joint PF & $\mathcal{O}\big(N_p(\bar{C}_{\Phi}+\bar{C}_{h}+C_{\Phi}+C_{h}+(n+d_\theta)^2+r^2)\big)$  \\
Joint UKF & $\mathcal{O}\big((n+d_{\theta})(\bar{C}_{\Phi}+\bar{C}_{h}+C_{\Phi}+C_h+r^2)+(n+d_{\theta})^3+(n+d_{\theta})^2r+r^3\big)$  \\
Joint EnKF & $\mathcal{O}\big(N_e(\bar{C}_{\Phi}+\bar{C}_{h}+C_{\Phi}+C_h+(n+d_{\theta})^2+(n+d_{\theta})r+ r^2)+(n+d_{\theta})r^2+r^3\big)$  \\
\hline
\end{tabular}
\end{table}

From Table~\ref{tab:computational_complexity}, we can see that the computational complexity of FBOVI depends strongly on users' choices, such as the sizes of the neural networks used to represent $m_k(\theta)$ and $c_k(\theta)$. These choices can be made according to the desired inference accuracy and the available computational resources. For example, when inference accuracy is prioritized, more expressive neural networks can be used, which generally increases the computational complexities associated with neural network evaluation, namely $C_m$ and $C_c$. Conversely, when computational efficiency is prioritized, simpler models can be used to represent $m_k(\theta)$ and $c_k(\theta)$, thereby reducing $C_m$ and $C_c$ and lowering the overall computational cost. 

We note here that the sizes of the required neural networks representing $m_k(\theta)$ and $c_k(\theta)$ to achieve high inference accuracy normally increase with the state dimension. This can lead to computational inefficiency when the state dimesion is large and high inference accuracy is required, which is a limitation of the proposed framework. Future work will explore learning an operator offline that takes the distribution $\nu_k(\theta)$, the data $y_k$, the previous mean function $m_{k-1}(\theta)$, and the previous covariance function $C_{k-1}(\theta)$ as inputs, and outputs the current mean function $m_k(\theta)$ and the current covariance function $C_k(\theta)$. This operator is trained offline, prior to the online estimation phase. During the online operation, the functions $m_k(\theta)$ and $C_k(\theta)$ can then be obtained directly from the pre-trained operator, substantially reducing the online computational cost.

We next compare the computational complexities of FBOVI, joint PF, joint UKF, and joint EnKF. When both the state dimension $n$ and the parameter dimension $d_{\theta}$ are small, joint UKF normally has the lowest computational complexity. When the computational complexities of evaluating the dynamical and observation models, $\bar{C}_{\Phi}$ and $\bar{C}_h$, are low, and the number of particles $N_p$ and the ensemble size $N_e$ are moderate, joint PF, joint UKF, and joint EnKF can be more computationally efficient than FBOVI. This is expected, as joint Gaussian filters are normally advantageous in terms of computational efficiency. However, when model evaluation is computationally expensive, as reflected by large values of $\bar{C}_{\Phi}$ and $\bar{C}_h$, the joint PF and joint EnKF can become less efficient than FBOVI if a large number of particles or ensemble members are required.

Finally, the inference speed depends not only on computational complexity but also on the practical implementation and available hardware. It is noteworthy that many computations required by FBOVI can be parallelized. For example, the terms in the computational complexities involving the training dataset size $M$ arise from computing the supervised learning targets in Stage 2, namely $\{m_k^*(\theta^{(i)})\}_{i=1}^M$ and $\{C_k^*(\theta^{(i)})\}_{i=1}^M$. These targets can be computed independently across different values of $\theta^{(i)}$, allowing the computations to be performed in parallel. If the algorithm is run on $Q$ parallel threads, the runtime required for this part can ideally be reduced by a factor of $1/Q$. Similarly, when computing the term $\E_{\theta \sim \nu_k(\cdot)}[\log I(\theta)]$ in the ELBO $\mathcal{L}_k(\nu_k)$ using the Monte Carlo approximation in Equation~\eqref{eq:approx_E_logI}, the evaluations of $\log I(\theta^{(i)})$ for different samples $\theta^{(i)}$ are independent and can therefore be performed in parallel. Since the computational cost of evaluating this term normally dominates the computational cost of evaluating the ELBO $\mathcal{L}_k(\nu_k)$, this parallelization can significantly reduce the time required for ELBO maximization. Many computations required by joint PF, joint UKF, and joint EnKF can also be parallelized. Therefore, the practical inference time is strongly influenced by the computational hardware and the degree of parallelization used in implementation.

\section{Numerical Experiments}
\label{sec:experiments}
In this section, we implement the proposed FBOVI algorithm across three numerical experiments and compare its performance against three online joint posterior approximation methods: the joint PF, joint UKF and joint EnKF. All these methods assign artificial dynamics to the unknown model parameters and augment the system state with them. Standard PF, UKF, and EnKF are then applied to the augmented system. The joint PF tested in this section employs the bootstrap PF. Details of the hyperparameters used for the joint PF, joint UKF, and joint EnKF in the experiments are provided in~\ref{app:hyper_param}. Each of these methods has its own strengths: the joint PF can provide accurate joint posterior approximations in low-dimensional systems; the joint EnKF is effective for large-scale problems; and the joint UKF is widely adopted for its good balance between inference performance and computational efficiency. The experimental setups are designed to capture several of the most common and significant challenges encountered in recent application scenarios of parameter–state estimation methods, such as DTs:
\begin{enumerate}
\item \textit{Observation-related challenges}: First, as previously discussed, observational data are typically noisy and often incomplete. Second, the observation model may be partially unknown, adding additional challenges to the inference process.
\item \textit{Model misspecification}: The dynamical models used by parameter-state estimation methods are often inaccurate due to limited knowledge or the inherent complexity of real-world physical systems. Furthermore, online parameter-state estimation methods frequently rely on simplified, low-fidelity models that differ from the true physical systems in order to reduce computational burden and allow rapid updates. 
\item \textit{Chaotic dynamics}: Many physical systems encountered in parameter–state estimation problems exhibit inherently chaotic behavior, wherein small perturbations in initial conditions or inputs can cause significant changes in system trajectories. This sensitivity poses a challenge for data-driven estimation and prediction.
\item \textit{High dimensionality}: parameter-state estimation methods must be scalable in order to provide reliable guidance for high-dimensional systems, a requirement that is particularly demanding for online inference algorithms.
\end{enumerate}

To evaluate the estimation and prediction capabilities of FBOVI under these conditions and illustrate how the Gaussian filtering method integrated into FBOVI can be chosen according to the problem setting, we present the following three experiments:

\begin{enumerate}
\item In Section~\ref{sec:low_dimension}, we consider a two-dimensional linear pendulum system with partially unknown dynamical and observation models. As the underlying system is linear, we use the Kalman filter within FBOVI to compute the filtering distribution conditioned on the parameters. 
\item In Section~\ref{sec:Lorentz}, we study a ten-dimensional chaotic Lorenz '96 system. FBOVI is implemented under two model settings, with and without model form error. In addition, we consider three noise settings: low-noise data, considerably noisy data, and highly noisy data.  The UKF is employed within FBOVI to balance computational efficiency and estimation accuracy. 
\item In Section~\ref{sec:Burger}, FBOVI is applied to a nonlinear convection–diffusion transport model. After spatial discretization, the system becomes 51-dimensional. To ensure scalability in this high-dimensional setting, we integrate EnKF into FBOVI. 
\end{enumerate}

All systems considered in these experiments are partially observable. Different evaluation metrics and validation strategies are employed to assess the estimation and prediction performance of FBOVI. The detailed experimental configurations, criteria, and results are provided in the subsequent subsections. 

\subsection{Linear single pendulum system}
\label{sec:low_dimension}
In this experiment, we consider a two-dimensional linear pendulum system governed by
\begin{align}
\begin{bmatrix}\dot{x}_1 \\ \dot{x}_2 \end{bmatrix} = 
\begin{bmatrix}x_2 \\ -\tfrac{g}{l}x_1 \end{bmatrix},
\end{align}
where $g=9.8$ is the gravitational acceleration and $l=1.2$ denotes the pendulum length. The initial state is chosen as $\mathbf{x}(0)=\begin{bmatrix}0.5 & 0.5\end{bmatrix}^T$. With a time step size of $\Delta t=0.1$, the resulting discrete-time system is 
\begin{equation}
X_{k+1} = \exp\bigg (\Delta t\begin{bmatrix}0 & 1 \\ -\frac{g}{l}&0\end{bmatrix}\bigg )X_k =\begin{bmatrix}0.9594 & 0.0986 \\ -0.8056 & 0.9594 \end{bmatrix}X_k, 
\label{eq:linear_pendulum_dt}
\end{equation}
where $X_k$ denotes the state at time $t=k\Delta t$. The reference trajectories are generated according to:
\begin{equation}
X_{k+1} = \begin{bmatrix}0.9594 & 0.0986 \\ -0.8056 & 0.9594 \end{bmatrix}X_k + \eta_k, \quad \eta_k \sim \mathcal{N}(0, 0.01I_{2\times2}),
\label{eq:linear_pendulum_dt_simulation}
\end{equation}
where the noise term $\eta_k$ is used to represent the reference process noise. Noisy measurements of the states are collected according to the model $Y_k = HX_k + V_k, \quad V_k \overset{\mathrm{i.i.d.}}{\sim} \mathcal{N}(0,0.01I_{r \times r})$. Throughout this section, the learning model does not have access to the true initial state. Instead, the prior distribution for $X_0$ is taken to be $\mathcal{N}([
    3, 4.5
]^T, 4I_{2\times2})$.
\subsubsection{System with partially unknown dynamical model}
\label{sec:pendulum_part1}
In this section, we focus on learning the parameters in a partially unknown dynamical model and learning partially observable states. We show that approximate posteriors obtained by FBOVI are close to the true posteriors at most time steps. 

The dynamical model used for learning is parameterized as:
\begin{equation}
X_{k+1} = \begin{bmatrix}\theta_1&0.0986\\ \theta_2&\theta_1\end{bmatrix}X_k + W_k, \qquad W_k \overset{\mathrm{i.i.d.}}{\sim} \mathcal{N}(0,0.01I_{2\times2}),
\end{equation}
where $\theta=(\theta_1,\theta_2)$ represent the unknown parameters with a prior ${\mathcal{N}}(0,I_{2\times2})$. The observation matrix $H=[1 \quad 0]$, the process noise covariance and the measurement noise covariance are known.
The simulation is conducted over a span of 50 time steps. Three online parameter-state estimation algorithms, FBOVI, joint PF, and joint UKF, are employed to approximate the joint posterior $p(X_k,\theta \mid \Y_k)$ at each time step $k$. FBOVI utilizes a Gaussian distribution to represent the distribution $\nu_k(\theta)$, while the joint PF utilizes $1\times 10^4$ particles to represent the joint posterior. The rest of the experiment settings can be found in~\ref{app:ep1}.

\begin{figure}
    \centering
    \begin{subfigure}[b]{0.98\linewidth}
        \centering
        \includegraphics[width=\textwidth]{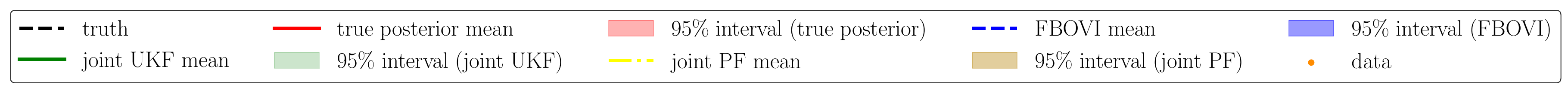}
    \end{subfigure}

    \vspace{0.8em}
     \begin{subfigure}[b]{0.24\linewidth}
         \centering
         \includegraphics[width=\textwidth]{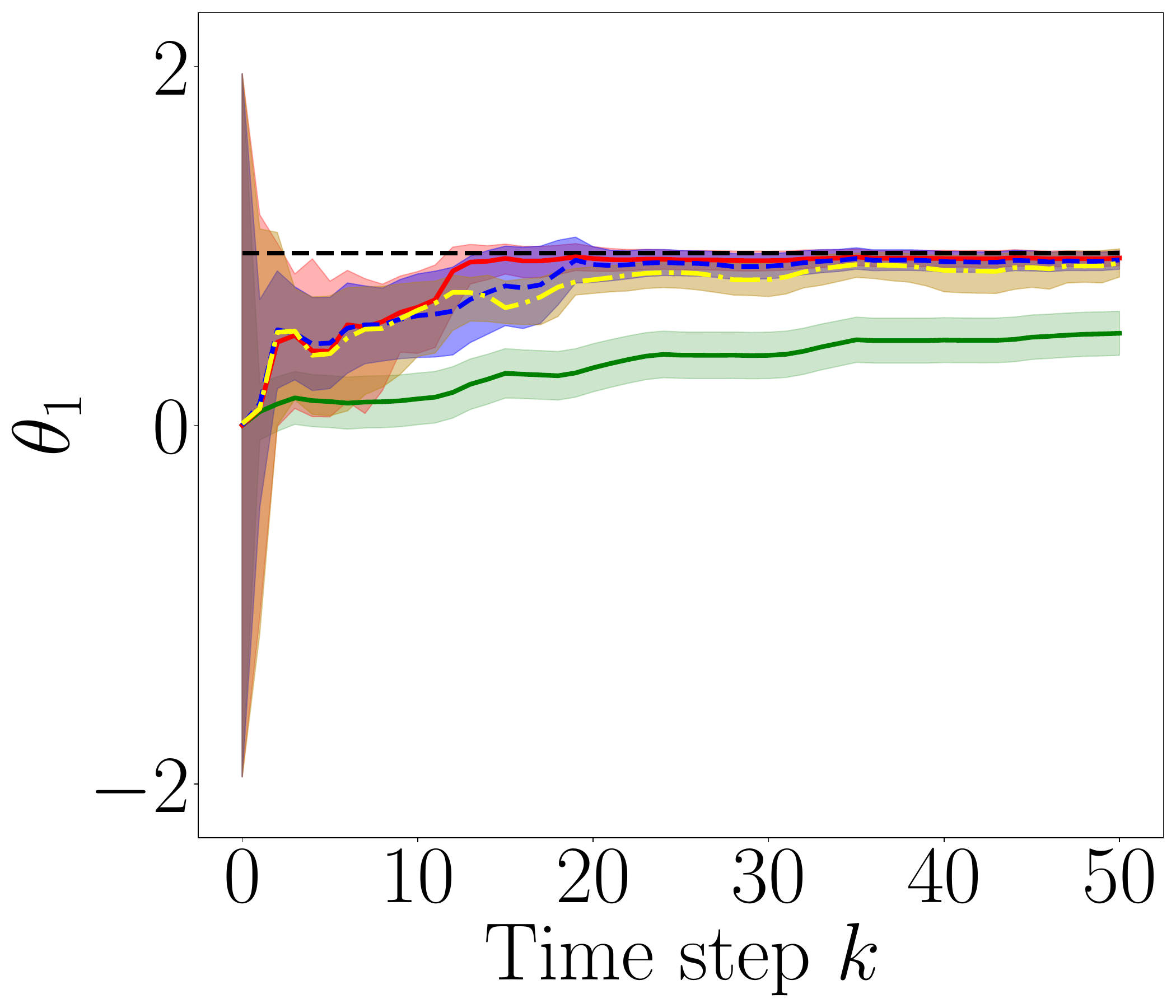}
         \caption{Distribution of $\theta_1$}
         \label{fig:dist_theta1}
     \end{subfigure}
     \hfill
     \begin{subfigure}[b]{0.24\linewidth}
         \centering
         \includegraphics[width=\textwidth]{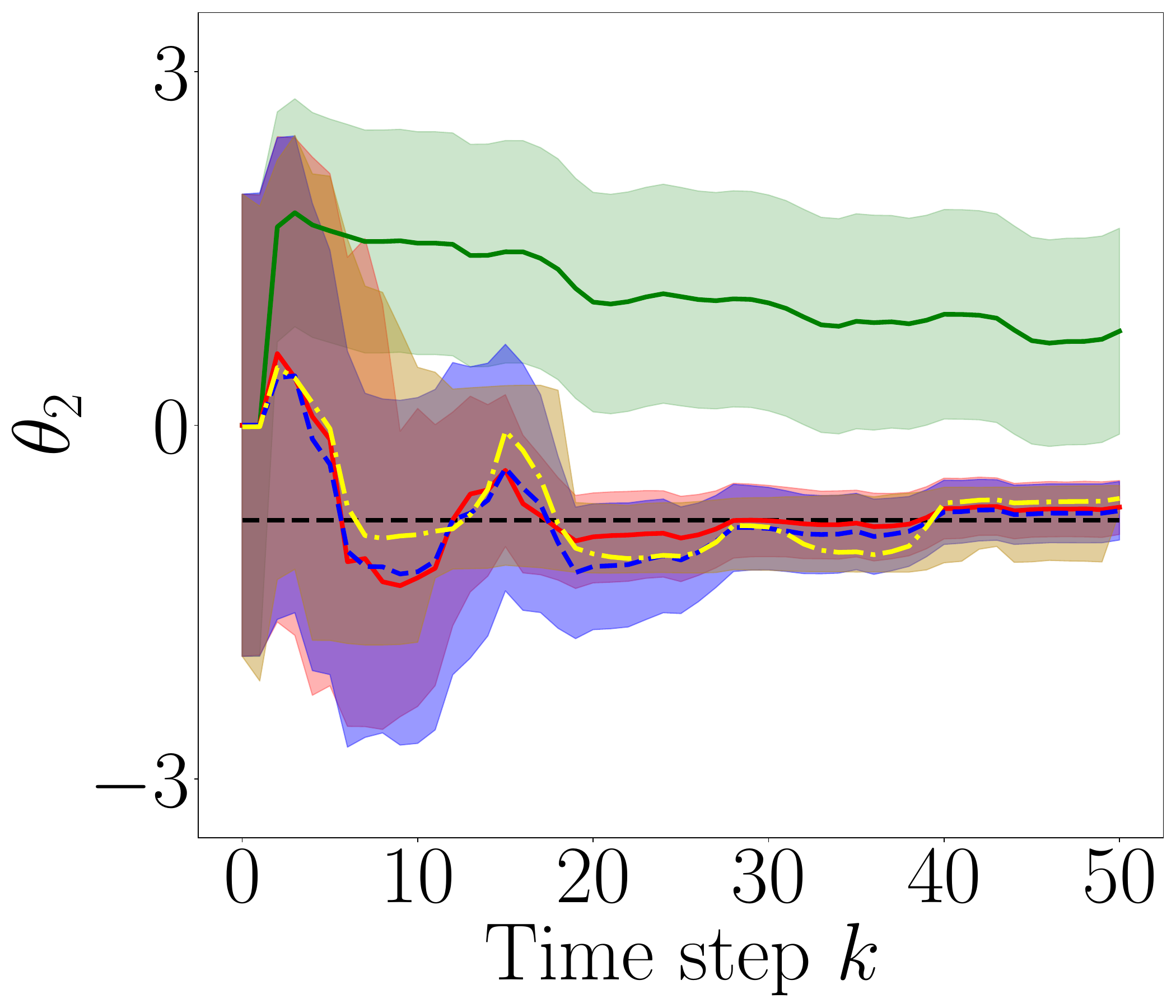}
         \caption{Distribution of $\theta_2$}
         \label{fig:dist_theta2}
     \end{subfigure}
     \hfill
     \begin{subfigure}[b]{0.24\linewidth}
         \centering
         \includegraphics[width=\textwidth]{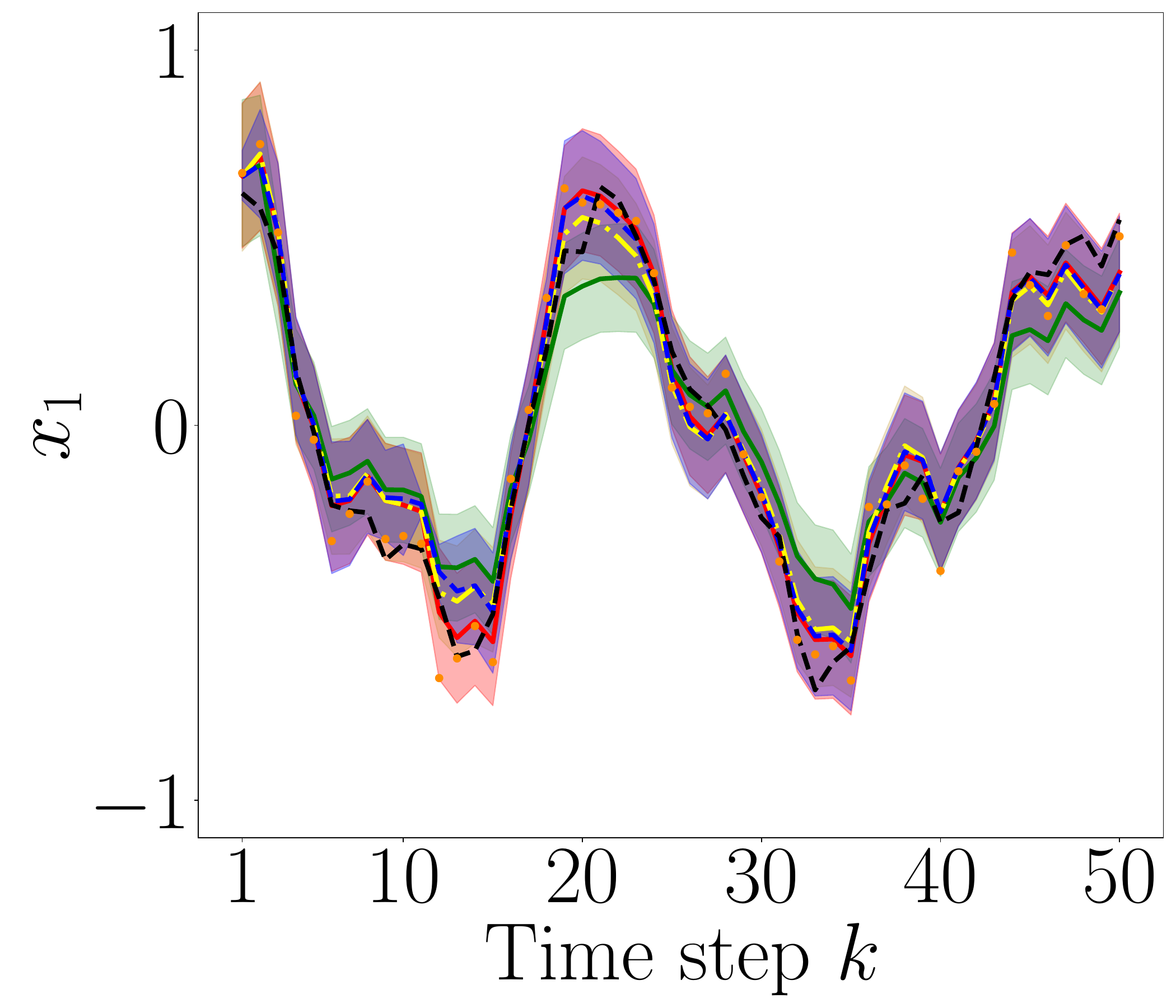}
         \caption{Distribution of $x_1(k)$}
         \label{fig:dist_x1}
     \end{subfigure}
     \hfill
     \begin{subfigure}[b]{0.24\linewidth}
         \centering
         \includegraphics[width=\textwidth]{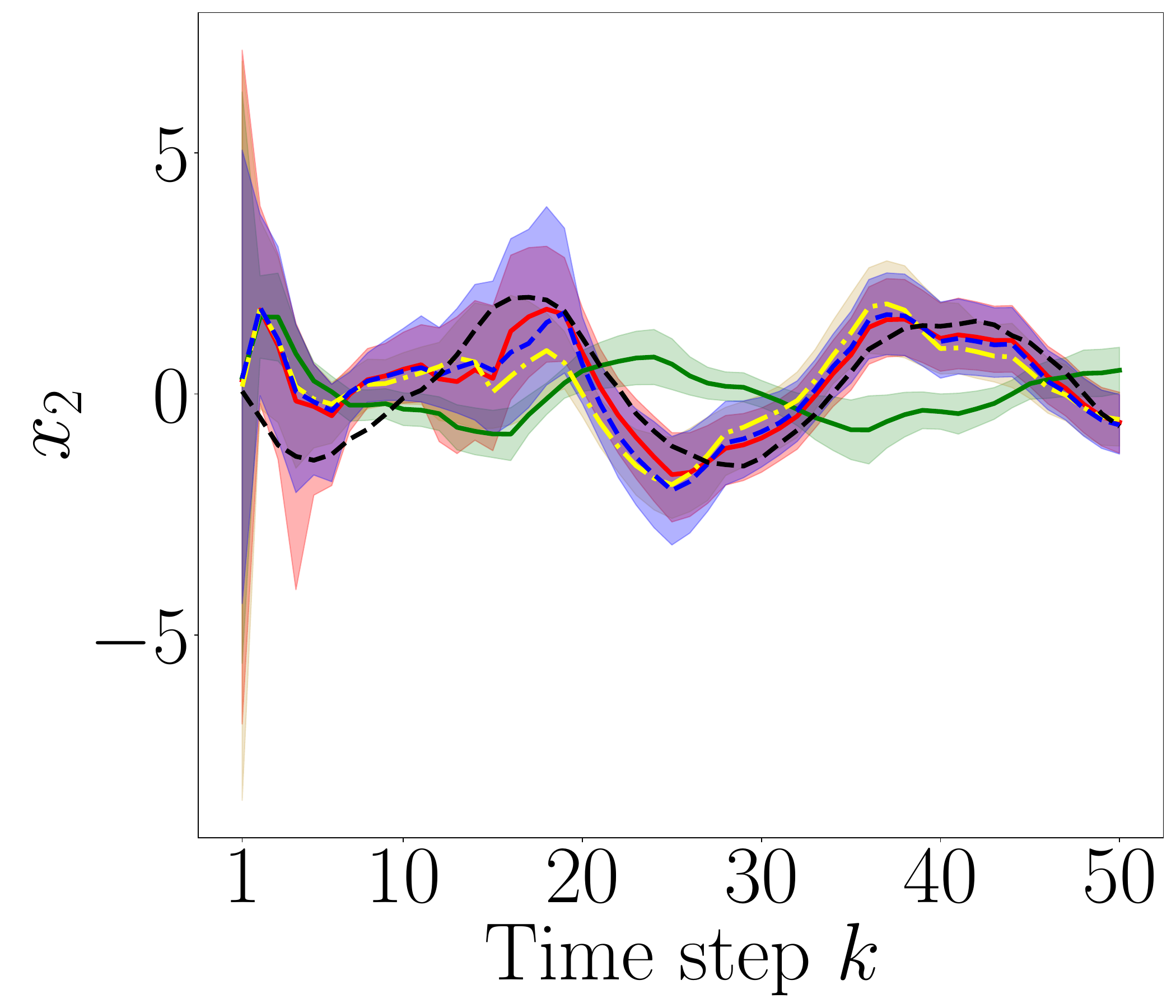}
         \caption{Distribution of $x_2(k)$ }
         \label{fig:dist_x2}
     \end{subfigure}
         \caption{\textbf{Single pendulum:} approximate marginal posterior distributions of the parameters and state components at each time step $k$. FBOVI closely follows the true posteriors at most time steps, although it overestimates the uncertainty of $\theta_2$ between steps $10$ and $28$.}
         \label{fig:marginal_dist_eg1}
\end{figure}

Figure~\ref{fig:marginal_dist_eg1} depicts the marginal posterior distributions of the parameters and state components produced by different algorithms at each time step for a single data realization. The results show that FBOVI yields approximate posterior distributions that closely track the true posteriors across most time steps. Although the joint PF uses a relatively large number of particles, the mean obtained by FBOVI are closer to the true posterior mean than that obtained by the joint PF for most time steps.

As time progresses, the means of the marginal distributions of parameters and states obtained by FBOVI gradually converge to their true values. However, during the time from around step $10$ to $28$, FBOVI tends to overestimate the uncertainty of $\theta_2$, as shown in Figure 9(b). This can be attributed to the fact that although the prior of $\theta$ is Gaussian, the posterior of $\theta$ is typically highly non-Gaussian in the early stages due to the limited amount of data incorporated. Consequently, a Gaussian distribution may not adequately capture the posterior of $\theta$. Moreover, as online inference progresses, approximation errors accumulate over time. As more data are assimilated, with linear dynamics and observations, the posterior becomes increasingly Gaussian, and FBOVI regains its ability to provide accurate uncertainty estimation. This highlights FBOVI's capability to yield an accurate posterior approximation online, starting from an initially inaccurate approximation, provided that the variational distribution family used for the parameters can adequately represent the parameter posterior. In contrast, the joint UKF fails to demonstrate this capability, with its approximate posterior diverging from the true posterior throughout the entire simulation period. This suggests that the joint UKF may struggle to provide a good approximation of the posterior when given an inaccurate prior, or that the approximation error of the joint UKF in the early stages is too substantial, and more data points especially noisy data points cannot provide sufficient information to mitigate the accumulated error within the framework of the joint UKF. 

Next, we evaluate the three online inference methods using $50$ independent data realizations. The estimation performance is assessed using the root mean square error (RMSE), which measures the average discrepancy between the true values of the states or parameters and the means of their corresponding approximate marginal posterior distributions over different data realizations. Specifically, the RMSE for estimating the $i$-th component of the parameter vector $\theta$ and the $i$-th component of the state $X_k$ at time step $k$ are defined as 
\begin{align}
RMSE_{\theta_i}(k) &:= \sqrt{\frac{1}{N}\sum_{j=1}^N\left(\hat{\theta}_{i}^{(j)}(k)-\bar{\theta}_i\right)^2}, 
\quad RMSE_{x_i}(k)  := \sqrt{\frac{1}{N}\sum_{j=1}^N\left(\hat{x}_{i}^{(j)}(k)-\bar{x}_{i}^{(j)}(k)\right)^2}, \label{eq:rmse_state}
\end{align}
where $N=50$ is the total number of data realizations. Here, $\hat{\theta}_{i}^{(j)}(k)$ and $\hat{x}_{i}^{(j)}(k)$ denote the means of the approximate marginal posterior distributions for the $i$-th elements of $\theta$ and $X_k$, respectively, obtained at time step $k$ by using the $j$-th data realization. The true parameter value is denoted by $\bar{\theta}_i$, and $\bar{x}_{i}^{(j)}(k)$ represents the true value of the $i$-th state component at time step $k$ for the $j$-th data realization.

\begin{figure}[t]
    \centering
    \begin{subfigure}[b]{0.4\linewidth}
        \centering
        \includegraphics[width=\textwidth]{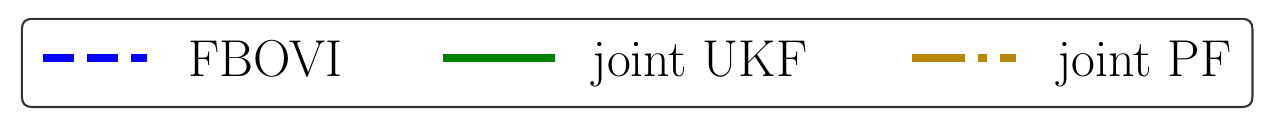}
    \end{subfigure}

    \vspace{0.8em}
     \begin{subfigure}[b]{0.24\linewidth}
         \centering
         \includegraphics[width=\textwidth]{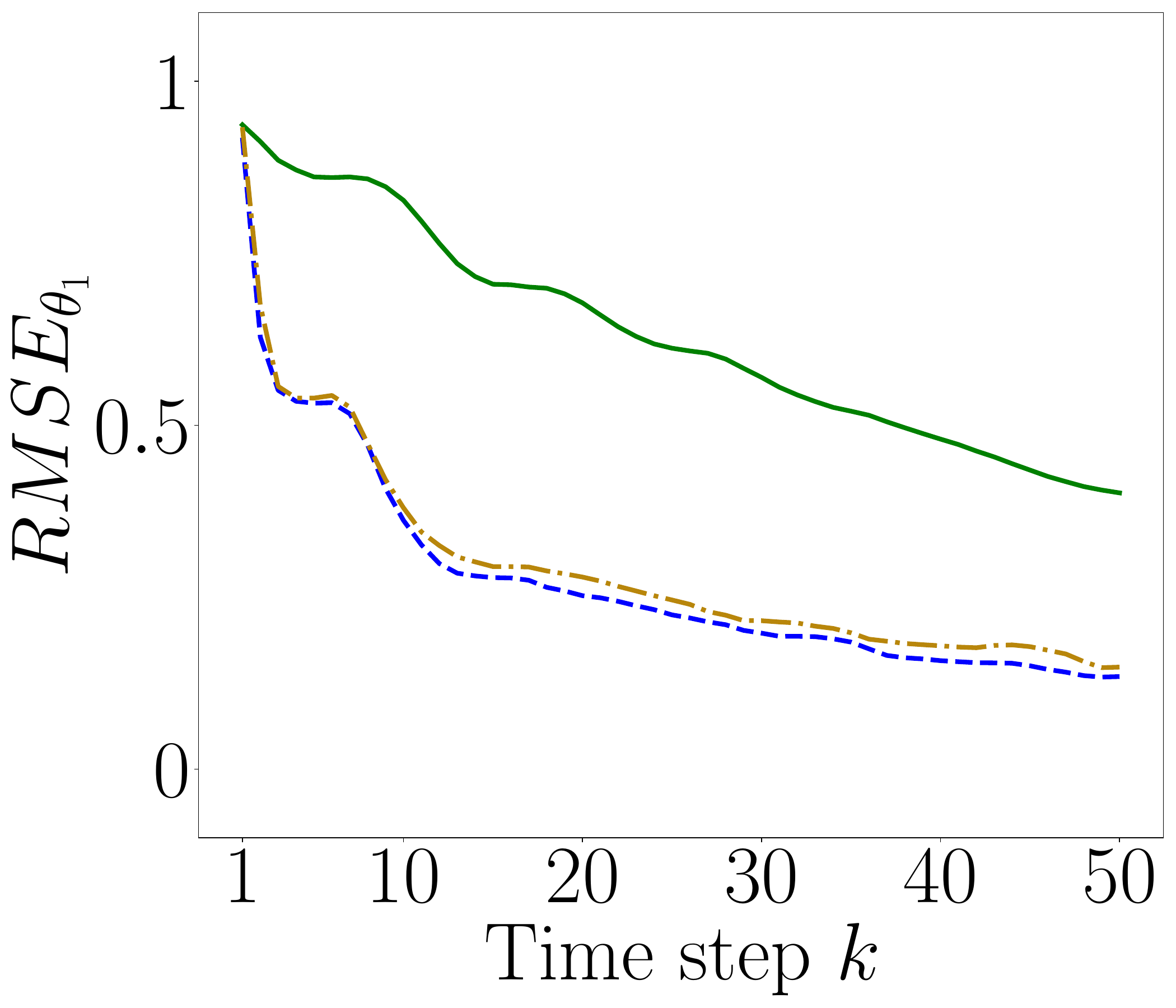}
         \caption{$RMSE_{\theta_1}$ vs. time}
         \label{fig:rmse_theta1}
     \end{subfigure}
     \hfill
     \begin{subfigure}[b]{0.24\linewidth}
         \centering
         \includegraphics[width=\textwidth]{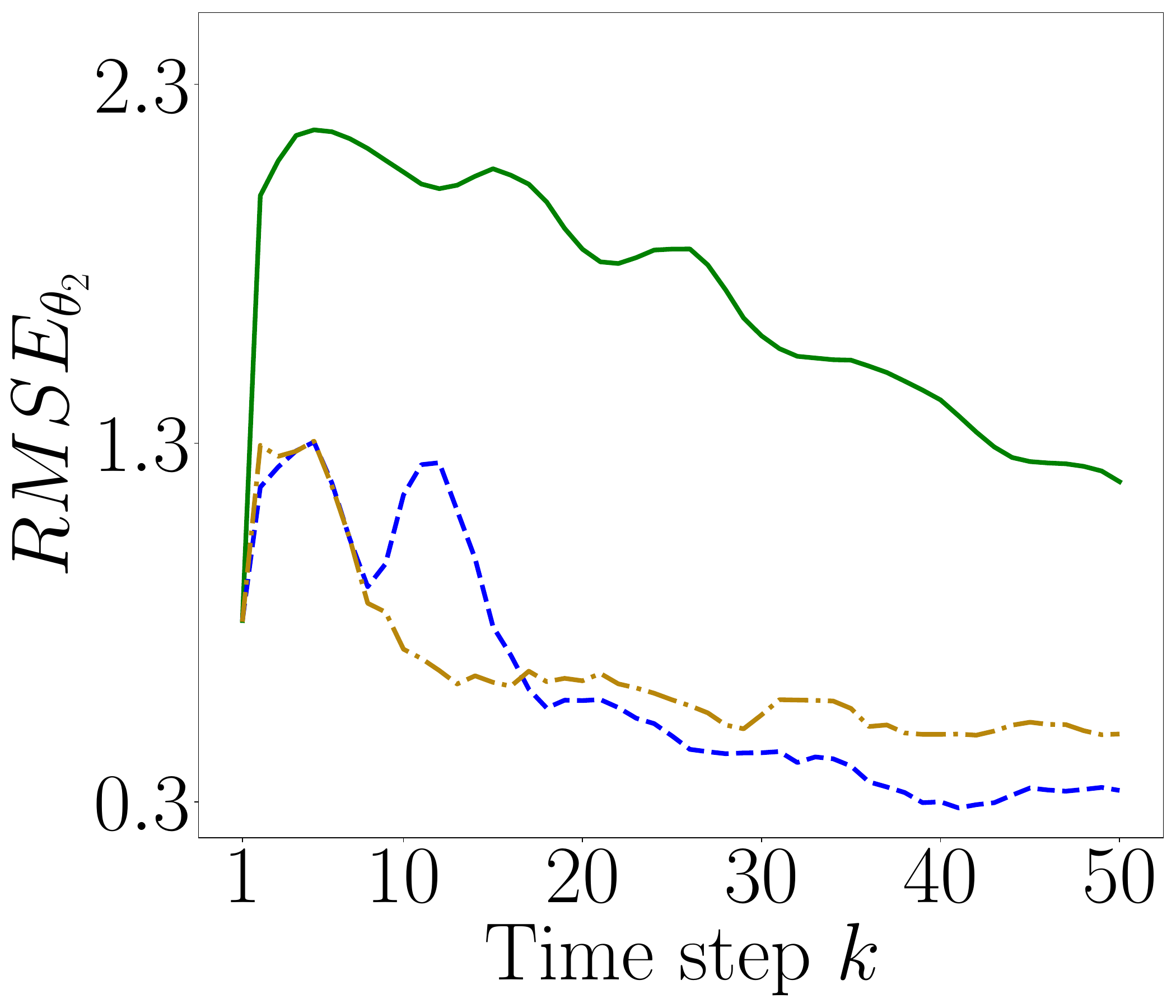}
         \caption{$RMSE_{\theta_2}$ vs. time}
         \label{fig:rmse_theta2}
     \end{subfigure}
     \hfill
     \begin{subfigure}[b]{0.24\linewidth}
         \centering
         \includegraphics[width=\textwidth]{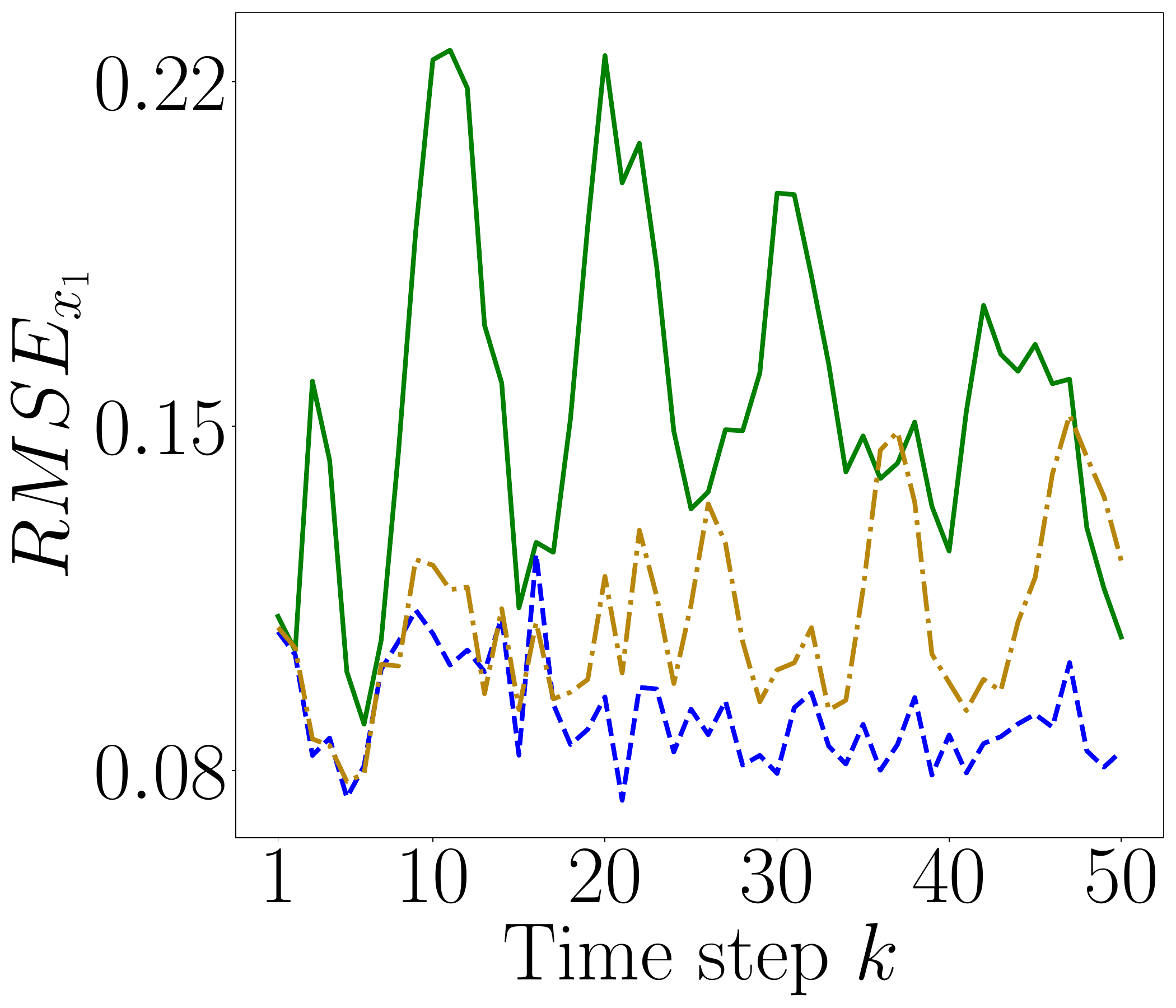}
         \caption{$RMSE_{x_1}$ vs. time}
         \label{fig:rmse_x1}
     \end{subfigure}
     \hfill
     \begin{subfigure}[b]{0.24\linewidth}
         \centering
         \includegraphics[width=\textwidth]{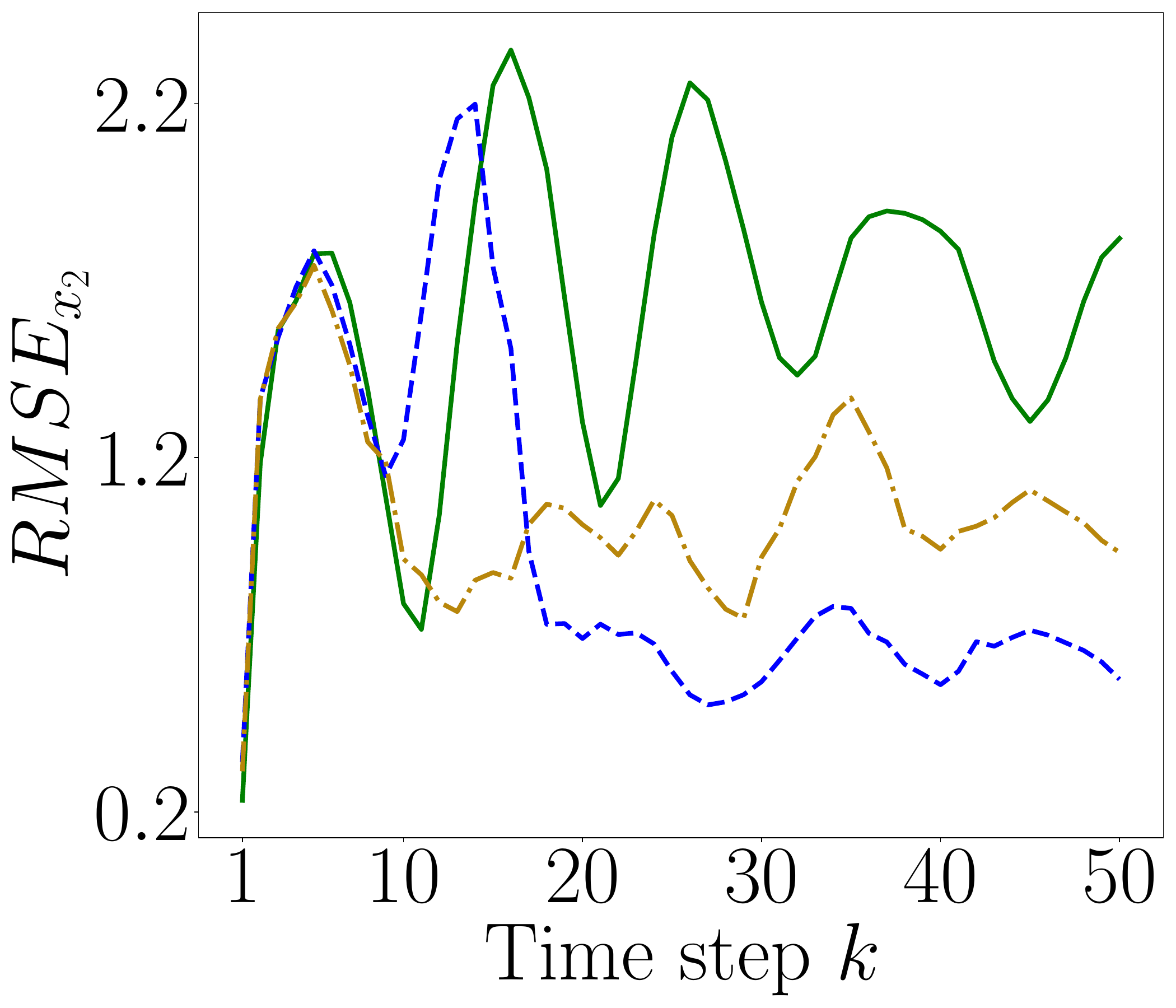}
         \caption{$RMSE_{x_2}$ vs. time}
         \label{fig:rmse_x2}
     \end{subfigure}
         \caption{\textbf{Single pendulum:} average estimation errors for states and parameters over $50$ independent data realizations at each time step $k$. The figure compares the performance of FBOVI, joint PF, and joint UKF. FBOVI achieves the lowest estimation errors for all parameters and states at most time steps.}
         \label{fig:RMSE}
\end{figure}

The resulting $RMSE_{\theta_i}(k)$ and $RMSE_{x_i}(k)$ for $i=1,2$ and $k=1,2,\ldots,50$ are shown in Figure~\ref{fig:RMSE}.
Overall, FBOVI achieves the lowest RMSE for both parameter and state estimation at most time steps, outperforming the joint PF and joint UKF after step $17$.

Next, we assess the predictive ability of the three online parameter–state estimation methods by computing the prediction RMSE over the same $N=50$ data realizations.
The prediction RMSE at time step $k$ is defined as
\begin{equation}
RMSE_{pred}(k) := \sqrt{\frac{1}{N}\sum_{j=1}^N\E_{(X_{k-1},\theta) \sim q_{k-1}^{(j)}\left(X_{k-1},\theta \right), W_{k-1} \sim \mathcal{N}\left(0,\Sigma(\theta)\right)}\left[\lVert (\Phi(X_{k-1};\theta)+W_{k-1}(\theta))-\bar{x}^{(j)}(k)\rVert _2^2\right]}, \label{eq:rmse_pred}
\end{equation}
where $q_{k-1}^{(j)}(X_{k-1},\theta)$ and $\bar{x}^{(j)}(k)$ denote the approximate joint posterior at time step $k-1$ and the true state at time step $k$, respectively, corresponding to the $j$-th data realization.

\begin{figure}
\centering
    \begin{subfigure}[b]{0.4\linewidth}
        \centering
        \includegraphics[width=\textwidth]{figures/pendulum/legends/legend_mse_e1_new.pdf}
    \end{subfigure}

    \vspace{0.4em}
     \includegraphics[width=0.3\textwidth]{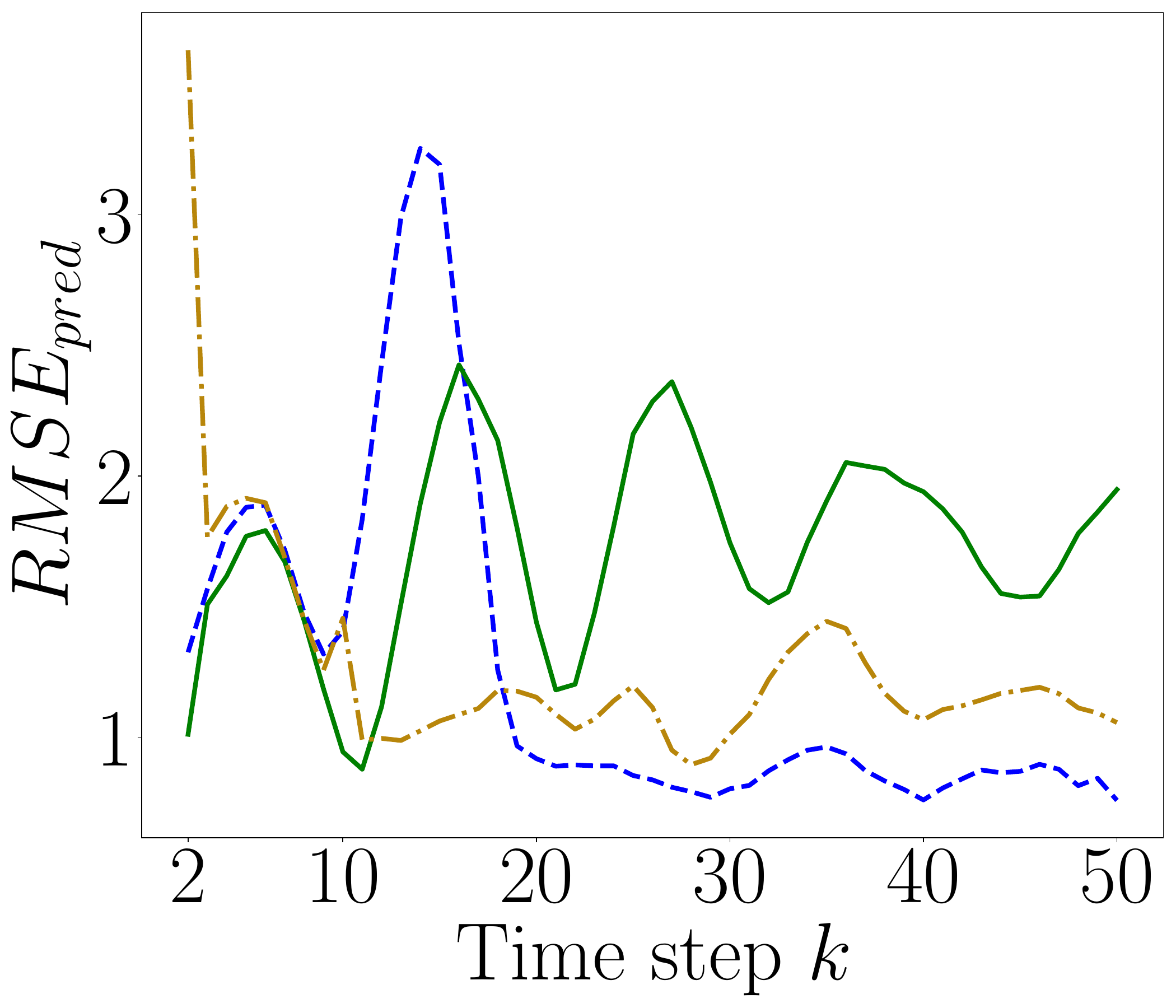}
     \caption{\textbf{Single pendulum:} time-step prediction error for the single pendulum system with a partially unknown dynamical model. The figure shows prediction errors calculated using results from FBOVI, joint PF and joint UKF. FBOVI provides the most accurate predictions among the three methods after step $18$.}
    \label{fig:rmse_pred}
\end{figure}

Figure~\ref{fig:rmse_pred} presents the prediction errors of different algorithms at each time step. FBOVI achieves the best predictive accuracy after time step $18$, although it underperforms the other two methods between steps $10$ and $16$. This temporary degradation can be partly attributed to its overestimation of uncertainty during that short period, as illustrated in Figure 9(b). Since the prediction error is computed based on the approximate joint posterior, FBOVI’s strong predictive performance demonstrates its ability to obtain accurate approximations of the joint posterior.

\subsubsection{System with partially unknown dynamical and observation models}
\label{sec:pendulum_part2}
In this section, we aim at learning the unknown parameters of the dynamical and observation models simultaneously. We show that the proposed method is able to yield good estimates of the dynamical and observation models as well as the states with both dynamical and observation models partially unknown.

The dynamical and observation models used by the learning agent are as follows:
\begin{equation*}
X_{k+1} = \begin{bmatrix}0.9594&\theta_1\\ -0.8056&0.9594\end{bmatrix}X_k + W_k, \quad W_k \sim \mathcal{N}(0, 0.01I_{2\times2}), \qquad Y_k = \begin{bmatrix}\theta_2&0\\0&1\end{bmatrix}X_k + V_k, \quad V_k \sim \mathcal{N}(0,0.01I_{2\times2}).
\end{equation*}
We assume that the parameters $\theta = (\theta_1,\theta_2)$ are unknown to the learning agent. The prior of $\theta$ is set to $\mathcal{N}([1 \quad 0]^T, I_{2\times 2})$. The covariances of the process and measurement noises are assumed to be known. 

To evaluate the efficacy of the proposed algorithm, data is generated with $H=I_{2\times2}$ for 100 steps. Next, FBOVI, joint PF and joint UKF are implemented. This problem is more challenging than the one discussed in Section~\ref{sec:pendulum_part1}. The joint PF with $10^4$ particles collapses and we increase the number of particles to $1\times 10^6$. 

\begin{figure}
     \centering
    \begin{subfigure}[b]{0.95\linewidth}
        \centering
        \includegraphics[width=\textwidth]{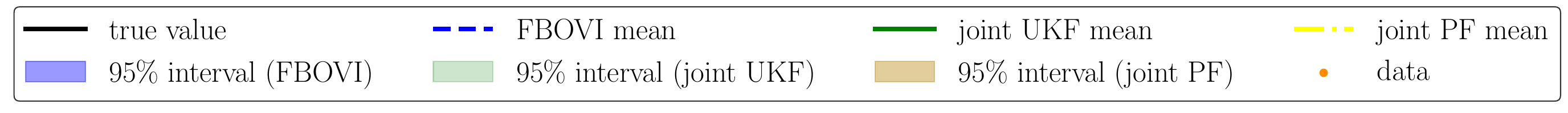}
    \end{subfigure}

    \vspace{0.8em}
     \begin{subfigure}[b]{0.24\linewidth}
         \centering
         \includegraphics[width=\textwidth]{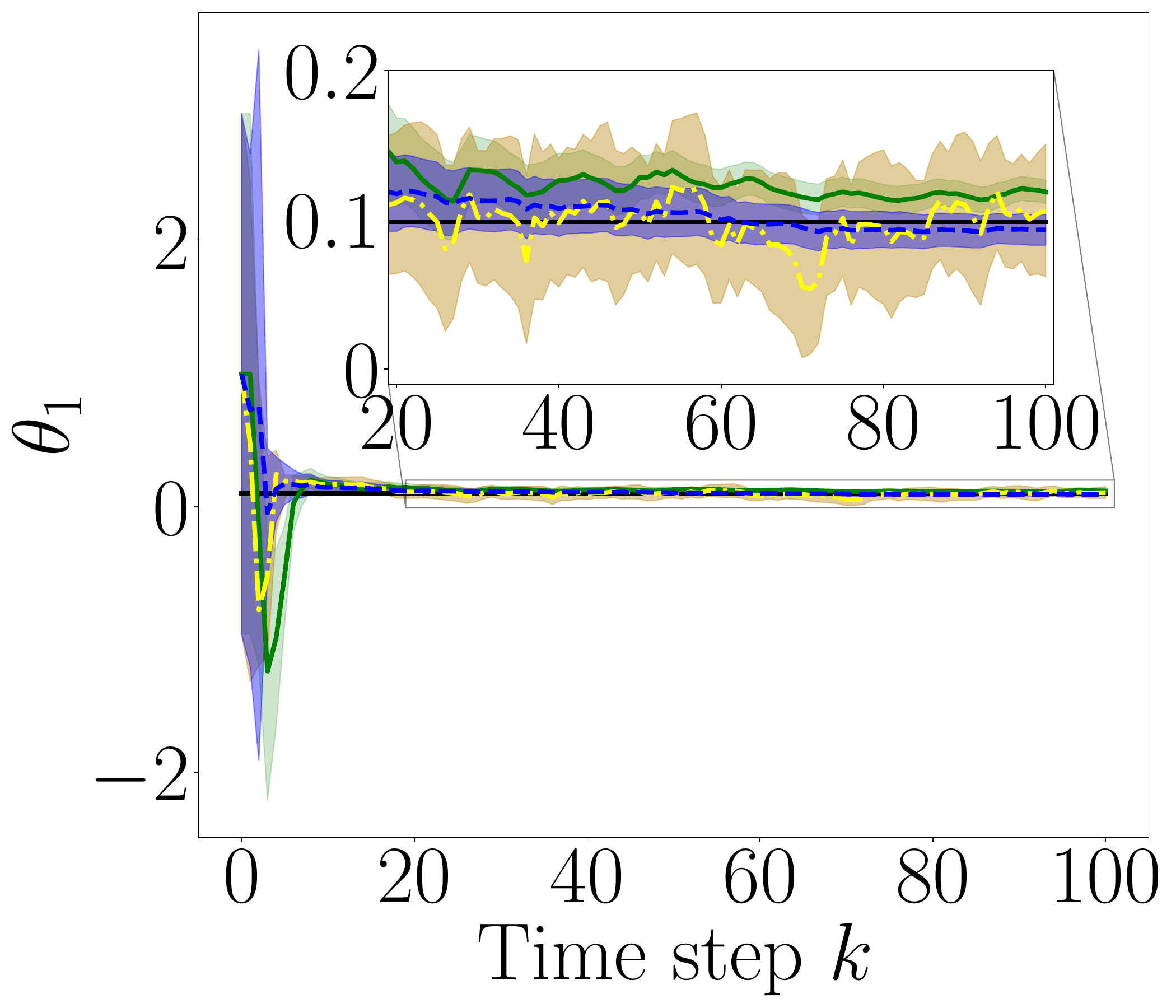}
         \caption{Distribution of $\theta_1$}
         \label{fig:dist_theta1_eg3}
     \end{subfigure}
     \hfill
     \begin{subfigure}[b]{0.24\linewidth}
         \centering
         \includegraphics[width=\textwidth]{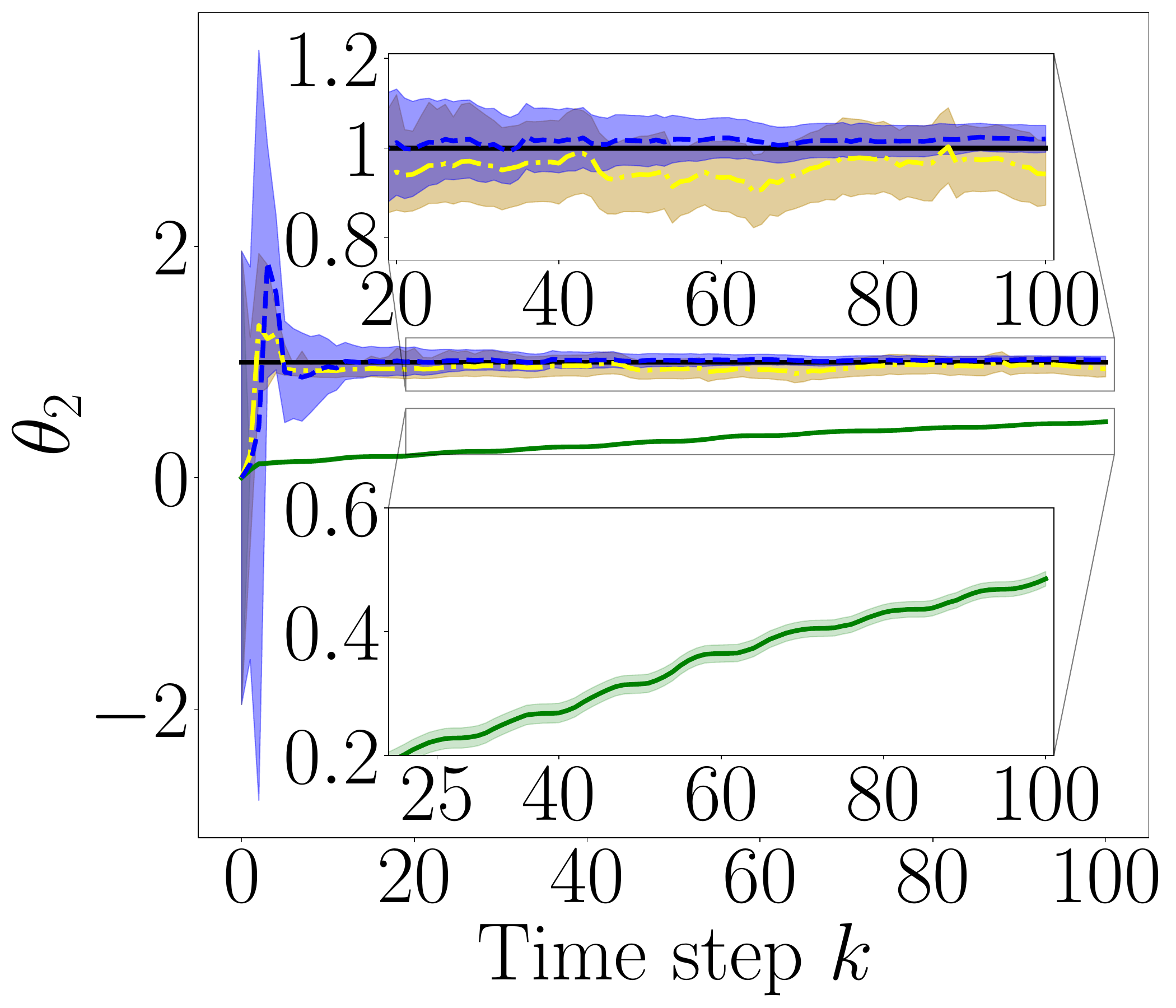}
         \caption{Distribution of $\theta_2$}
         \label{fig:dist_theta2_eg3}
     \end{subfigure}
     \hfill
     \begin{subfigure}[b]{0.24\linewidth}
         \centering
         \includegraphics[width=\textwidth]{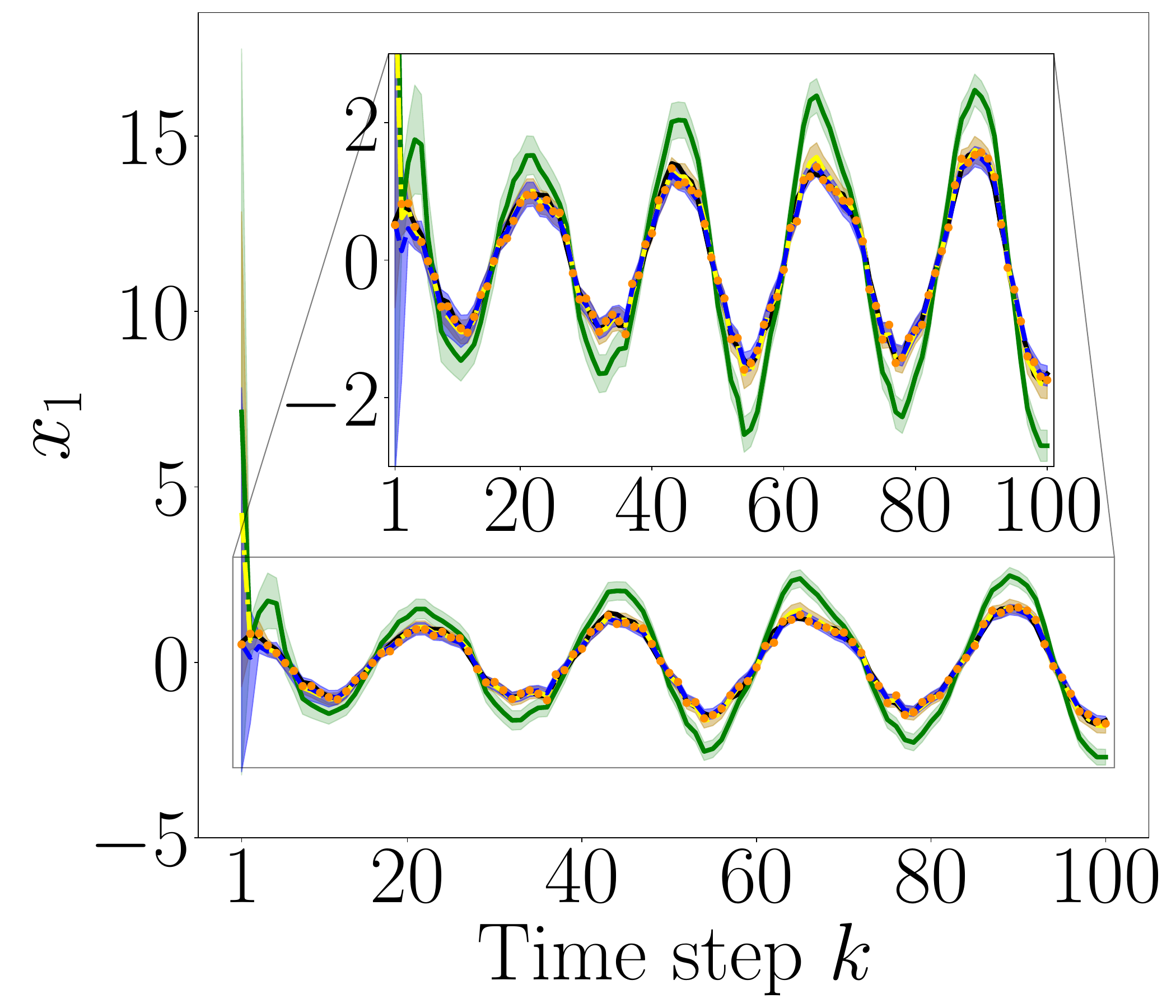}
         \caption{Distribution of $x_1(k)$}
         \label{fig:dist_x1_eg3}
     \end{subfigure}
     \hfill
     \begin{subfigure}[b]{0.24\linewidth}
         \centering
         \includegraphics[width=\textwidth]{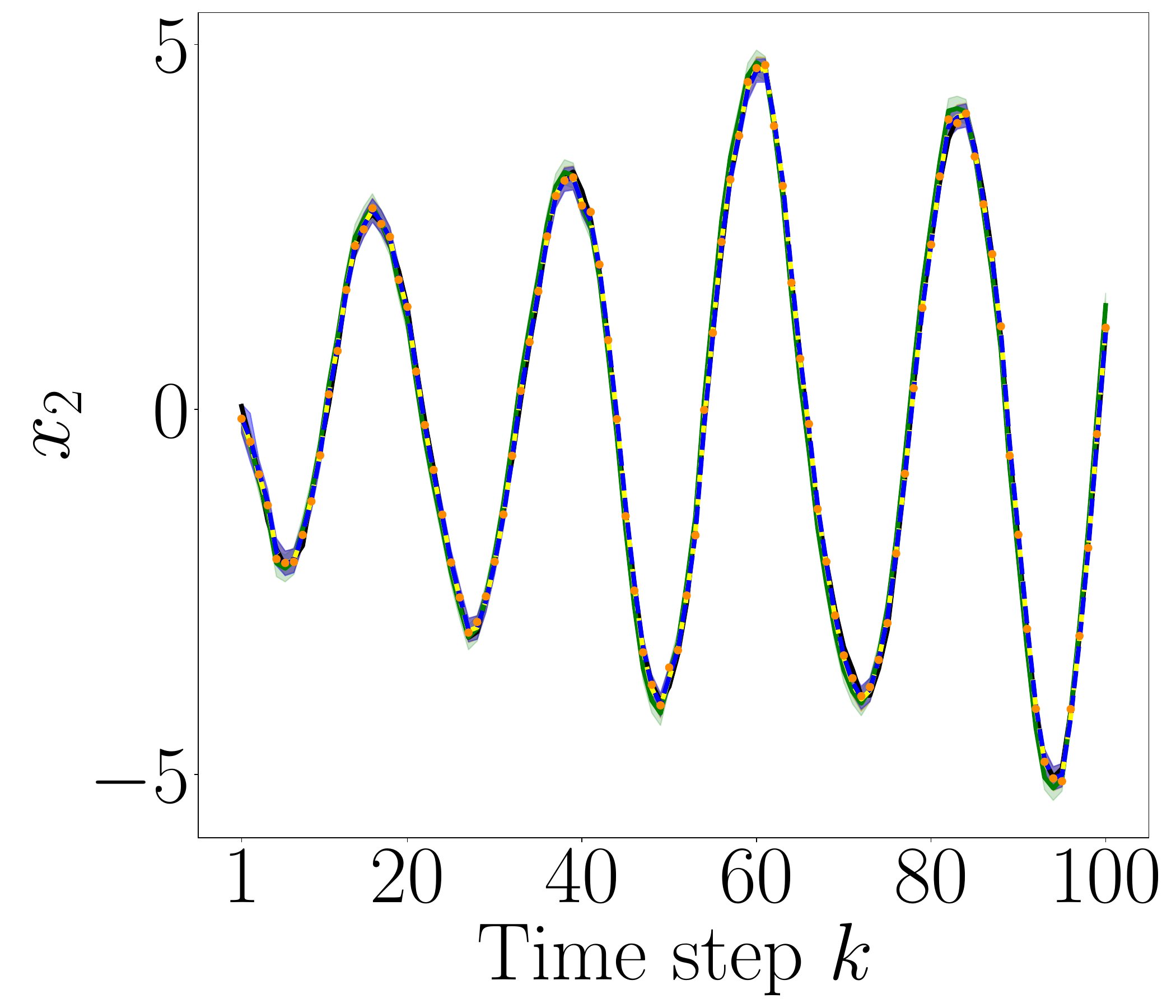}
         \caption{Distribution of $x_2(k)$}
         \label{fig:dist_x2_eg3}
     \end{subfigure}
         \caption{\textbf{Single pendulum:} approximate marginal posterior distributions of parameters and states for the single pendulum system with partially unknown dynamics and partially unknown observation model. For detailed analysis, the zoom-in views are provided in (a), (b), and (c). The estimates produced by FBOVI and the joint PF converge to the truth rapidly.}
         \label{fig:marginal_dist_eg3}
\end{figure}

The marginal distributions of $\theta$ and states obtained by different algorithms at each time step are shown in Figure \ref{fig:marginal_dist_eg3}. Remarkably, the means of FBOVI and joint PF for all parameters converge to the truth rapidly despite the inherent difficulty in learning $\theta_2$ as it constitutes an entry in the observation matrix $H$. In contrast, the entire distribution of $\theta_2$ obtained by the joint UKF deviates significantly from the truth for most of the duration. It can be seen that FBOVI and joint PF are both able to track the true states closely while the joint UKF tends to overestimate the amplitude of the trajectory of $x_1$, corresponding to an underestimation of $\theta_2$.

\subsection{Chaotic system: Lorenz '96 model}
\label{sec:Lorentz}
In this example, we evaluate the performance of the proposed method for parameter-state estimation in a chaotic, partially observable dynamical system, where learning is performed using both accurate models and models subject to structural errors. 

We consider a generalized Lorenz '96 model~\cite{Lorentz96_generalization}\footnote{The standard Lorenz '96 model is a special case of this generalized formulation.} with the state vector $x=[x_{1},x_{2},\cdots, x_{M}]^T$. The governing equations are given by
\begin{equation}
\dot{x_i} = \alpha x_{i-1}\left(x_{i+1}-x_{i-2}\right)-\beta x_{i}+F, \quad \forall i\in [1,M], 
\label{eq:Lorentz_dynamics_ct}
\end{equation}
where $x_{-1}:=x_{M-1}$, $x_{0}:=x_{M}$, and $x_{M+1}:=x_{1}$. The parameters $\alpha=1.1$ and $\beta=0.9$ are assumed to be unknown. The external forcing $F=8$ is known to the learning agent and the system dimension is $M=10$. Following~\cite{Lorentz96_initial_condition}, the initial condition is set as $x_i = \cos(2\pi i/M), \; \forall i \in [1,M]$, which is unknown to the learning agent. 
The system is simulated using a time step $\Delta t=0.01$. Noisy measurements of the odd-numbered state components $x_1,x_3,x_5,x_7$, and $x_9$ are collected every 5 simulation steps. The measurement noise follows a Gaussian distribution with zero mean and covariance $I_{5\times5}$. This measurement model is known to the learning agent. To ensure that the learning process is performed when the system is operating in the chaotic regime, the learning process begins at the $3005$th simulation step. The 4th-order runge-kutta (RK4) method is used to generate reference solution trajectories and data.

 Let $\theta = (\alpha, \beta)$ denote the vector of unknown parameters. We assign $\theta$ a Gaussian prior $\mathcal{N}(0_2, I_{2\times2})$. The discrete-time dynamical model used by the learning agent is as follows:
\begin{equation}
X_{k+1} = \Phi(X_k; \theta) + W_k, \quad W_k \sim \mathcal{N}(0, \sigma^2 I_{M \times M}),
\label{eq:learning_model}
\end{equation}
where the integer $k \geq 0$ represents the time instance, and each time step corresponds to a duration of $5\Delta t$, i.e., one learning step corresponds to $5$ simulation steps. The function $\Phi$ is a discrete-time approximation of the continuous-time dynamics in Equation~\eqref{eq:Lorentz_dynamics_ct}, obtained by applying the forward Euler method or RK4 over a time interval of length $5\Delta t$, with simulation time step $\Delta t$. We present the results of our method applying both the forward Euler method and RK4 method to obtain the function $\Phi$, as detailed in Section~\ref{sec:correct_model} and Section~\ref{sec:wrong_model}. The joint PF with $10^6$ particles collapses in this example due to the increasing dimensionality, and we therefore compare the performance of FBOVI with the joint UKF.
\subsubsection{Correct model}
\label{sec:correct_model}
In this case, the RK4 method is used to obtain the function $\Phi(X_k;\theta)$. Therefore, the deterministic part of the model used for learning, $\Phi$, matches the dynamical system which is used to generate the reference solutions. The variance of the process noise is chosen to be $\sigma^2=0.5$. 

\begin{figure}
     \centering
    \begin{subfigure}[b]{0.9\linewidth}
        \centering
        \includegraphics[width=\textwidth]{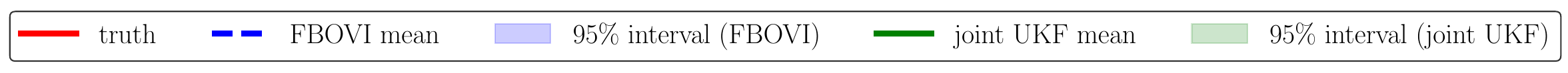}
    \end{subfigure}

    \vspace{0.8em}
   
     \begin{subfigure}[b]{0.27\linewidth}
         \centering
         \includegraphics[width=\textwidth]{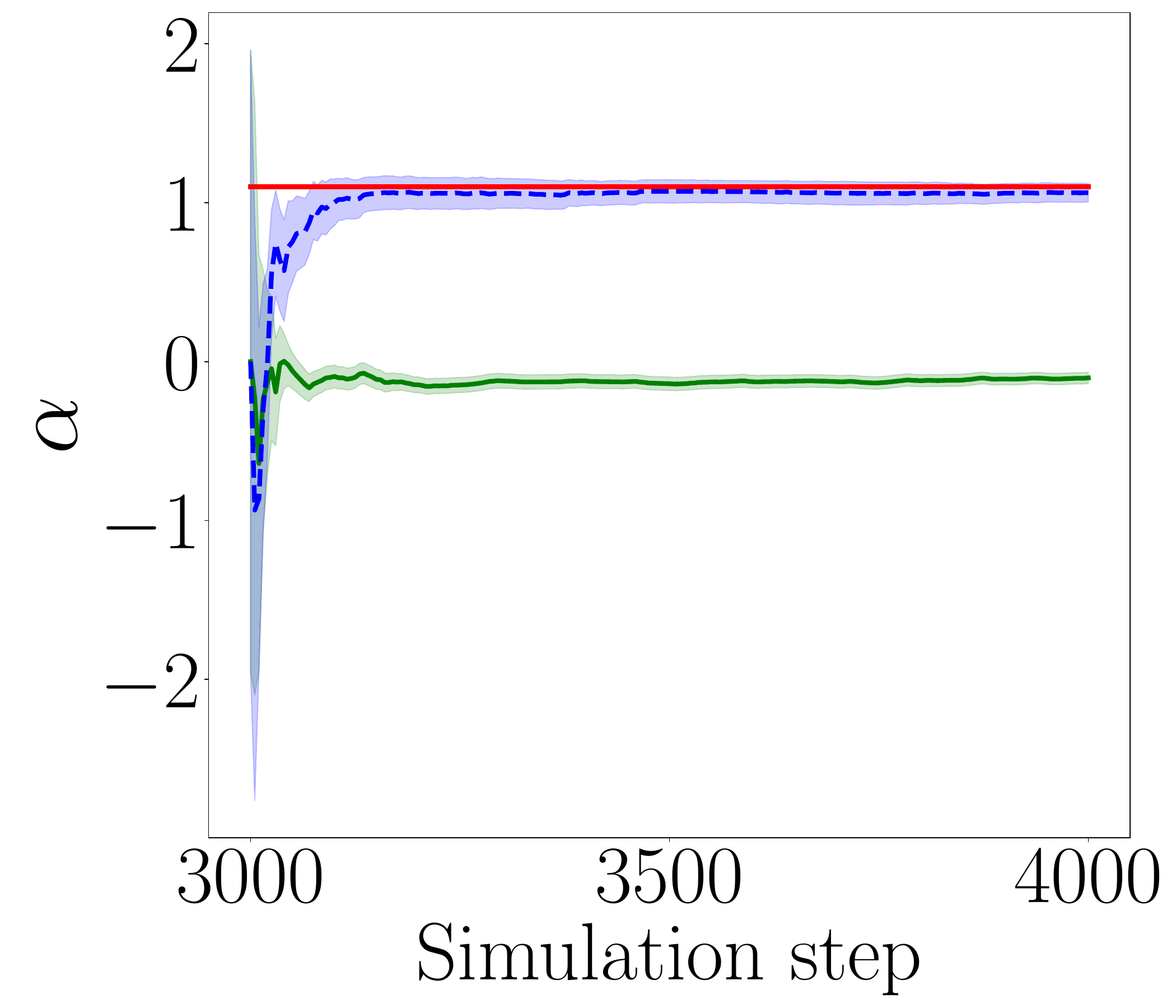}
         \caption{Approximate marginal posterior distribution of $\alpha$}
         \label{fig:alpha_e2_1}
     \end{subfigure}
     \hspace{0.1\linewidth}
     \begin{subfigure}[b]{0.27\linewidth}
         \centering
         \includegraphics[width=\textwidth]{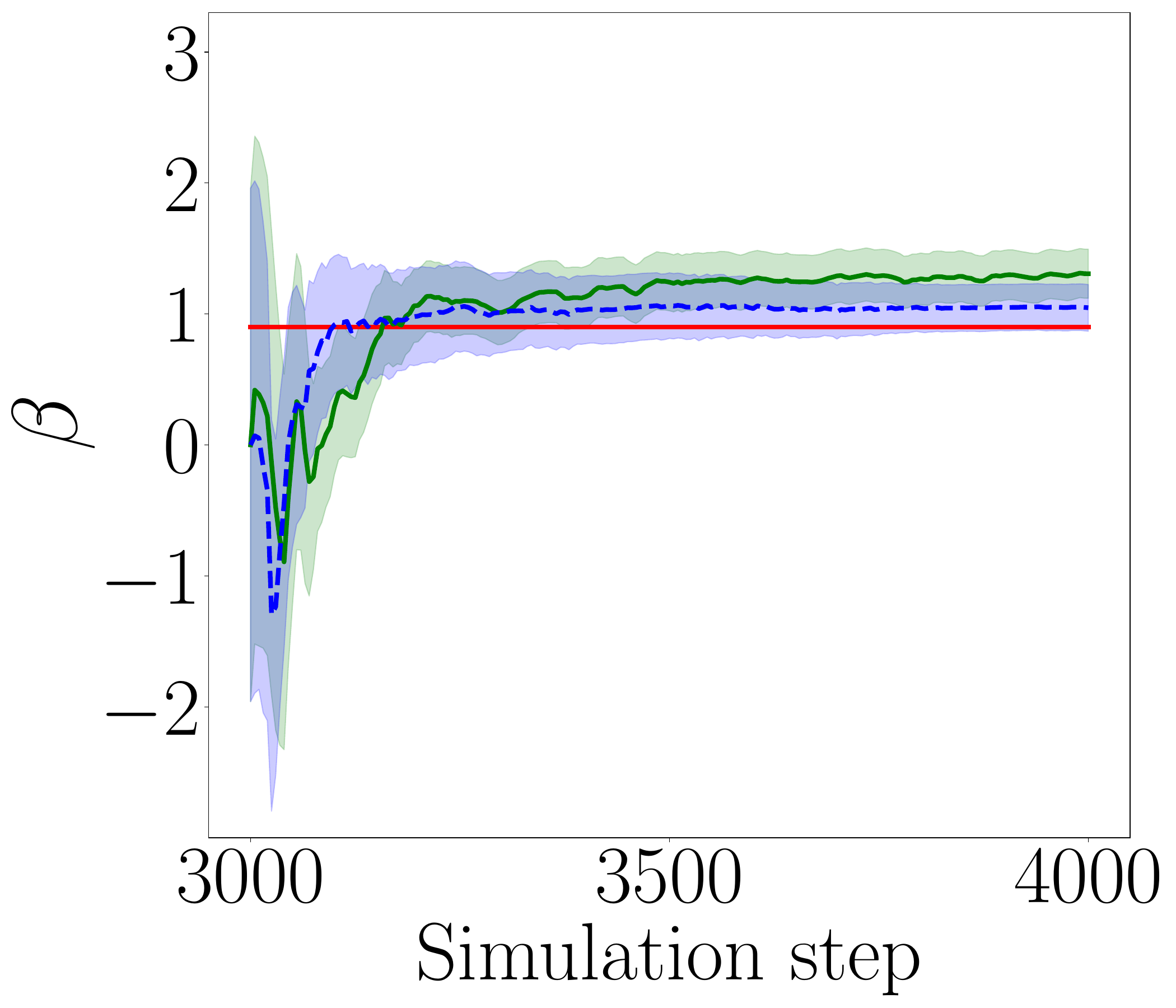}
         \caption{Approximate marginal posterior distribution of $\beta$}
         \label{fig:beta_e2_1}
     \end{subfigure}
     \caption{\textbf{Lorentz '96 system with correct model:}
     approximate marginal posterior distributions of the parameters $\alpha$ and $\beta$ when the deterministic part of the dynamical model used for inference matches the true model. FBOVI achieves substantially higher estimation accuracy than joint UKF.}
     \label{fig:param_accurateModel}
\end{figure}

\begin{figure}[t]
     \centering
    \begin{subfigure}[b]{1.0\linewidth}
        \centering
        \includegraphics[width=\textwidth]{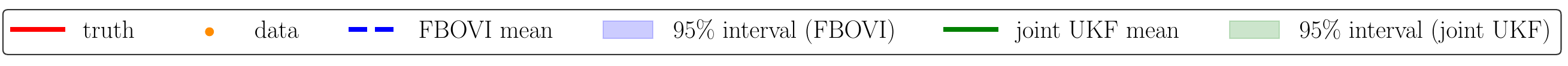}
    \end{subfigure}

     \vspace{0.4em}
     \includegraphics[width=1.0\textwidth]{figures/Lorentz/accurateModel/estimation/estimation_accurateModel.pdf}
         \caption{\textbf{Lorentz '96 system with correct model:} approximate marginal posterior distributions of the state components obtained by FBOVI and joint UKF when the deterministic part of the dynamical model used for inference matches the true model. Top row, from left to right: the approximate marginal posterior distributions of $x_1$, $x_2$, $x_3$, $x_4$, and $x_5$; bottom row, from left to right: the approximate marginal posterior distributions of $x_6$, $x_7$, $x_8$, $x_9$, and $x_{10}$. Shaded regions represent the $95\%$ credible intervals. FBOVI accurately tracks both the observed and unobserved state components.}
         \label{fig:state_accurateModel}
\end{figure}

Figure~\ref{fig:param_accurateModel} presents the approximate marginal posterior distributions of the parameters obtained by FBOVI and joint UKF. The FBOVI estimate for the parameter $\alpha$ converges rapidly to the true value, while joint UKF fails to learn $\alpha$. For the estimation of the parameter $\beta$, although the FBOVI mean converges to a value slightly different from the true value, FBOVI still significantly outperforms the joint UKF. The state estimation results are depicted in Figure~\ref{fig:state_accurateModel}. FBOVI provides accurate estimates for both the observed and unobserved state components for most time steps, whereas the distributions obtained by joint UKF are often far from the truth. The accurate estimation of unobserved state components is particularly noteworthy, given that only half of the state components are observed and the model is partially unknown. This challenge is further exacerbated by the system's operation in a chaotic regime.

\begin{figure}[t]
     \centering
    \begin{subfigure}[b]{0.9\linewidth}
        \centering
        \includegraphics[width=\textwidth]{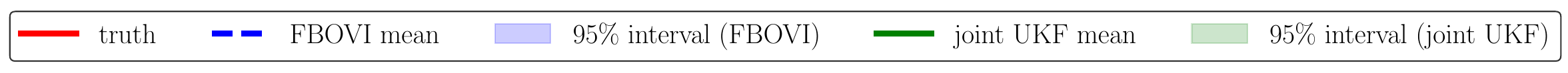}
    \end{subfigure}

     \vspace{0.4em}
     \includegraphics[width=1.0\textwidth]{figures/Lorentz/accurateModel/prediction/prediction_accurateModel_3steps_3200.pdf}
         \caption{\textbf{Lorentz '96 system with correct model (prediction):} the distributions of the predicted state components. The prediction of $X_k$ is based on the approximate joint posterior obtained at the previous learning step $k-3$. The deterministic part of the dynamical model used for estimation and \textit{prediction} matches the true system. Top row, from left to right: the predictive distributions of $x_1$, $x_2$, $x_3$, $x_4$, and $x_5$; bottom row, from left to right: the predictive distributions of $x_6$, $x_7$, $x_8$, $x_9$, and $x_{10}$. Shaded regions indicate the $95\%$ credible intervals. FBOVI substantially outperforms the joint UKF and provides accurate predictions for most time steps.}
         \label{fig:state_pred_accurateModel}
\end{figure}

Figure~\ref{fig:state_pred_accurateModel} shows the predictive distribution of the state $X_k$, generated using the approximate joint posterior obtained at learning step $k-3$. Specifically, we draw $N=1\times 10^5$ samples $\{(X_{k-3}^{(i)}, \theta^{(i)})\}_{i=1}^N$ from the approximate joint posterior $q_{k-3}(X_{k-3},\theta)$. For each sample $(X_{k-3}^{(i)}, \theta^{(i)})$, we compute a predicted state $\hat{X}_k^{(i)}$ by iteratively applying the model in Equation~\eqref{eq:learning_model} with $\theta=\theta^{(i)}$:
\begin{equation*}
\hat{X}_{j+1}^{(i)} = \Phi(\hat{X}_{j}^{(i)};\theta^{(i)}) + V_{j}^{(i)}, 
\quad 
V_{j}^{(i)}\sim \mathcal{N}(0,\sigma^2 I_{M \times M}), 
\quad 
\text{for } j = k-2, k-1,
\end{equation*}
where $\hat{X}_{k-2}^{(i)} := \Phi(X_{k-3}^{(i)};\theta^{(i)}) + V_{k-3}^{(i)}, 
\quad 
V_{k-3}^{(i)}\sim \mathcal{N}(0,\sigma^2 I_{M \times M})$. The resulting set $\{\hat{X}_k^{(i)}\}_{i=1}^N$ represents the predictive distribution of $X_k$, which we compute for $k \geq 40$. For a chaotic system, using the approximate joint posterior at the learning step $k-3$ to predict the state at step $k$ is challenging, because small parameter-state estimation errors can be amplified by the chaotic dynamics and lead to substantially larger prediction errors. Despite this difficulty, FBOVI achieves accurate predictions for most time steps, as illustrated in Figure~\ref{fig:state_pred_accurateModel}.

\begin{figure}[t]
     \centering
    \begin{subfigure}[b]{0.8\linewidth}
        \centering
        \includegraphics[width=\textwidth]{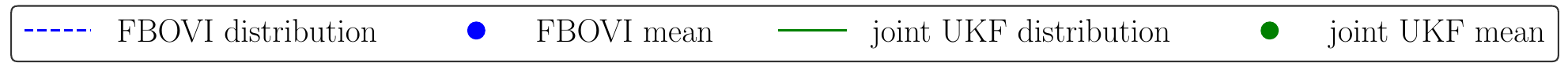}
    \end{subfigure}
    
     \begin{subfigure}[b]{0.32\linewidth}
         \centering
         \includegraphics[width=\textwidth]{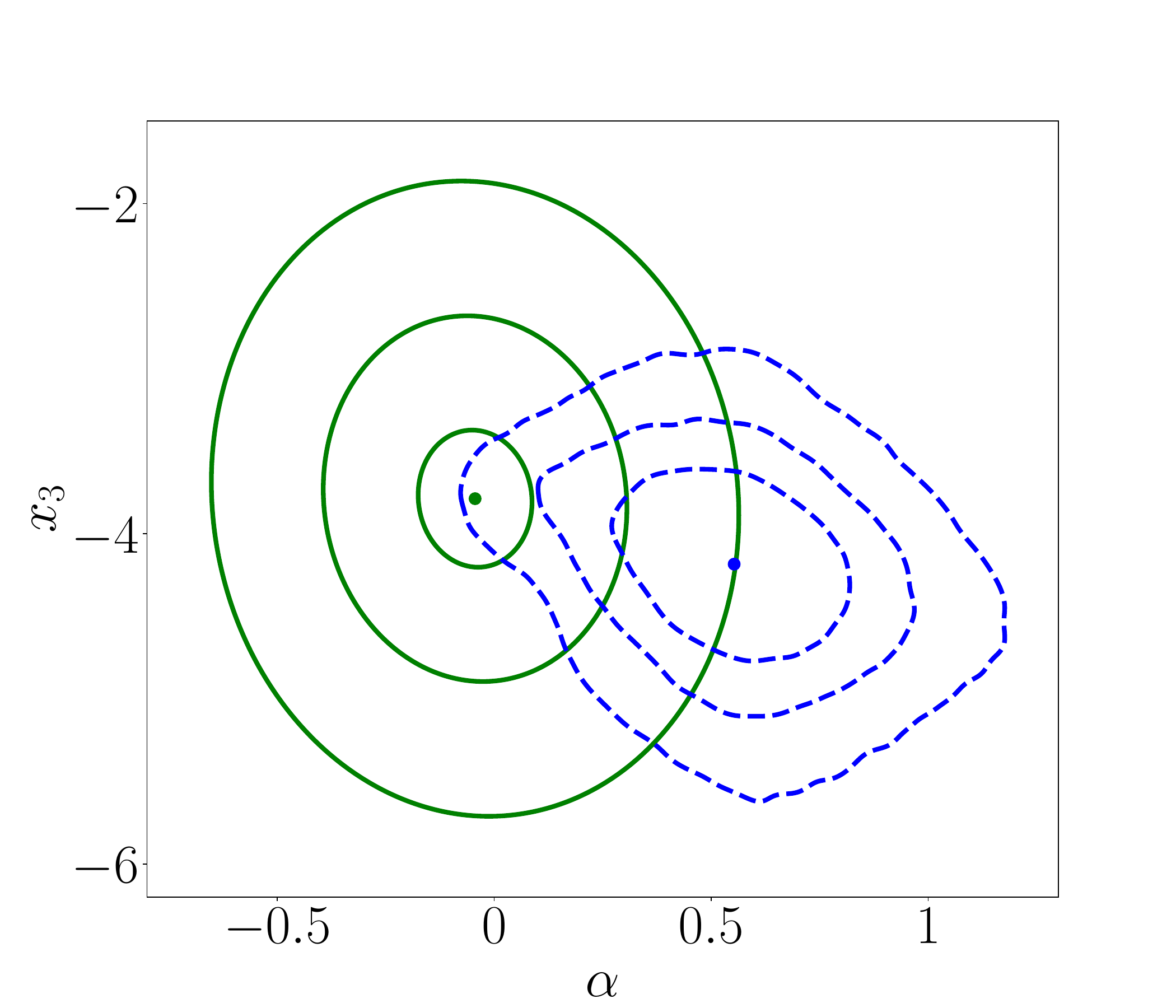}
         \caption{ joint distribution of $x_3$ and $\alpha$}
         \label{fig:alpha_x3_alpha}
     \end{subfigure}
     \hspace{0.005\linewidth}
     \begin{subfigure}[b]{0.32\linewidth}
         \centering
         \includegraphics[width=\textwidth]{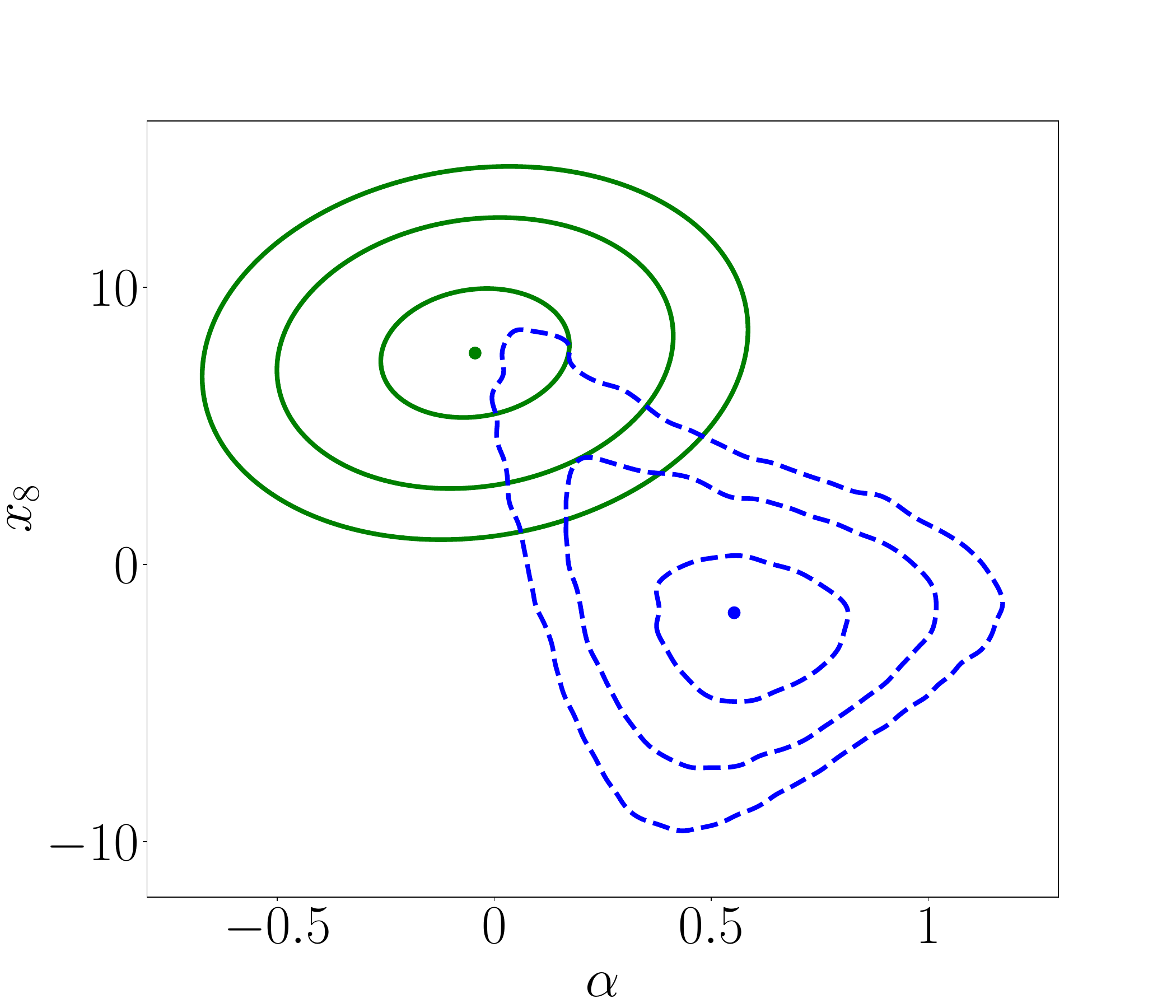}
         \caption{joint distribution of $x_8$ and $\alpha$}
         \label{fig:alpha_x2_alpha}
     \end{subfigure}
     \hspace{0.005\linewidth}
     \begin{subfigure}[b]{0.32\linewidth}
         \centering
         \includegraphics[width=\textwidth]{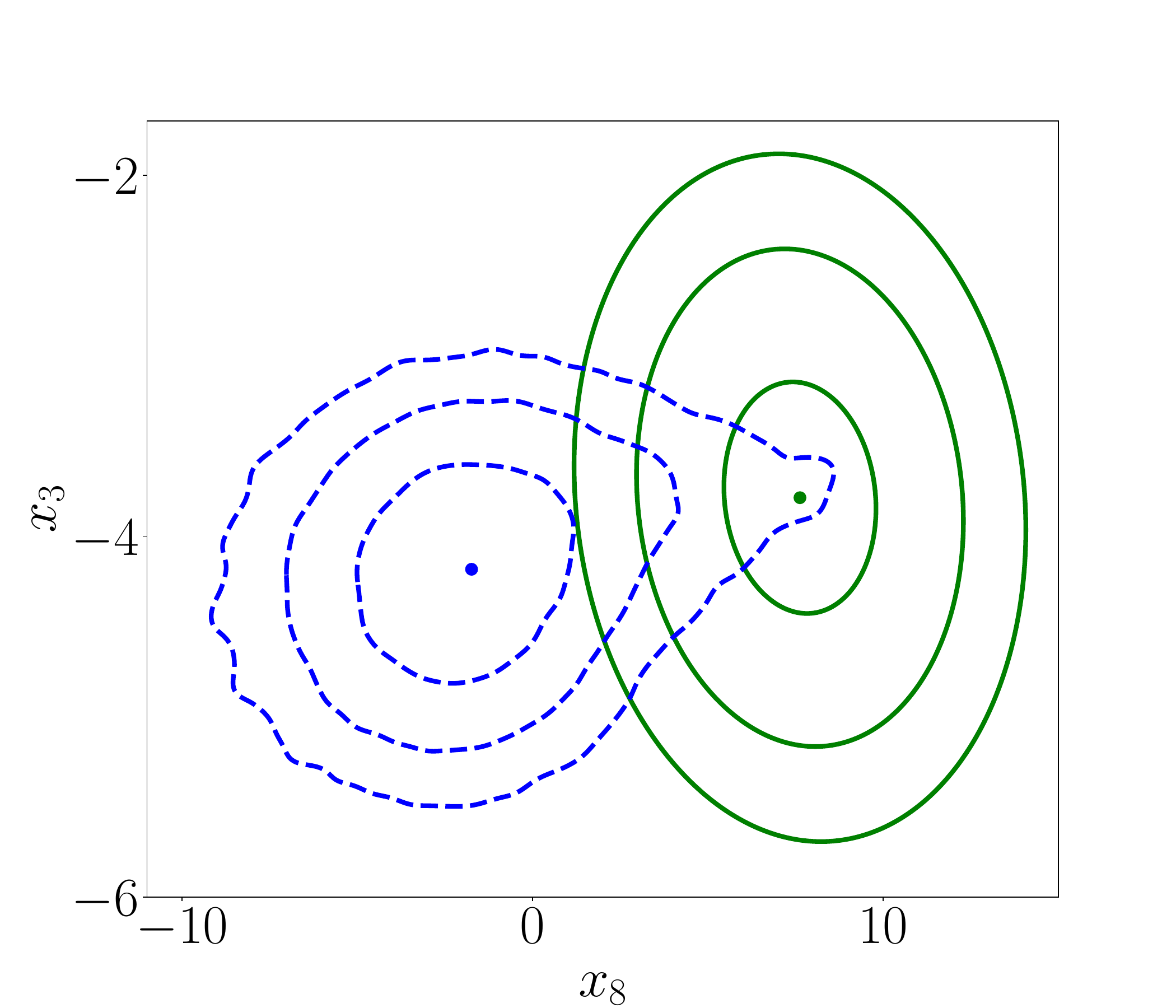}
         \caption{joint distribution of $x_3$ and $x_8$}
         \label{fig:beta_x10_alpha}
     \end{subfigure}
     \caption{\textbf{Lorentz '96 system with correct model:} representative examples of approximate joint posterior distributions at the simulation step $3025$ for the Lorenz '96 system under the correct model. Contour lines represent density levels of the joint distributions. FBOVI captures the non-Gaussian characteristics in both parameter-state and state-state joint distributions. Notably, although Gaussian filtering is used to approximate the conditional distributions of the state given parameters, the overall resulting joint distributions remain expressive because the framework avoids imposing mean-field assumptions.}
         \label{fig:joint_dist_e2}
\end{figure}

In the early stages, the joint posterior in nonlinear systems is often non-Gaussian. To demonstrate that FBOVI is able to capture such non-Gaussian features, we present the approximate joint posteriors obtained by FBOVI and the joint UKF. Figure~\ref{fig:joint_dist_e2} shows representative examples of these approximate joint posterior distributions at the simulation step $3025$, i.e., the learning step $k=5$. The results show that FBOVI is capable of capturing significant non-Gaussian features in both parameter-state and state-state joint distributions. Importantly, although FBOVI employs Gaussian filtering to approximate the conditional distribution of the state given the parameters, not imposing a mean-field assumption allows the resulting joint distributions to retain the inherent non-Gaussian dependence structure between the parameters and the state. As a result, the marginal posterior distributions of the states, after integrating out the parameters, can still exhibit rich, non-Gaussian behavior.

\subsubsection{Incorrect model}
\label{sec:wrong_model}
Next, we study whether the proposed method can still provide accurate estimation and prediction when the model used for learning deviates from the true system dynamics. Specifically, we use the forward Euler method to compute the deterministic component of the dynamical model used for learning, namely $\Phi$. Consequently, $\Phi$ differs from the true model. To compensate for the model-form error, the process noise variance is increased to $\sigma^2=2$. In this setting, the notion of ``true values'' for the parameters no longer applies, as the model form used for learning does not match the true dynamics. We evaluate the performance of the proposed method based on its ability to estimate the current states, with particular focus on the unobserved ones, and its predictive accuracy for the state of the subsequent step.

\begin{figure}[t]
     \centering
    \begin{subfigure}[b]{1.0\linewidth}
        \centering
        \includegraphics[width=\textwidth]{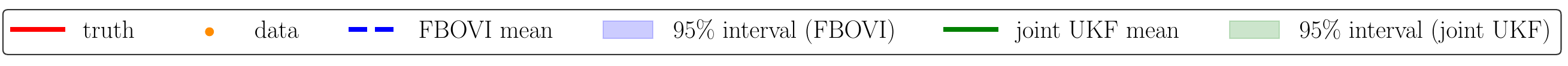}
    \end{subfigure}
    \vspace{0.4em}
     \includegraphics[width=1.0\textwidth]{figures/Lorentz/inaccurateModel/estimation/estimation_inaccurateModel.pdf}
         \caption{\textbf{Lorentz '96 system with model form error:} approximate marginal posterior distributions of the states when the deterministic part of the model used for learning differs from the true system. Top row, from left to right: the approximate marginal posterior distributions of $x_1$, $x_2$, $x_3$, $x_4$, and $x_5$; bottom row, from left to right: approximate marginal posterior distributions of $x_6$, $x_7$, $x_8$, $x_9$, and $x_{10}$. The shaded regions represent the $95\%$ credible intervals. Under model-form inaccuracies, FBOVI provides relatively robust estimates for all state components.}
         \label{fig:state_inaccurateModel}
\end{figure}

Figure~\ref{fig:state_inaccurateModel} shows the approximate marginal posterior distributions of the state components. Despite the use of an inaccurate model form, FBOVI is able to track all states reasonably well. However, in comparison to the results obtained under the correct model form, the estimation quality around some peaks and troughs of the trajectories degrades. This degradation is likely attributable to the limited expressiveness of the incorrect model used for learning. 

\begin{figure}[t]
     \centering
    \begin{subfigure}[b]{0.9\linewidth}
        \centering
        \includegraphics[width=\textwidth]{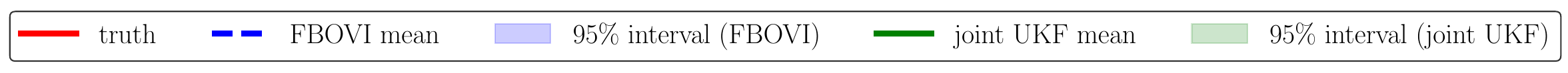}
    \end{subfigure}

\vspace{0.4em}
     \includegraphics[width=1.0\textwidth]{figures/Lorentz/inaccurateModel/prediction/prediction_inaccurateModel_3steps_3200.pdf}
     \caption{\textbf{Lorentz '96 system with model form error (prediction):} distributions of the predicted state components when the deterministic part of the dynamical model used for estimation and \textit{prediction} differs from the true system. The prediction for the state at each learning step $k$ is based on the approximate joint posterior obtained at the previous learning step $k-3$. Top row, from left to right: the predictive distributions of $x_1$, $x_2$, $x_3$, $x_4$, and $x_5$; bottom row, from left to right: the predictive distributions of $x_6$, $x_7$, $x_8$, $x_9$, and $x_{10}$. Shaded regions denote the $95\%$ credible intervals. }
         \label{fig:state_pred_inaccurateModel}
\end{figure}

Figure~\ref{fig:state_pred_inaccurateModel} presents the distributions of the predicted states, where the predicted values are computed in the same way as described in Section~\ref{sec:correct_model}. Under model-form error, prediction becomes more challenging because structural model inaccuracies introduce additional propagation errors, and the chaotic dynamics can amplify errors in the estimated states and parameters. As shown in Figure~\ref{fig:state_pred_inaccurateModel}, FBOVI still yields reasonably satisfactory predictive performance. Compared to the results under the correct model form shown in Figure~\ref{fig:state_pred_accurateModel}, the predictive accuracy around some peaks and troughs is noticeably lower. This lower predictive accuracy is expected, as structural model inaccuracies adversely affect the joint posterior approximation at the previous learning step and inherently limit the predictive capability of the learned model.

We note here that the proposed framework is not designed to solve the general problem of model misspecification. In this section, we demonstrate that our approach is robust under small to moderate deviations from the true model form. However, its performance may degrade significantly under severe model misspecification, such as when the model used for inference is time-invariant while the true model changes substantially over time.

\subsubsection{Highly noisy measurements}
\begin{figure}
     \centering
    \begin{subfigure}[b]{0.3\linewidth}
        \centering
        \includegraphics[width=\textwidth]{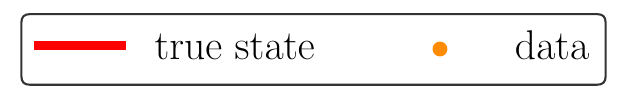}
    \end{subfigure}

    \vspace{0.8em}
   
     \begin{subfigure}[b]{0.28\linewidth}
         \centering
         \includegraphics[width=\textwidth]{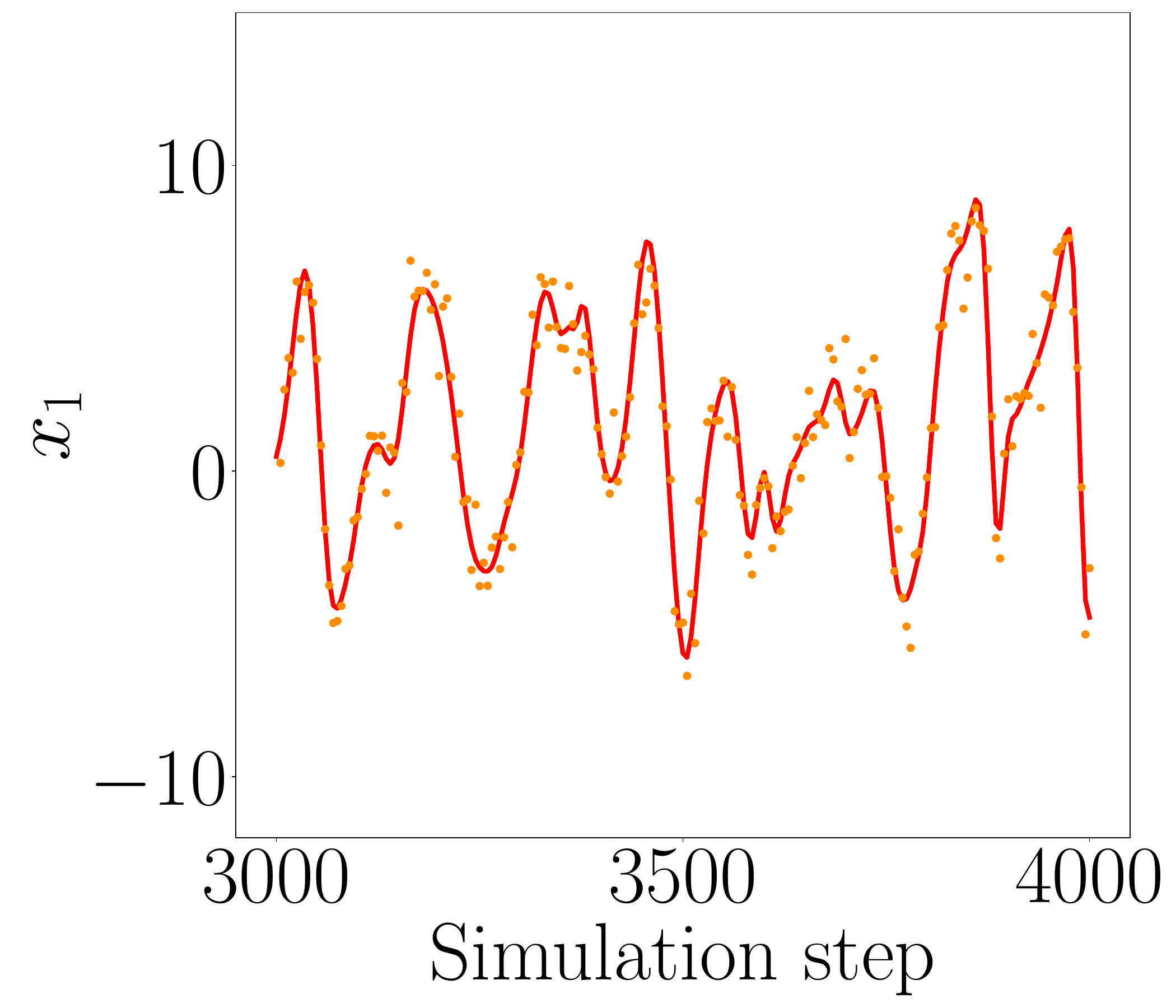}
         \caption{$\gamma=1$}
         \label{fig:data_e2_highNoise_1}
     \end{subfigure}
     \hspace{0.05\linewidth}
     \begin{subfigure}[b]{0.28\linewidth}
         \centering
         \includegraphics[width=\textwidth]{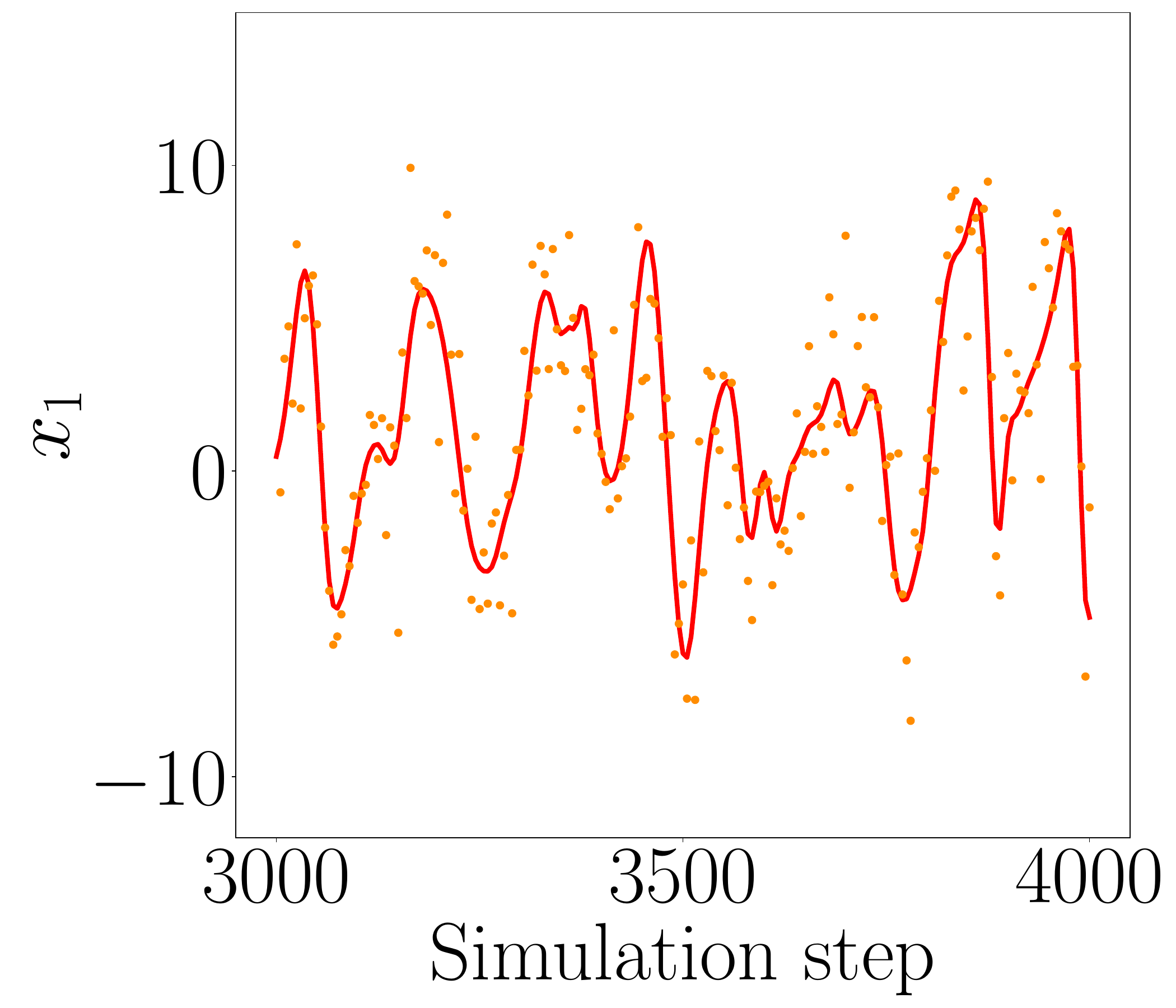}
         \caption{$\gamma=5$}
         \label{fig:data_e2_highNoise_5}
     \end{subfigure}
     \hspace{0.05\linewidth}
     \begin{subfigure}[b]{0.28\linewidth}
         \centering
         \includegraphics[width=\textwidth]{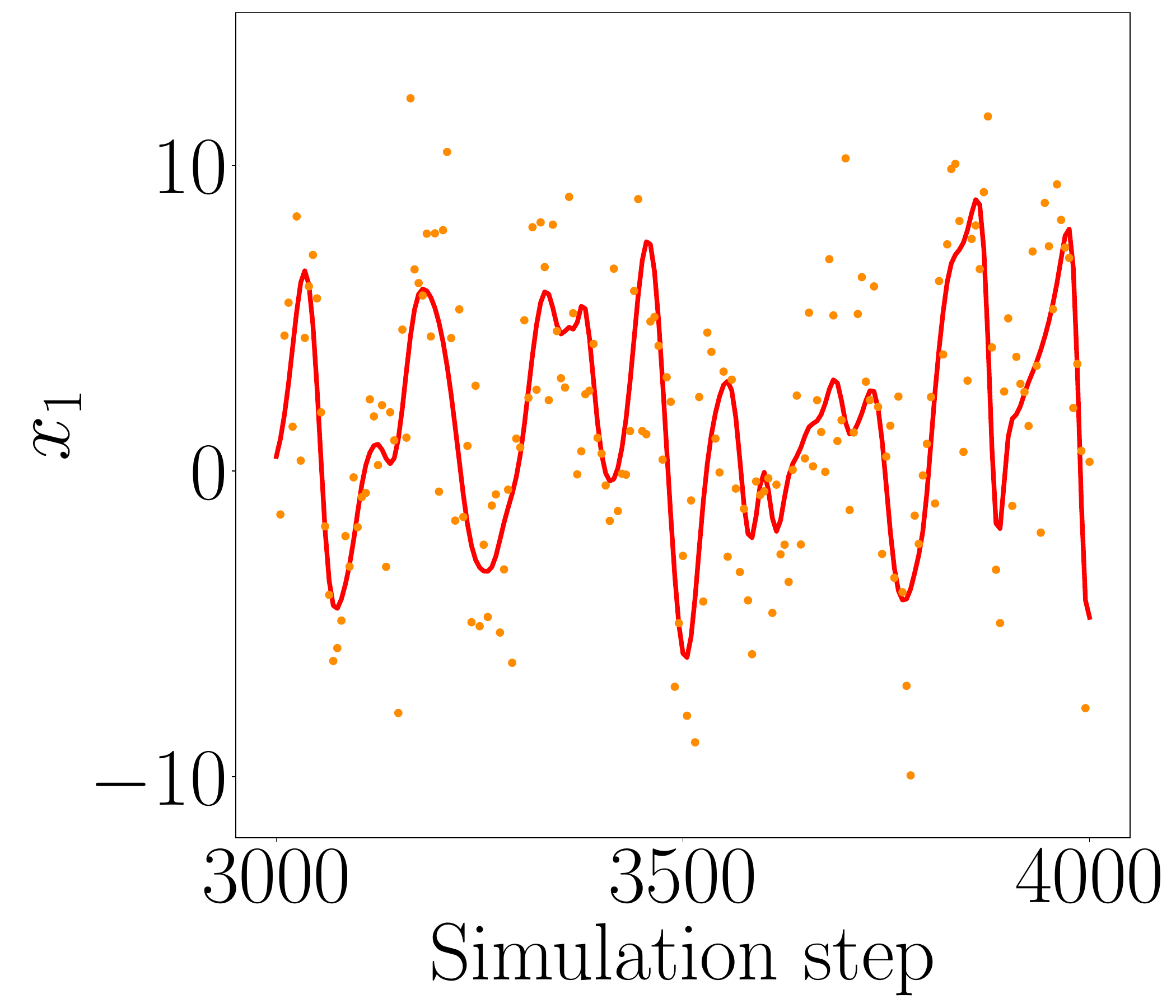}
         \caption{$\gamma=10$}
         \label{fig:data_e2_highNoise_10}
     \end{subfigure}
     \caption{
     \textbf{True trajectory of state component $x_1$ and its measurements under different measurement noise levels:}
     when $\gamma=1$, the data noise level is moderate. When $\gamma=5$, the data are significantly noisier. When $\gamma=10$, the data are highly noisy, which can make the inference problem particularly challenging.}
     \label{fig:data_accurateModel_highNoise}
\end{figure}

In this section, we study the robustness of FBOVI to high measurement noise. The settings are identical to those in the experiment presented in Section~\ref{sec:correct_model}, except that the measurement noise $V_k$ now satisfies
\begin{equation*}
V_k \sim \mathcal{N}(0, \gamma I_{5 \times 5}),
\end{equation*}
where $\gamma$ is increased from $1$ which is used in Section~\ref{sec:correct_model}, to $5$ and $10$. We collect two additional sets of measurements, one with $\gamma=5$ and the other with $\gamma=10$. The measurements and true values of the state component $x_1$ for $\gamma=1,5,10$ are presented in Figure~\ref{fig:data_accurateModel_highNoise}. As shown in Figure~\ref{fig:data_accurateModel_highNoise}, the measurement data are considerably noisy when $\gamma=5$ and highly noisy when $\gamma=10$.

\begin{figure}
     \centering
    \begin{subfigure}[b]{0.5\linewidth}
        \centering
        \includegraphics[width=\textwidth]{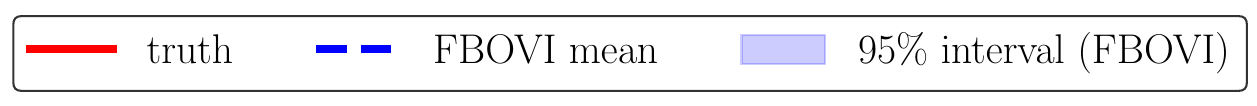}
    \end{subfigure}

    \vspace{0.8em}
\begin{subfigure}[b]{0.163\linewidth}
    \centering
    \includegraphics[width=\textwidth]{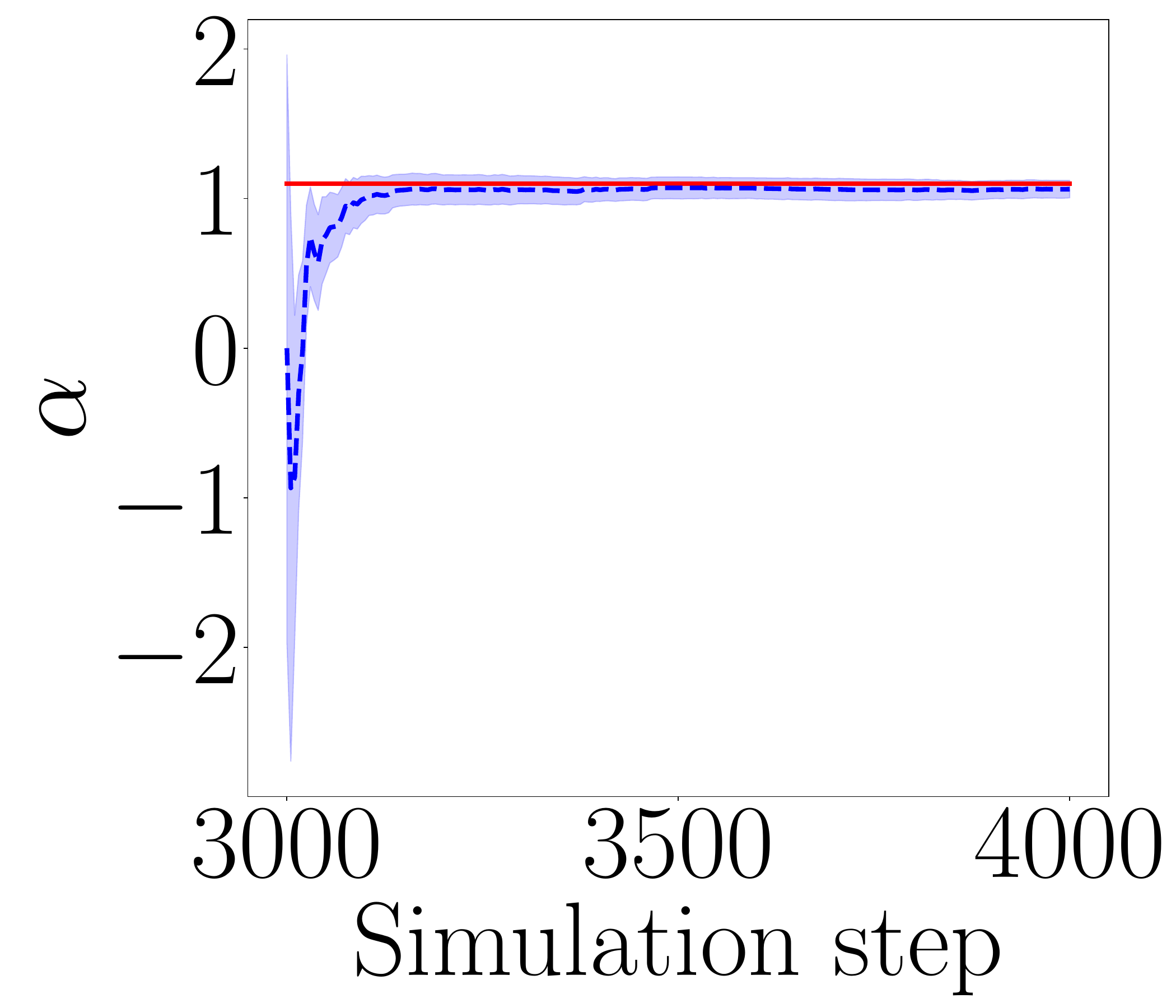}
    \caption{$\gamma=1$}
    \label{fig:alpha_e2_1_highNoise_1}
\end{subfigure}%
\hspace{0.003\linewidth}%
\begin{subfigure}[b]{0.163\linewidth}
    \centering
    \includegraphics[width=\textwidth]{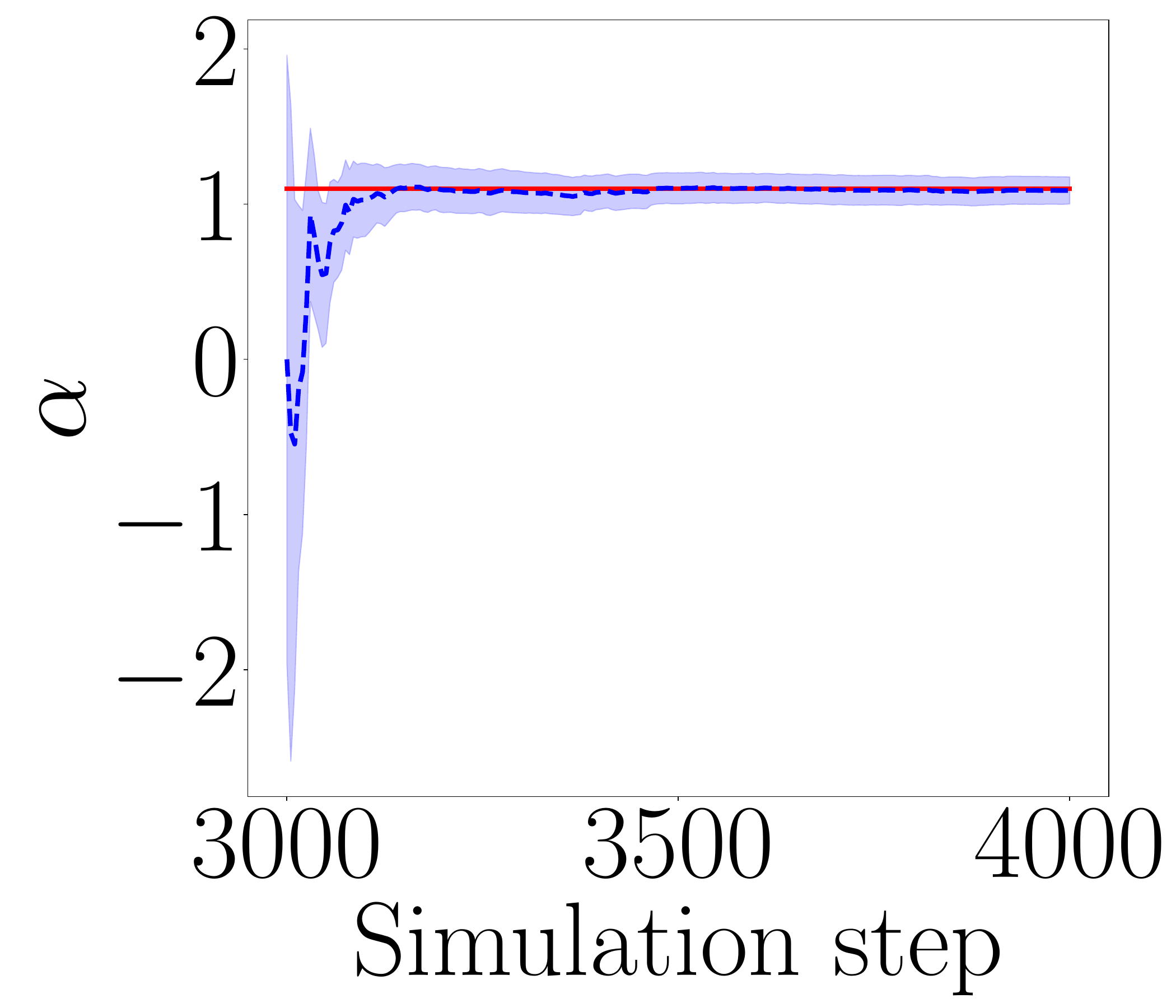}
    \caption{$\gamma=5$}
    \label{fig:alpha_e2_1_highNoise_5}
\end{subfigure}%
\hspace{0.003\linewidth}%
\begin{subfigure}[b]{0.163\linewidth}
    \centering
    \includegraphics[width=\textwidth]{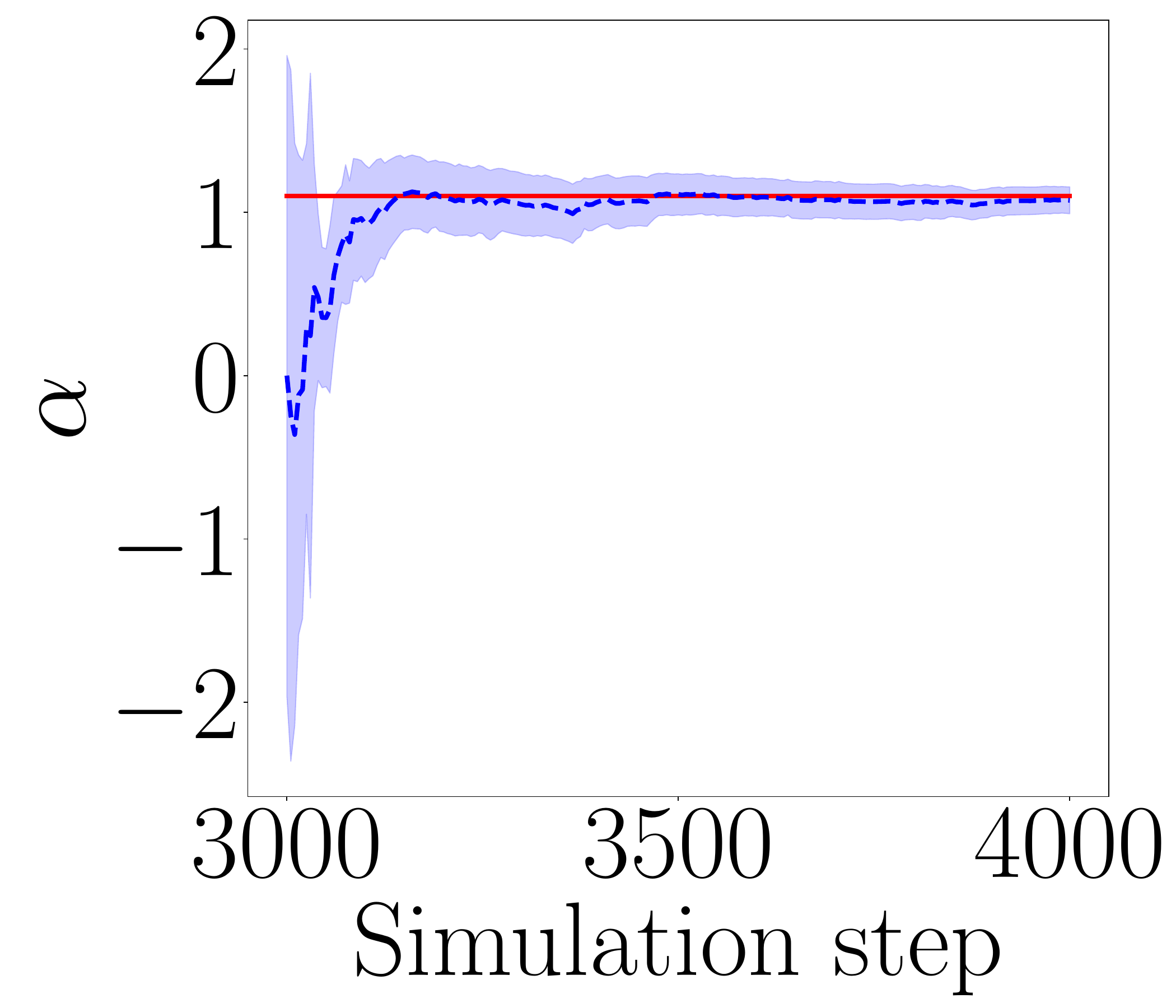}
    \caption{$\gamma=10$}
    \label{fig:alpha_e2_1_highNoise_10}
\end{subfigure}%
\hspace{0.003\linewidth}%
\begin{subfigure}[b]{0.163\linewidth}
    \centering
    \includegraphics[width=\textwidth]{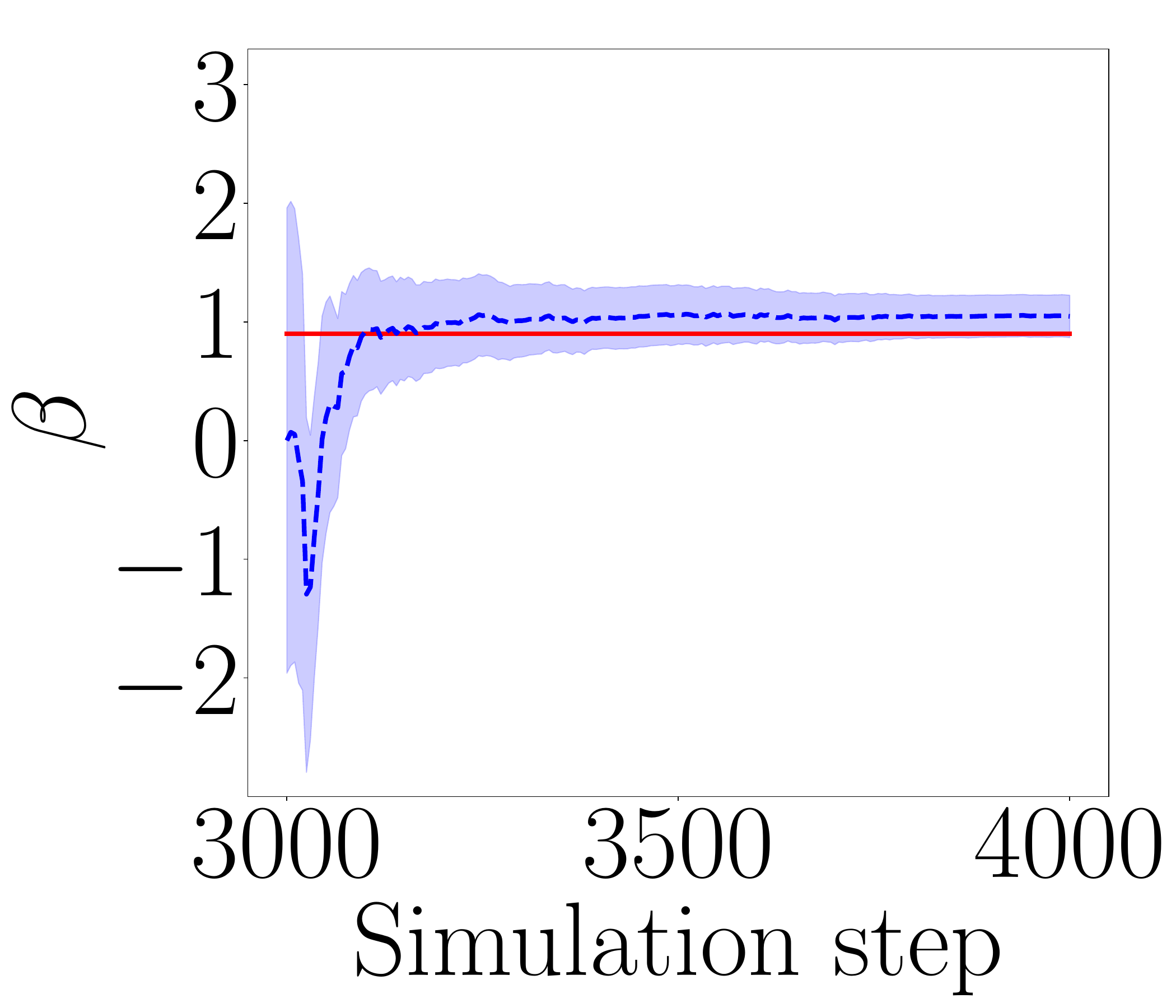}
    \caption{$\gamma=1$}
    \label{fig:beta_e2_1_highNoise_1}
\end{subfigure}%
\hspace{0.003\linewidth}%
\begin{subfigure}[b]{0.163\linewidth}
    \centering
    \includegraphics[width=\textwidth]{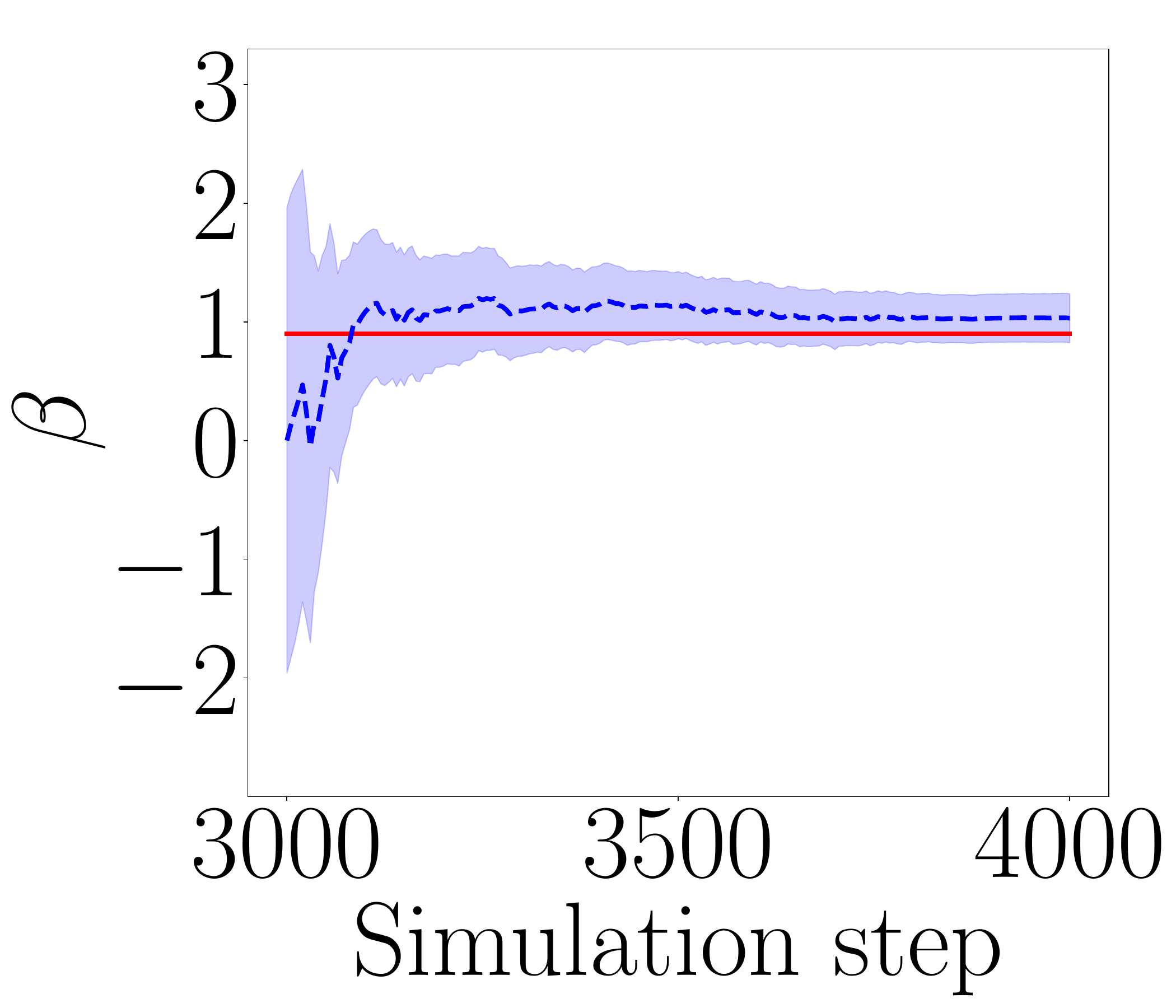}
    \caption{$\gamma=5$}
    \label{fig:beta_e2_1_highNoise_5}
\end{subfigure}%
\hspace{0.003\linewidth}%
\begin{subfigure}[b]{0.163\linewidth}
    \centering
    \includegraphics[width=\textwidth]{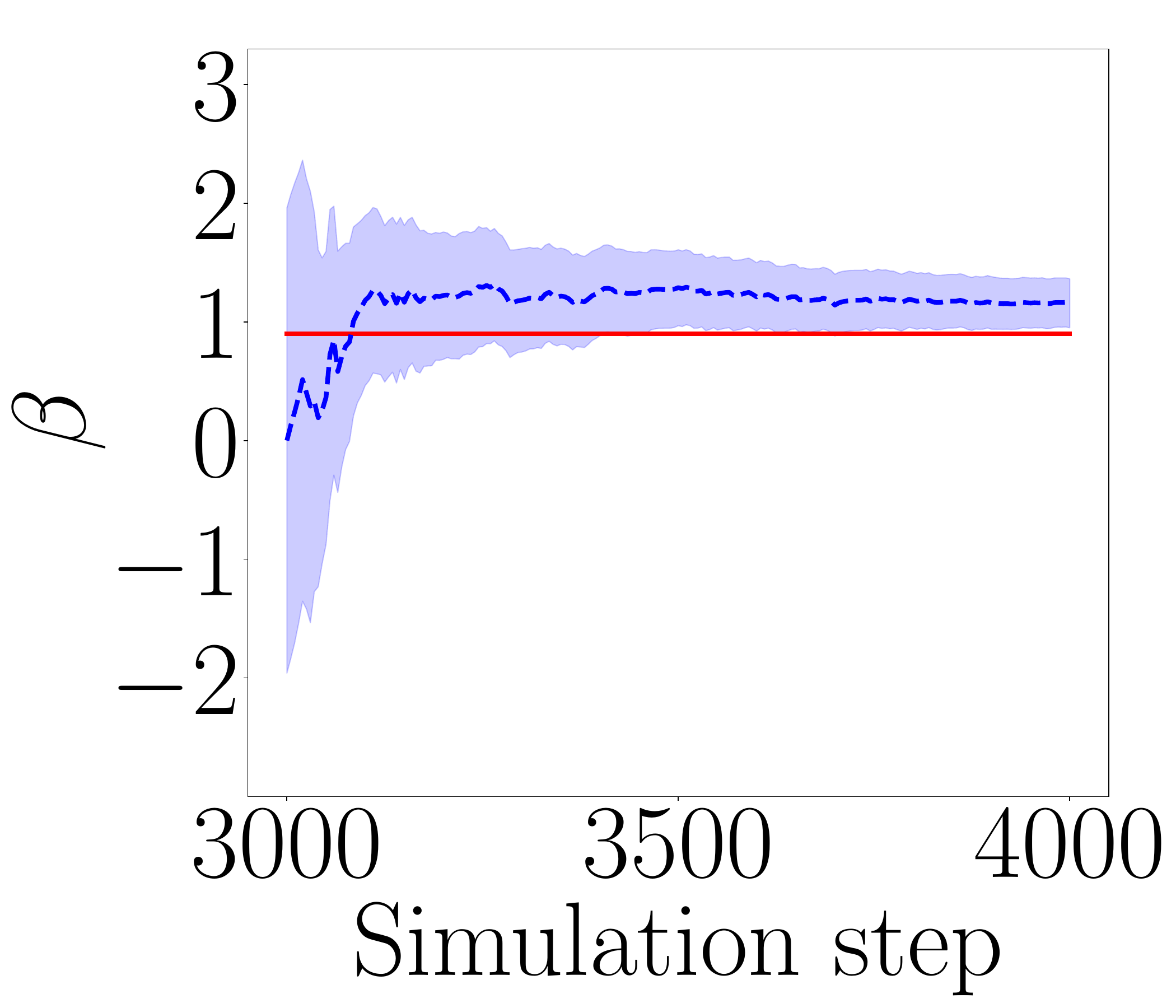}
    \caption{$\gamma=10$}
    \label{fig:beta_e2_1_highNoise_10}
\end{subfigure}
     \caption{\textbf{Lorenz '96 system with the correct model and different measurement noise levels:}
     approximate marginal posterior distributions of the parameters $\alpha$ and $\beta$ obtained by FBOVI with measurement noise variance $\gamma=1,5,10$. FBOVI provides accurate estimation of the parameter $\alpha$ in all three cases. For the parameter $\beta$, the mean converges to a value slightly larger than the true value when the measurements are considerably noisy ($\gamma=5$). When the measurements are highly noisy, corresponding to $\gamma=10$, the deviation of the mean from the true value after step $3500$ is larger than those when $\gamma=1$ and $\gamma=5$.}
     \label{fig:param_accurateModel_highNoise}
\end{figure}

During the inference procedure, the true observation model is assumed to be known. Figure~\ref{fig:param_accurateModel_highNoise} depicts the approximate marginal posterior distributions of the parameters $\alpha$ and $\beta$ obtained by FBOVI using data collected with measurement noise variance $\gamma=1,5,10$. The results show that the mean of the approximate marginal posterior distribution for the parameter $\alpha$ converges rapidly to the true value in all three cases. For the parameter $\beta$, there is a very small deviation of the converged mean from the true value when $\gamma=1$ and $\gamma=5$. This deviation becomes larger when $\gamma=10$.

\begin{figure}[t]
     \centering
    \begin{subfigure}[b]{0.5\linewidth}
        \centering
        \includegraphics[width=\textwidth]{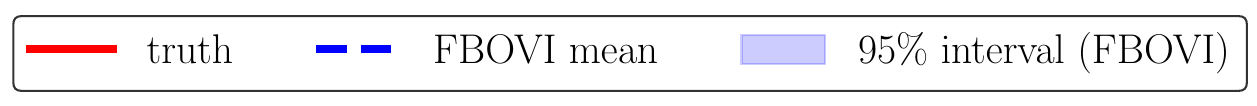}
    \end{subfigure}
     \vspace{0.4em}
     \includegraphics[width=1.0\textwidth]{figures/Lorentz/highNoise/estimation/estimation_highNoise.pdf}
         \caption{\textbf{Lorenz '96 system with the correct model and different measurement noise levels:} approximate marginal posterior distributions of the unobserved state components obtained by FBOVI in the cases with measurement noise variance $\gamma=1,5,10$. Top row, from left to right: distributions of $x_2$, $x_4$, $x_6$, $x_8$, and $x_{10}$ when $\boldsymbol{\gamma=1}$; middle row, from left to right: distributions of $x_2$, $x_4$, $x_6$, $x_8$, and $x_{10}$ when $\boldsymbol{\gamma=5}$; bottom row, from left to right: distributions of $x_2$, $x_4$, $x_6$, $x_8$, and $x_{10}$ when $\boldsymbol{\gamma=10}$. FBOVI estimates the state accurately at most time steps when considerably noisy data ($\gamma=5$) are used. When the data are highly noisy ($\gamma=10$), the estimation performance are reasonably good except when the state components are near some peaks or troughs of the trajectories.}
         \label{fig:state_accurateModel_highNoise}
\end{figure}

We present the estimation performance of FBOVI for the unobserved state components in the cases with $\gamma=1,5,$ and $10$ in Figure~\ref{fig:state_accurateModel_highNoise}. Figure~\ref{fig:state_accurateModel_highNoise} shows that increasing the measurement noise level mainly affects the estimation near some peaks and troughs of the trajectories. As $\gamma$ increases from $1$ to $5$ and then to $10$, the estimates of the state components not around the peaks and troughs remain satisfactory while the estimates near some peaks and troughs become less accurate.

\begin{figure}[t]
     \centering
    \begin{subfigure}[b]{0.5\linewidth}
        \centering
        \includegraphics[width=\textwidth]{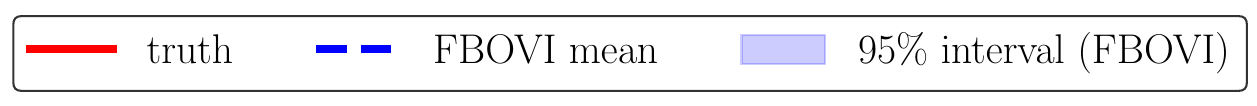}
    \end{subfigure}
    \vspace{0.4em}
     \includegraphics[width=1.0\textwidth]{figures/Lorentz/highNoise/prediction/prediction_highNoise_3steps_3200.pdf}
         \caption{\textbf{Lorenz '96 system with the correct model and different measurement noise levels (prediction):} predictive distributions of the state components $x_2$, $x_4$, $x_6$, $x_8$, and $x_{10}$ in the cases with $\gamma=1,5,10$. Top row, from left to right: predictive distributions of $x_2$, $x_4$, $x_6$, $x_8$, and $x_{10}$ when $\boldsymbol{\gamma=1}$; middle row, from left to right: predictive distributions of $x_2$, $x_4$, $x_6$, $x_8$, and $x_{10}$ when $\boldsymbol{\gamma=5}$; bottom row, from left to right: predictive distributions of $x_2$, $x_4$, $x_6$, $x_8$, and $x_{10}$ when $\boldsymbol{\gamma=10}$. FBOVI provides relatively accurate predictions at most time steps when $\gamma=5$. When the data become highly noisy ($\gamma=10$), the prediction performance remains reasonably satisfactory when the state components are not around the peaks or troughs of the trajectories, while the predictions near some peaks and troughs are less accurate.}
         \label{fig:state_accurateModel_highNoise_pred}
\end{figure}

The prediction results for the state components $x_2$, $x_4$, $x_6$, $x_8$, and $x_{10}$ obtained by FBOVI are provided in Figure~\ref{fig:state_accurateModel_highNoise_pred}. Similar behavior is observed in the prediction results: FBOVI provides reliable predictions for the state components not around the trajectory peaks and troughs across all measurement noise levels, whereas larger values of $\gamma$ lead to increased prediction errors near these extrema.

\subsection{Nonlinear convection--diffusion transport model}
\label{sec:Burger}
In this section, we consider a one-dimensional nonlinear convection--diffusion transport model given by
\begin{equation}
\frac{\partial u}{\partial t} + \alpha u \frac{\partial u}{\partial x} = \nu \frac{\partial^2 u}{\partial x^2},
\label{eq:convection-diffusion}
\end{equation}
for $u = u(x,t)$ with $x \in [0,1]$. Here, $\alpha$ denotes the nonlinear convection coefficient and $\nu$ the diffusion coefficient. We set $\alpha = 1$\footnote{Equation~\eqref{eq:convection-diffusion} reduces to Burgers’ equation when $\alpha = 1$.} and $\nu = 0.01$. The initial condition is specified as
\begin{equation*}
u(x,0) = \sin(2\pi x) + 0.5\cos(4\pi x) + 10.
\end{equation*}
We impose the boundary conditions $u(0,t) = u(1,t)$ and $\frac{\partial u}{\partial x}(0,t) = \frac{\partial u}{\partial x}(1,t)$.

The reference solution is computed using the forward Euler method with time interval $\Delta t = 0.001$ and a uniform grid with the mesh size of $0.02$. This discretization yields 51 equally spaced grid points, with the first at $x=0$ and the last at $x=1$. Based on this reference solution, measurements are taken at the $(3n+1)$-th and $(3n+2)$-th grid points for $n=0,1,\ldots,16$. At each time step $t_k = k \Delta t$, the measurements are corrupted by a Gaussian noise $V_k \sim \mathcal{N}(0, I_{34 \times 34})$, so that noisy observations are available at $2/3$ of the spatial grid points.

For online inference, we employ the following $51$-dimensional discrete-time dynamical model:
\begin{equation}
X_{k+1} = \Phi(X_k;\theta) + W_k, 
\quad W_k \sim \mathcal{N}(0, 0.01 I_{51 \times 51}), 
\label{eq:dynamical_model_e3}
\end{equation}
where the state $X_k \in \mathbb{R}^{51}$ represents the solution values at the 51 grid points at time $t_k$. The mapping $\Phi$ denotes the deterministic numerical propagation from $X_k$ to $X_{k+1}$, obtained using the same temporal and spatial discretization scheme as that used to generate the reference solution. It is important to note that the numerical method used to generate the reference solution differs from that used for $\Phi$. For the reference solution, the upwind scheme is employed to compute the spatial derivative $\frac{\partial u}{\partial x}$, since this scheme is stable when the viscosity $\nu$ is small, which is is the case here. In contrast, the model $\Phi$ uses the simple central difference scheme for computing the derivative $\frac{\partial u}{\partial x}$, which is less stable under this condition. The parameter vector $\theta = (\alpha, \nu)$ is treated as unknown, with prior distribution $\mathcal{N}([0,1]^T, 4I_{2 \times 2})$. Because of the model-form error, there are no “true” parameter values. The initial state $X_0$ is assigned a prior $\mathcal{N}(\mu_0, 4I_{51 \times 51})$, where the prior mean $\mu_0$ is randomly drawn according to $\mu_0 \sim \mathcal{N}(\mathbf{10}_{51}, 9I_{51 \times 51})$, with $\mathbf{10}_{51}$ denoting the 51-dimensional all-10 vector.

As the dynamical system~\eqref{eq:dynamical_model_e3} used for learning is high-dimensional, we employ the EnKF,  which is effective for large-scale problems, within the framework of FBOVI. To demonstrate the effectiveness of FBOVI, we compare its performance with that of joint EnKF, which is commonly applied to large-scale state–parameter estimation problems.  

\begin{figure}[t]
     \centering
     \begin{subfigure}[b]{0.3\linewidth}
         \centering
         \includegraphics[width=\textwidth]{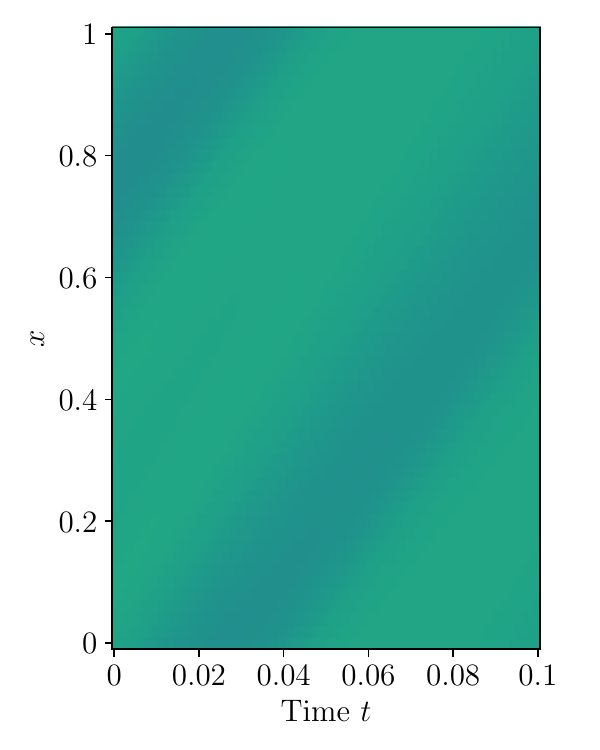}
         \caption{Reference solution $u(x,t)$}
         \label{fig:refer_e3}
     \end{subfigure}
     \hfill
     \begin{subfigure}[b]{0.3\linewidth}
         \centering
         \includegraphics[width=\textwidth]{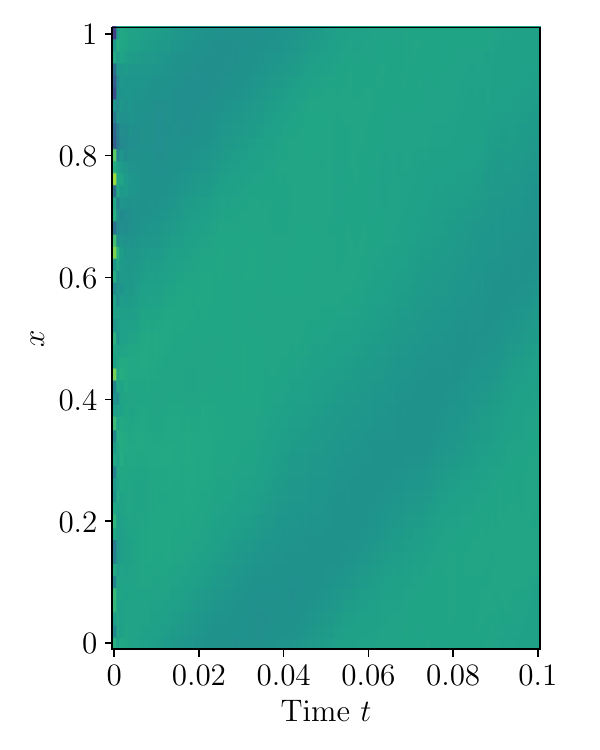}
         \caption{FBOVI mean for $u(x,t)$}
         \label{fig:fbovi_e3}
     \end{subfigure}
     \hfill
     \begin{subfigure}[b]{0.3\linewidth}
         \centering
         \includegraphics[width=\textwidth]{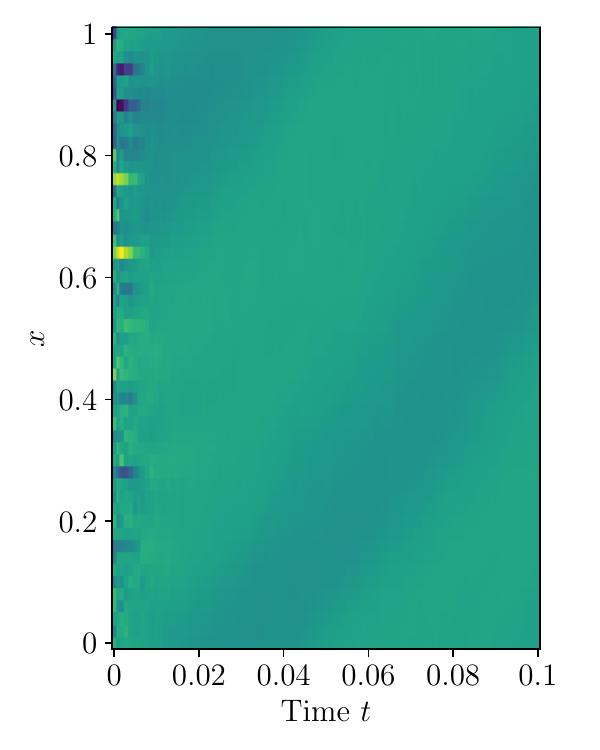}
         \caption{Joint EnKF mean for $u(x,t)$}
         \label{fig:enkf_e3}
     \end{subfigure}
     \hfill
     \begin{subfigure}[b]{0.052\linewidth}
         \centering
         \includegraphics[width=\textwidth]{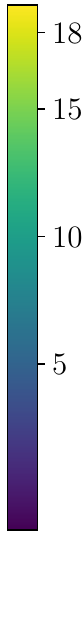}
     \end{subfigure}
         \caption{\textbf{Nonlinear convection--diffusion transport model}: reference solution $u(x,t)$ at different spatial locations $x$ and times $t$, together with the means of the approximate marginal posterior distributions of $u(x,t)$ obtained by FBOVI and joint EnKF. The color of each cell in this plot represents the value of the solution $u(x,t)$ at a time $t$ and a spatial position $x$. The coordinates of the \textbf{center} of each cell correspond to that time $t$ and spatial location $x$. FBOVI mean rapidly approaches the reference solution (from $t \approx 0.003$ onward), while the joint EnKF mean are significantly biased during the early stage.}
         \label{fig:state_e3}
\end{figure}

The means of the marginal posterior distributions of the solution $u(x,t)$ obtained by FBOVI and joint EnKF, along with the reference solution, are shown in Figure~\ref{fig:state_e3}. We can see that the mean computed by FBOVI closely follows the reference solution. This demonstrates that, although the prior for the initial solution value is far from the reference, FBOVI quickly adapts and begins to track the reference within a short time. In contrast, the discrepancy between the reference and the joint EnKF mean is considerably larger during the early stage ($t < 0.02$).  

\begin{figure}
     \centering
     \begin{subfigure}[b]{0.3\linewidth}
         \centering
         \includegraphics[width=\textwidth]{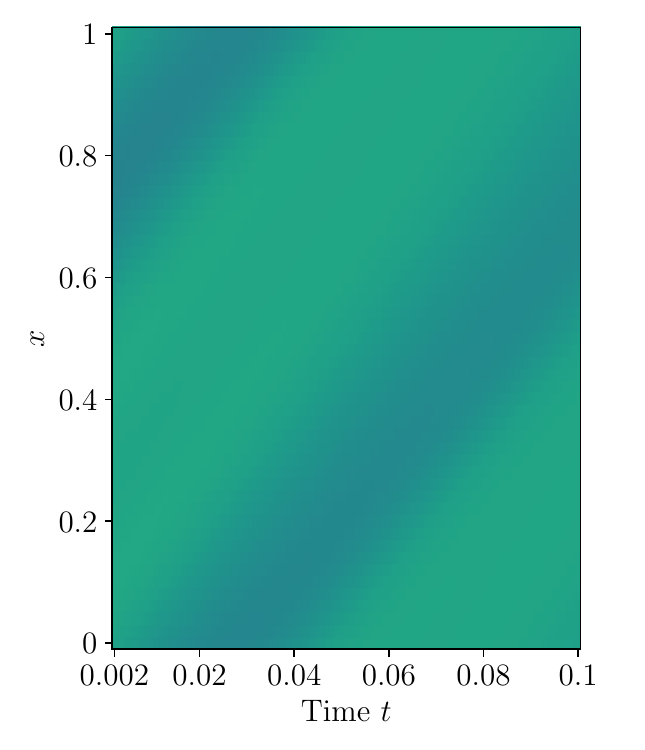}
         \caption{Reference solution $u(x,t)$}
         \label{fig:refer_pred_e3}
     \end{subfigure}
     \hfill
     \begin{subfigure}[b]{0.3\linewidth}
         \centering
         \includegraphics[width=\textwidth]{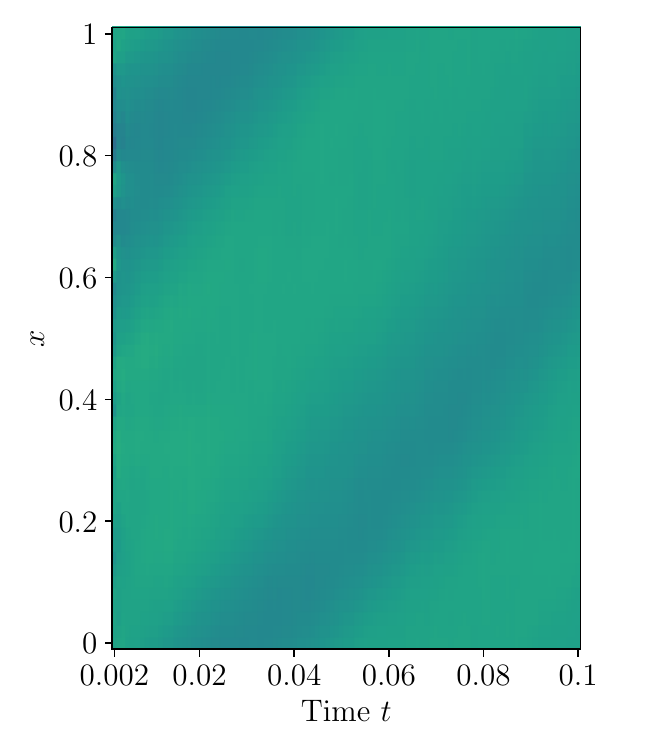}
         \caption{Mean of FBOVI prediction for $u(x,t)$}
         \label{fig:fbovi_pred_e3}
     \end{subfigure}
     \hfill
     \begin{subfigure}[b]{0.3\linewidth}
         \centering
         \includegraphics[width=\textwidth]{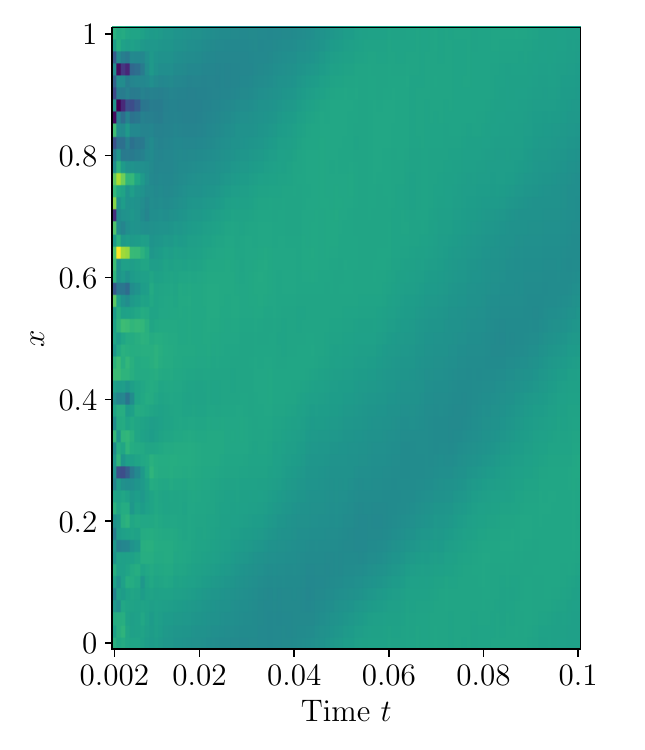}
         \caption{Mean of joint EnKF prediction for $u(x,t)$}
         \label{fig:enkf_pred_e3}
     \end{subfigure}
     \hfill
     \begin{subfigure}[b]{0.047\linewidth}
         \centering
         \includegraphics[width=\textwidth]{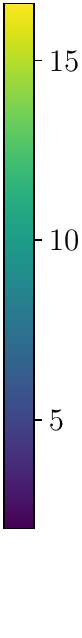}
     \end{subfigure}
         \caption{\textbf{Nonlinear convection--diffusion transport model (prediction)}: reference solution $u(x,t)$ at different spatial locations $x$ and times $t$, along with the means of the predictive distributions obtained by FBOVI and joint EnKF. The color of each cell in this plot represents the predicted value of the solution $u(x,t)$ at a time $t$ and a spatial position $x$. The coordinates of the \textbf{center} of each cell correspond to that time $t$ and spatial location $x$. The earliest time for which prediction is performed is $t=0.002$. FBOVI provides accurate predictions, while joint EnKF predictions deviate substantially from the reference when $t<0.01$. }
         \label{fig:pred_e3}
\end{figure}

Figure~\ref{fig:pred_e3} presents the means of the predictive distributions produced by FBOVI and the joint EnKF. For each $k\geq 2$, corresponding to $t \geq 0.002$, the predictive distribution of $X_k$ is computed using the approximate joint posterior obtained at time step $k-1$ and the dynamical system~\eqref{eq:dynamical_model_e3}. As show in Figure~\ref{fig:pred_e3}, the prediction generated by FBOVI is relatively accurate, reflecting the high quality of its joint posterior approximation. On the contrary, the Gaussian assumption for the joint posterior used by the joint EnKF limits its predictive capability. This deficiency is most obvious in the early stage, when only a small number of observations have been assimilated and the true joint posterior tend to be highly non-Gaussian.  

\section{Conclusions}
\label{sec:conclusion}
In this paper, we have presented an online variational inference (VI) framework for joint parameter-state estimation of dynamical systems, particularly under conditions of noisy and incomplete data. The core idea of our approach is to approximate the joint posterior distribution of model parameters and system states through a structured factorization: a variational distribution over parameters and a conditional distribution over states given those parameters. Unlike approaches that rely on restrictive assumptions about the joint posterior structure, our method preserves the structure between the state and parameters, leading to improved estimation and predictive accuracy. The modular design of the framework allows it to incorporate different filtering techniques, making it applicable to a broad range of problems. A theoretical error bound for the approximation has been established, which not only quantifies inference accuracy but also informs algorithmic design.

Through a series of numerical experiments, we demonstrated that the proposed method maintains strong performance in both low- and high-dimensional settings and remains robust under noisy, partial observations and model discrepancies. Together, these results highlight the reliability, scalability, and adaptability of the approach, underscoring its potential as a powerful tool in challenging application scenarios, including DT systems.

\section*{Acknowledgements}
This work was supported by an NSF CAREER Award, grant number CMMI-2238913.

\section*{Declaration of generative AI and AI-assisted technologies in the manuscript preparation process}

During the preparation of this work the authors used OpenAI’s ChatGPT in order to perform minor grammar polishing and spellchecking. After using this tool/service, the authors reviewed and edited the content as needed and take full responsibility for the content of the published article.

\bibliographystyle{elsarticle-num} 
\bibliography{ref}

\newpage

\appendix
\section{Metrics of difference between probability measures}
\label{app:metrics_dist}
In this section, we introduce two widely used metrics for quantifying differences between probability distributions: the total variation (TV) distance and the Hellinger distance. 

Let $\mu_1$ and $\mu_2$ be two probability measures with corresponding probability density functions $f_1(x)$ and $f_2(x)$. The total variation distance between $\mu_1$ and $\mu_2$ is defined as:
\begin{equation}
d_{TV}(\mu_1,\mu_2) := \frac{1}{2}\int \big \lvert f_1(x) - f_2(x) \big \rvert dx.
\end{equation}
The Hellinger distance between $\mu_1$ and $\mu_2$ is defined by:
\begin{equation}
d_{H}(\mu_1,\mu_2) := \sqrt{\frac{1}{2}\int \left( \sqrt{f_1(x)} - \sqrt{f_2(x) }\right)^2 dx}.
\end{equation}

\section{Proof of Theorem~\ref{thm:error_bound}}
\label{app:proof_thm}
Let $q_k^*(X_k,\theta)$ denote the resulting joint posterior distribution $\pdense(X_k,\theta \mid \Y_k)$ when the previous joint posterior $\pdense(X_{k-1},\theta \mid \Y_{k-1})$ is replaced with its approximation $q_{k-1}(X_{k-1},\theta)$. The following theorem established in our previous work~\cite{wang_gorodetsky_2025_lipschitz_stability} is used for the proof of Theorem~\ref{thm:error_bound}.
\begin{theorem}[Theorem 16 in~\cite{wang_gorodetsky_2025_lipschitz_stability}]
\label{thm:learning_bound}
For any step $k \geq 2$, given a sequence of data $\Y_k=(y_1,y_2,\cdots,y_k)$, suppose that for any step $i \in [2,k]$, we have
\begin{equation}
\bar{C}(y_i;i) := \sup_{X_{i-1},\theta}\int p(y_i \mid X_i,\theta)p(X_i \mid X_{i-1},\theta)dX_i < \infty.
\label{ieq:assumption}
\end{equation}
The total variation distance between $p(X_k,\theta \mid \Y_k)$ and $q_k(X_k,\theta)$, namely $d_{TV}\left(p(X_k,\theta \mid \Y_k), q_k(X_k,\theta)\right)$, satisfies:
\begin{align}
d_{TV}\left(p(X_k,\theta \mid \Y_k), q_k(X_k,\theta)\right) &\leq \sum_{j=1}^{k-1}\frac{\prod_{i=j+1}^k\bar{C}(y_i;i)}{p(\Y_{j+1:k}\mid \Y_{1:j})}d_{TV}\left(q_j, q_j^*\right) + d_{TV}\left(q_k, q_k^*\right).
\end{align}
The Hellinger distance between $p(X_k,\theta \mid \Y_k)$ and $q_k(X_k,\theta)$, namely $d_{H}(p(X_k,\theta \mid \Y_k), q_k(X_k,\theta))$, satisfies:
\begin{align}
d_{H}(p(X_k,\theta \mid \Y_k), q_k(X_k,\theta)) &\leq \sum_{j=1}^{k-1}\frac{2^{(k-j)}\prod_{i=j+1}^k\sqrt{\bar{C}(y_i;i)}}{\sqrt{p(\Y_{j+1:k}\mid \Y_{1:j})}}d_{H}\left(q_j, q_j^*\right) + d_{H}\left(q_k, q_k^*\right).
\end{align}

\end{theorem}
We now prove Theorem~\ref{thm:error_bound}.
\begin{proof} 
We begin by confirming that the assumption in Theorem~\ref{thm:learning_bound} holds.

For any $i \in [2, k]$, if $\inf_{\theta} |\Gamma(\theta)| = \tilde{C} > 0$, then
\begin{align*}
\pdense(y_i \mid X_i,\theta) &= \pdense_{\mathcal{N}}(y_i ; h(X_i;\theta), \Gamma(\theta)) = (2\pi)^{-\frac{r}{2}}\big \lvert \Gamma(\theta)\big \rvert^{-\frac{1}{2}} \exp \left(-\frac{1}{2}\left(y_i - h(X_i;\theta)\right)^T\Gamma(\theta)^{-1}\left(y_i - h(X_i;\theta)\right)\right) \nonumber \\
&\leq (2\pi)^{-\frac{r}{2}} \big \lvert \Gamma(\theta) \big \rvert^{-\frac{1}{2}} \leq (2\pi)^{-\frac{r}{2}}\tilde{C}^{-\frac{1}{2}}.
\end{align*}
Thus,
\begin{align}
\bar{C}(y_i;i) &=\sup_{X_{i-1},\theta}\int p(y_i \mid X_i,\theta)p(X_i \mid X_{i-1},\theta)dX_i \nonumber \\
&\leq (2\pi)^{-\frac{r}{2}}\tilde{C}^{-\frac{1}{2}}\sup_{X_{i-1},\theta}\int p(X_i \mid X_{i-1},\theta)dX_i  \nonumber \\
&= (2\pi)^{-\frac{r}{2}}\tilde{C}^{-\frac{1}{2}} < \infty,
\label{ieq:vi:type1:C}
\end{align}
and hence the assumption of Theorem~\ref{thm:learning_bound} is satisfied.

We now derive an upper bound on the total variation distance between $q_i(X_i,\theta)$ and $q^*_i(X_i,\theta)$. Note that $q^*_i(X_i,\theta)$ can be written as 
\begin{equation*}
q^*_i(X_i,\theta) = \rho_i^*(X_i\mid \theta)\nu^*_i(\theta).
\end{equation*}
The KL divergence of the distribution $q_i$ with respect to the distribution $q^*_i$ can be written as
\begin{align}
\KL(q_i \lVert q^*_i) &= \E_{(X_i,\theta) \sim q_i(\cdot)}\left[\log \left(\frac{q_i(X_i,\theta)}{q^*_i(X_i,\theta)}\right) \right] \nonumber \\
&= \E_{(X_i,\theta) \sim q_i(\cdot)}\left[\log \left(\frac{\rho_i(X_i \mid \theta)\nu_i(\theta)}{\rho^*_i(X_i \mid \theta)\nu^*_i(\theta)}\right)\right] \nonumber \\
&= \E_{(X_i,\theta) \sim q_i(\cdot)}\left[\log \left(\frac{\rho_i(X_i \mid \theta)}{\rho^*_i(X_i \mid \theta)}\right)\right] + \E_{(X_i,\theta) \sim q_i(\cdot)}\left[\log \left(\frac{\nu_i(\theta)}{\nu^*_i(\theta)}\right)\right].
\label{eq:kl}
\end{align}
The first term on the right hand side in Equation~\eqref{eq:kl} can be expressed as
\begin{equation}
\E_{(X_i,\theta) \sim q_i(\cdot)}\left[\log \left(\frac{\rho_i(X_i \mid \theta)}{\rho^*_i(X_i \mid \theta)}\right)\right] = \E_{\theta \sim \nu_i(\cdot)}\left[\E_{X_i \sim \rho_i(\cdot \mid \theta)}\left[\log \left(\frac{\rho_i(X_i \mid \theta)}{\rho^*_i(X_i \mid \theta)}\right)\right]\right] = \E_{\theta \sim \nu_i(\cdot)}\left[\KL\left(\rho_i(\cdot \mid \theta) \lVert \rho_i^*(\cdot \mid \theta) \right) \right].
\label{eq:term1_kl}
\end{equation}
Since the term $\log \left(\frac{\nu_i(\theta)}{\nu^*_i(\theta)}\right)$ does not depend on $X_i$, the second term on the right hand side in Equation~\eqref{eq:kl} satisfies
\begin{equation}
\E_{(X_i,\theta) \sim q_i(\cdot)}\left[\log \left(\frac{\nu_i(\theta)}{\nu^*_i(\theta)}\right)\right] = \E_{\theta \sim \nu_i(\cdot)}\left[\log \left(\frac{\nu_i(\theta)}{\nu^*_i(\theta)}\right)\right] = \KL\left(\nu_i \lVert \nu_i^*\right).
\label{eq:term2_kl}
\end{equation}
Plugging Equations~\eqref{eq:term1_kl} and ~\eqref{eq:term2_kl} into Equation~\eqref{eq:kl} yields
\begin{align}
\KL\left(q_i \lVert q^*_i \right) = \E_{\theta \sim \nu_i(\cdot)}\left[\KL\left(\rho_i(\cdot \mid \theta) \lVert \rho_i^*(\cdot \mid \theta) \right) \right] +\KL\left(\nu_i \lVert \nu_i^*\right).
\label{eq:kl_rewritten}
\end{align}
Therefore, we get
\begin{align}
d_{TV}(q_i,q_i^*) & \leq \frac{1} {\sqrt{2}}\sqrt{\KL\left(q_i \lVert q^*_i\right)} \nonumber \\
&= \frac{1}{\sqrt{2}}\sqrt{\E_{\theta \sim \nu_i(\cdot)}\left[\KL\left(\rho_i(\cdot\mid \theta) \lVert \rho_i^*(\cdot \mid \theta) \right) \right] +\KL\left(\nu_i \lVert \nu_i^*\right)}.
\label{ieq:vi:type1:tv_approx}
\end{align}
An analogous inequality holds for the Hellinger distance:
\begin{align}
d_{H}(q_i,q_i^*) & \leq \frac{1}{\sqrt{2}}\sqrt{\KL\left(q_i \lVert q^*_i\right)} \nonumber \\
& = \frac{1}{\sqrt{2}}\sqrt{\E_{\theta \sim \nu_i(\cdot)}\left[\KL\left(\rho_i(\cdot \mid \theta) \lVert \rho_i^*(\cdot \mid \theta) \right) \right] +\KL\left(\nu_i \lVert \nu_i^*\right)}.
\label{ieq:vi:type1:H_approx}
\end{align}
Lastly, combining Theorem~\ref{thm:learning_bound} with
Equations~\eqref{ieq:vi:type1:C}, \eqref{ieq:vi:type1:tv_approx}, and \eqref{ieq:vi:type1:H_approx}
completes the proof of Theorem~\ref{thm:error_bound}.
\end{proof}

\section{Proof of Theorem~\ref{thm:error_bound_linear}}
\label{app:proof_thm_linear}
The following theorem established in our prior work~\cite{wang_gorodetsky_2025_lipschitz_stability} is used for the proof of Theorem~\ref{thm:error_bound_linear}.
\begin{theorem}[Theorem 16 in~\cite{wang_gorodetsky_2025_lipschitz_stability}]
\label{thm:learning_bound_z}
For any step $k \geq 2$, given a sequence of data $\Y_k=(y_1,y_2,\cdots,y_k)$, suppose that for any step $i \in [2,k]$, we have
\begin{equation}
\bar{C}(y_i;i) := \sup_{X_{i-1},\theta}\int p(y_i \mid X_i,\theta)p(X_i \mid X_{i-1},\theta)dX_i < \infty.
\label{ieq:assumption_z}
\end{equation}
The total variation distance between $p(X_k,\theta \mid \Y_k)$ and $q_k(X_k,\theta)$, namely $d_{TV}(p(X_k,\theta \mid \Y_k), q_k(X_k,\theta))$, satisfies:
\begin{align}
d_{TV}\left(p(X_k,\theta \mid \Y_k), q_k(X_k,\theta)\right) &\leq \sum_{j=1}^{k-1}\frac{\prod_{i=j+1}^k\bar{C}(y_i;i)}{\prod_{i=j+1}^k \tilde{Z}_i(q_{i-1})}d_{TV}\left(q_j, q_j^*\right) + d_{TV}\left(q_k, q_k^*\right).
\end{align}
The Hellinger distance between $p(X_k,\theta \mid \Y_k)$ and $q_k(X_k,\theta)$, namely $d_{H}(p(X_k,\theta \mid \Y_k), q_k(X_k,\theta))$, satisfies:
\begin{align}
d_{H}(p(X_k,\theta \mid \Y_k), q_k(X_k,\theta)) &\leq \sum_{j=1}^{k-1}\frac{2^{(k-j)}\prod_{i=j+1}^k\sqrt{\bar{C}(y_i;i)}}{\prod_{i=j+1}^k\sqrt{ \tilde{Z}_i(q_{i-1})}}d_{H}\left(q_j, q_j^*\right) + d_{H}\left(q_k, q_k^*\right).
\end{align}
Here $\tilde{Z}_i(q_{i-1})$ is defined as $\tilde{Z}_i(q_{i-1}):= \int p(y_i \mid X_i,\theta)p(X_i \mid X_{i-1},\theta)q_{i-1}(X_{i-1},\theta)dX_{i-1}dX_id\theta$.
\end{theorem}
Next, we provide the proof of Theorem~\ref{thm:error_bound_linear}.
\begin{proof}
In the proof of Theorem~\ref{thm:error_bound}, we showed that the assumption in Theorem~\ref{thm:learning_bound_z} holds under the condition $\inf_{\theta} \lvert \Gamma(\theta) \rvert = \tilde{C} > 0$. Here $\bar{C}(y_i; i)$ satisfies 
\begin{equation}
\bar{C}(y_i;i) \leq (2\pi)^{-\frac{r}{2}}\tilde{C}^{-\frac{1}{2}}.
\label{eq:C_thm2}
\end{equation}
Next, we prove that $\tilde{Z}_i(q_{i-1})= \E_{\theta \sim \nu_{i-1}(\cdot)}[\pdense_{\mathcal{N}}(y_i;H(\theta)m_i^{*-}(\theta), H(\theta)C_i^{*-}(\theta)H(\theta)^T+\Gamma(\theta))]$. According to the definition of $\tilde{Z}_{i}(q_{i-1})$, it can be written as 
\begin{align}
\tilde{Z}_{i}(q_{i-1}) &= \int p_\mathcal{N}(y_i; H(\theta)X_i, \Gamma(\theta))p_\mathcal{N}(X_i; A(\theta)X_{i-1},\Sigma(\theta))p_\mathcal{N}(X_{i-1};m_{i-1}(\theta), C_{i-1}(\theta))\nu_{i-1}(\theta)dX_{i-1}dX_id\theta \nonumber \\
&= \E_{\theta \sim \nu_{i-1}(\cdot)}\left[\int p_\mathcal{N}(y_i; H(\theta)X_i, \Gamma(\theta))p_\mathcal{N}(X_i; A(\theta)X_{i-1},\Sigma(\theta))p_\mathcal{N}(X_{i-1};m_{i-1}(\theta), C_{i-1}(\theta))dX_{i-1}dX_i\right] \nonumber \\
&= \E_{\theta \sim \nu_{i-1}(\cdot)}[\pdense_{\mathcal{N}}(y_i;H(\theta)m_i^{*-}(\theta), H(\theta)C_i^{*-}(\theta)H(\theta)^T+\Gamma(\theta))].
\label{eq:z}
\end{align}
By Equations~\eqref{ieq:vi:type1:tv_approx} and ~\eqref{ieq:vi:type1:H_approx}, the terms $d_{TV}(q_j,q_j^*)$ and $d_{H}(q_j,q_j^*)$ satisfy
\begin{align}
d_{TV}(q_j,q_j^*) & \leq \frac{1} {\sqrt{2}}\sqrt{\KL\left(q_j \lVert q^*_j\right)} = \frac{1}{\sqrt{2}}\sqrt{\E_{\theta \sim \nu_j(\cdot)}\left[\KL\left(\rho_j(\cdot\mid \theta) \lVert \rho_j^*(\cdot \mid \theta) \right) \right] +\KL\left(\nu_j \lVert \nu_j^*\right)},
\label{ieq:vi:type1:tv_approx_2}
\end{align}
and
\begin{align}
d_{H}(q_j,q_j^*) & \leq \frac{1}{\sqrt{2}}\sqrt{\KL\left(q_j \lVert q^*_j\right)} = \frac{1}{\sqrt{2}}\sqrt{\E_{\theta \sim \nu_j(\cdot)}\left[\KL\left(\rho_j(\cdot \mid \theta) \lVert \rho_j^*(\cdot \mid \theta) \right) \right] +\KL\left(\nu_j \lVert \nu_j^*\right)}.
\label{ieq:vi:type1:H_approx_2}
\end{align}
According to Equations~\eqref{eq:nu_star},~\eqref{eq:kl_step1} and~\eqref{eq:elbo_step1}, the KL divergence $\KL\left(\nu_j \lVert \nu_j^*\right)$ can be written as
\begin{equation*}
\KL\left(\nu_j \lVert \nu_j^*\right) = \log \hat{Z}_j - \mathcal{L}_j(\nu_j),
\end{equation*}
where 
\begin{equation*}
\hat{Z}_j =  \int \pdense_\mathcal{N}(y_j; h(X_j;\theta), \Gamma(\theta))\, \rho_j^{*-}(X_j \mid \theta)\,\nu_{j-1}(\theta)dX_jd\theta.
\end{equation*}
Recall that the probability density $\pdense_\mathcal{N}(y_j; h(X_j;\theta), \Gamma(\theta))$
satisfies
\begin{align*}
\pdense_\mathcal{N}(y_j; h(X_j;\theta), \Gamma(\theta)) \leq (2\pi)^{-\frac{r}{2}}\tilde{C}^{-\frac{1}{2}}, \quad \forall X_j, \theta
\end{align*}
The KL divergence $\KL\left(\nu_j \lVert \nu_j^*\right)$ satisfies
\begin{align}
\KL(\nu_j  \| \nu_j^*) 
&\leq \log \left((2\pi)^{-\frac{r}{2}}\tilde{C}^{-\frac{1}{2}}\int \rho_j^{*-}(X_j \mid \theta)\nu_{j-1}(\theta) \, dX_jd\theta \right) - \mathcal{L}_j(\nu_j) \nonumber \\
& =\log \left((2\pi)^{-\frac{r}{2}}\tilde{C}^{-\frac{1}{2}}\right) - \mathcal{L}_j(\nu_j) \nonumber \\
& \leq -\frac{r}{2}\log(2\pi) - \frac{1}{2}\log \tilde{C} - \epsilon_j.
\label{ieq:kl_theta}
\end{align}

We next derive the expression of the KL divergence $\KL\left(\rho_j(\cdot \mid \theta) \lVert \rho_j^*(\cdot \mid \theta) \right)$.
The analytical expression of $\rho_j^*(X_j \mid \theta)$ is given by Kalman filter:
\begin{equation}
\rho_j^*(X_j \mid \theta) = \pdense_{\mathcal{N}}\bigg (X_j; F_m(m_{j-1}(\theta), C_{j-1}(\theta),\theta), F_c(m_{j-1}(\theta), C_{j-1}(\theta),\theta)\bigg).
\end{equation}
Recall that $\rho_j(X_j \mid \theta) = \pdense_{\mathcal{N}}(X_j; m_j(\theta), C_j(\theta))$. By employing the formula for the KL divergence of one Gaussian distribution with respect to another Gaussian distribution, we obtain
\begin{equation}
\KL\left(\rho_j(\cdot \mid \theta) \lVert \rho_j^*(\cdot \mid \theta) \right) = \Psi_j(C_j(\theta), m_j(\theta), C_{j-1}(\theta), m_{j-1}(\theta), \theta). 
\label{eq:kl_state}
\end{equation}
Finally, combining Equations~\eqref{eq:C_thm2},~\eqref{eq:z}, \eqref{ieq:vi:type1:tv_approx_2}, \eqref{ieq:vi:type1:H_approx_2}, \eqref{ieq:kl_theta}, and \eqref{eq:kl_state}, together with Theorem~\ref{thm:learning_bound_z}, completes the proof.
\end{proof}

\section{Guidance on variational family and Gaussian filter selection}
\label{sec:guidance}
In this section, we discuss how the choices of the variational family for the approximate parameter posterior $\nu_k(\theta)$ and the Gaussian filter incorporated into FBOVI may influence the resulting method, and provide guidance for making these choices in practice.

The choice of the variational family for $\nu_k(\theta)$ influences the approximation accuracy, stability, and computational efficiency of FBOVI. In practice, this choice can be made by balancing expressive power, optimization stability, and computational efficiency. A common starting point is a Gaussian variational family because of its simple parameterization, its compatibility with stable and efficient gradient based optimization methods, and efficient sampling from Gaussian distributions. When higher approximation accuracy is required, or when domain knowledge suggests that the parameter posterior may exhibit more complex structures such as multimodality, more expressive families, such as Gaussian mixture models (GMMs) or families defined by normalizing flows, can be adopted. These richer families enable more flexible and accurate approximations of the online parameter posterior $\nu_k^*(\theta)$, although this normally comes with increased computational complexity. Conversely, if computational speed is the primary concern, additional structural assumptions may be imposed on the variational family, which can further reduce computational cost but may lead to inaccurate estimation of uncertainty.

The choice of the incorporated Gaussian filter affects the approximation accuracy of $\rho_k(X_k \mid \theta)=\mathcal{N}(X_k;m_k(\theta),C_k(\theta))$ with respect to the online conditional filtering distribution $\rho_k^*(X_k \mid \theta)$. According to Theorem~\ref{thm:error_bound}, this approximation accuracy can further influence the accuracy of the joint posterior approximation. Recall that we first approximate $\rho_k^*(X_k \mid \theta)$ with a Gaussian distribution $\mathcal{N}(X_k;m_k^*(\theta),C_k^*(\theta))$, where $m_k^*(\theta)$ and $C_k^*(\theta)$ are functions mapping the parameters $\theta$ to the corresponding mean and covariance generated by the chosen Gaussian filter. We then use the neural-network representations $m_k(\theta)$ and $C_k(\theta)$ to approximate the target functions $m_k^*(\theta)$ and $C_k^*(\theta)$. Therefore, the approximation accuracy of $\rho_k(X_k \mid \theta)$ with respect to $\rho_k^*(X_k \mid \theta)$ is influenced by two factors: the accuracy of the Gaussian filtering approximation $\mathcal{N}(X_k;m_k^*(\theta),C_k^*(\theta))$ for $\rho_k^*(X_k \mid \theta)$, and the accuracy with which the neural networks approximate the target functions $m_k^*(\theta)$ and $C_k^*(\theta)$. The first factor is determined by the inference accuracy of the chosen Gaussian filter for standard filtering problems. The second factor depends on the complexities of the target functions $m_k^*(\theta)$ and $C_k^*(\theta)$, as well as the expressiveness of the neural networks used to approximate them. For neural networks with fixed architectures, more complex target functions are generally harder to learn accurately. In general, the complexities of the target functions $m_k^*(\theta)$ and $C_k^*(\theta)$ tend to increase as the Gaussian filter captures higher-order nonlinear effects. Under the same underlying model, the target functions generated by KF are typically the simplest, followed by those generated by ExKF, UKF, and GHKF.

The stability of the target-generation process performed by the chosen Gaussian filter directly affects the stability of Stage~2. When applicable, the KF typically provides the most stable generation of the target values $m^*_k(\theta^{(i)})$ and $C^*_k(\theta^{(i)})$. ExKF can also provide stable target generation for mildly nonlinear models with well-behaved Jacobians. In contrast, UKF may suffer from numerical instability and produce problematic covariance matrices $C_k^*(\theta)$, especially in high-dimensional problems. This can lead to instability of the training procedure for the neural network representing the covariance function $C_k(\theta)$. For a given parameter sample $\theta^{(i)}$, the KF, ExKF, UKF, and GHKF generate deterministic target values, which is favorable for the stability of supervised learning. In contrast, the EnKF generates stochastic target values for a given $\theta^{(i)}$, due to the finite-ensemble sampling. This introduces noise into the training data and may reduce the stability of supervised learning in Stage~2.

The computational cost of Stage~2 in FBOVI grows linearly with the computational cost of the incorporated Gaussian filter. Among all Gaussian filters, KF is normally the most computationally efficient when applicable. For nonlinear systems, UKF is typically slightly more computationally expensive than ExKF~\cite{bayesian_filter}. The computational cost of the GHKF increases rapidly with the problem dimension, making it computationally expensive, especially for high-dimensional problems.

\section{Non-Gaussian approximation for $\rho_k(X_k \mid \theta)$}
\label{app:non_Gaussian}
In this section, we introduce how the marginal distribution $\nu_k(\theta)$ and the conditional distribution $\rho_k(X_k \mid \theta)$ can be computed in the setting where $\rho_k(X_k \mid \theta)$ is represented by a non-Gaussian distribution and the corresponding filtering method which is integrated into our framework does not belong to the family of Gaussian filtering.

\subsection{Stage 1: Computation of $\nu_k(\theta)$}
\label{sec:step1_eg1}
When a non-Gaussian distribution is used to represent $\rho_k(X_k \mid \theta)$, the following procedure can be applied to compute the integral $I(\theta)$ in Equation~\eqref{eq:elbo_step1}, for evaluating the ELBO $\mathcal{L}_k(\nu_k)$.  

For a given $\theta$, draw $N$ samples from the previous conditional distribution $\rho_{k-1}(X_{k-1} \mid \theta)$ and $N$ samples of the process noise:
\begin{equation*}
X_{k-1}^{(i)} \overset{\mathrm{i.i.d.}}{\sim} \rho_{k-1}(X_{k-1} \mid \theta), \qquad 
\xi^{(i)} \overset{\mathrm{i.i.d.}}{\sim} \mathcal{N}(0,\Sigma(\theta)). 
\end{equation*}
The samples of the predictive distribution $\rho_k^{*-}(X_k \mid \theta)$ can then be constructed by
\begin{equation*}
X_k^{(i)-} = \Phi(X_{k-1}^{(i)};\theta) + \xi^{(i)}.
\end{equation*}
As a result, the integral 
\(I(\theta)\) can be approximated by
\begin{equation}
I(\theta)
\approx \frac{1}{N}\sum_{i=1}^N \pdense_{\mathcal{N}}(y_k;h(X_k^{(i)-};\theta), \Gamma(\theta)).
\label{eq:elbo_eg1}
\end{equation}

\subsection{Stage 2: Computation of $\rho_k(X_k \mid \theta)$}
Following Stage 1, we first employ a filtering method to approximate the online conditional filtering distribution $\rho_k^*(X_k \mid \theta)$ with a distribution $\hat{\rho}_k(X_k \mid \theta)$. The parametric family used to represent $\rho_k(X_k \mid \theta)$ is chosen to be the same with that of $\hat{\rho}_k(X_k \mid \theta)$. Let $\phi_k(\theta)$ and $\phi_k^*(\theta)$ denote the corresponding hyperparameters used to represent the distributions $\rho_k(X_k \mid \theta)$ and $\hat{\rho}_k(X_k \mid \theta)$, respectively. If we directly set $\rho_k(X_k \mid \theta) = \hat{\rho}_k(X_k \mid \theta)$, i.e., $\phi_k(\theta) = \phi_k^*(\theta)$, then the complexity of the functional representation of $\phi_k(\theta)$ will grow with time step $k$. To maintain constant computational and storage costs over time, we therefore re-approximate $\hat{\rho}(X_k \mid \theta)$ with $\rho(X_k \mid \theta)$ by minimizing the expected KL divergence $\E_{\theta \sim \nu_k(\cdot)}[\KL\left(\rho_k(\cdot \mid \theta) \lVert \hat{\rho}_k(\cdot \mid \theta)\right)]$. For a given $\theta$, the KL divergence $\KL\left(\rho_k(\cdot \mid \theta) \lVert \hat{\rho}_k(\cdot \mid \theta)\right)$ can be approximated using the Monte Carlo method with samples generated from $\rho_k(X_k \mid \theta)$:
\begin{equation*}
\KL\left(\rho_k(\cdot \mid \theta) \lVert \hat{\rho}_k(\cdot \mid \theta)\right) \approx \frac{1}{N_x}\sum_{i=1}^{N_x} \left(\log \rho_k(X_k^{(i)} \mid \theta) - \log \hat{\rho}_k(X_k^{(i)} \mid \theta) \right), \qquad X_k^{(i)} \overset{\mathrm{i.i.d.}}{\sim} \rho_k(\cdot \mid \theta).
\end{equation*}  
When computational resources are highly constrained, the training loss can be simplified to  
\begin{equation*}
\frac{1}{N_{\theta}}\sum_{i=1}^{N_{\theta}} \big\lVert \phi_k(\theta^{(i)}) - \phi_k^*(\theta^{(i)}) \big\rVert_2^2, 
\qquad \theta^{(i)} \overset{\mathrm{i.i.d.}}{\sim} \nu_k(\cdot).
\end{equation*}
Finally, the distribution $\nu_k(\theta)$ together with the function $\phi_k(\theta)$ are passed forward to the next time step $k+1$.

\section{Hyperparameters for the baseline methods in Section~\ref{sec:experiments}}
\label{app:hyper_param}
The hyperparameters of the unscented transform used by the joint UKF in Section~\ref{sec:low_dimension} and Section~\ref{sec:Lorentz} are set to $\alpha = 0.5$, $\beta = 2$, and $\kappa = 0$, following the settings in~\cite{joint_ukf2}. Specifically, $\alpha$ is typically chosen within the range $[10^{-4}, 1]$~\cite{joint_ukf2}. For Gaussian distributions, which are employed by joint UKF, $\beta = 2$ is generally regarded as an optimal choice~\cite{julier2002scaled}. It is also common practice to set $\kappa = 0$.

The joint PF, joint UKF, and joint EnKF assign random-walk dynamics with zero-mean Gaussian noise to the model parameters. For the joint PF, the covariance of this Gaussian noise is set to $10^{-4}I_{d_\theta \times d_\theta}$ to balance estimation accuracy and the mitigation of weight degeneracy. The joint UKF and joint EnKF use a smaller noise covariance of $10^{-8}I_{d_\theta \times d_\theta}$. Here $d_\theta$ represents the dimension of the parameters. The ensemble size of the joint EnKF in Section~\ref{sec:Burger} is $1000$.

The ground truth of the posterior $\pdense(X_k,\theta \mid \mathcal{Y}_k)$ in Section~\ref{sec:pendulum_part1} is obtained using delayed rejection adaptive Metropolis (DRAM), a highly effective Markov chain Monte Carlo (MCMC) method. MCMC operates in an offline manner: for each time step $k$, it assimilates the entire history of observations, $\mathcal{Y}_k=\{y_1,y_2,\cdots,y_k\}$, at once. The hyperparameters of DRAM are set as follows. The threshold after which adaptation begins in adaptive Metropolis is set to $100$. We use the two-stage delayed rejection method, with the scaling factor for the proposal covariance in the second stage set to $0.5$. For each time step $k$, DRAM initially generates $3\times 10^6$ samples. After discarding the first $50\%$ of the samples as burn-in, half of the remaining samples, i.e., $7.5\times 10^5$ samples, are used to represent the posterior $\pdense(X_k,\theta \mid \Y_k)$.

\section{Detailed training settings}
\label{app:ep1}
\subsection{Training hyperparameters}
The neural networks representing the mean function $m_k(\theta)$ and the function $c_k(\theta)$, are trained using the Adam optimizer with a base learning rate of $0.01$. This training configuration is used for all experiments presented in Section~\ref{sec:experiments}. 

The total number of optimization steps for training is set to $10000$, $2500$, $5000$, and $5000$ for the experiments in Section~\ref{sec:pendulum_part1}, Section~\ref{sec:pendulum_part2}, Section~\ref{sec:Lorentz}, and Section~\ref{sec:Burger}, respectively.

\subsection{Neural network architectures}

\paragraph{Linear pendulum system (Section~\ref{sec:low_dimension})}
The neural networks representing the mean function $m_k(\theta)$ and the function $c_k(\theta)$, share the same architecture. Each consists of three fully connected layers with dimensions
\[
(\text{input size}, 32), \quad (32, 64), \quad (64, \text{output size}).
\]
Each of the first two layers is followed by a ReLU activation function.

\paragraph{Lorenz 96 system (Section~\ref{sec:Lorentz})}
For the mean function $m_k(\theta)$, the neural network consists of four fully connected layers with dimensions
\[
(\text{input size}, 32), \quad (32, 64), \quad (64, 64), \quad (64, \text{output size}),
\]
where each of the first three layers is followed by a ReLU activation function. For the function $c_k(\theta)$, the neural network consists of four fully connected layers with dimensions
\[
(\text{input size}, 128), \quad (128, 256), \quad (256, 256), \quad (256, \text{output size}),
\]
with a ReLU activation function applied after each of the first three layers.

\paragraph{Burgers' equation (Section~\ref{sec:Burger})}
The neural network representing the mean function $m_k(\theta)$ consists of four fully connected layers with dimensions
\[
(\text{input size}, 128), \quad (128, 256), \quad (256, 256), \quad (256, \text{output size}),
\]
where each of the first three layers is followed by a ReLU activation function. The neural network representing $c_k(\theta)$ consists of five fully connected layers with dimensions
\[
(\text{input size}, 256), \quad (256, 512), \quad (512, 512), \quad (512, 256), \quad (256, \text{output size}).
\]
Each of the first three layers is followed by layer normalization, a GeLU activation function, and dropout. The fourth layer is followed by layer normalization and a GeLU activation function.

\subsection{Choice of the number of ensemble members $N_{e,F}$ and training dataset size $M$}
In Section~\ref{sec:Burger}, we use $N_{e,F}=1000$ ensemble members for the EnKF incorporated into FBOVI, for a good balance between computational efficiency and inference accuracy. The total number of parameter samples used for training, equivalently the training dataset size, is $M=5\times 10^4$.

The choices of $N_{e,F}$ and $M$ both play important roles in the overall inference performance. If $N_{e,F}$ is too small, for example $10$ or $50$, the EnKF component within FBOVI may provide an inaccurate approximation of the conditional filtering distribution, which can degrade the accuracy of the joint posterior approximation. Increasing $N_{e,F}$ can improve the conditional filtering approximation accuracy and enhance the stability of FBOVI, but it also increases the computational and memory costs.

The choice of $M$ affects how accurately $m_k(\theta)$ and $C_k(\theta)$ approximate the target functions $m_k^*(\theta)$ and $C_k^*(\theta)$. If $M$ is too small, the learned neural networks may poorly approximate their targets, leading to a large discrepancy between the approximate conditional filtering distribution $\rho_k(X_k \mid \theta)$ and the target conditional filtering distribution. Conversely, although a very large $M$ may provide improvements in approximation accuracy, it can also substantially increase computational and memory costs and the additional benefit over a moderate value of $M$ is often limited.

\section{Nonlinear Pendulum with a Nonlinear Observation Model}
\label{sec:nonlinear_observations}
In this section, we consider a nonlinear single pendulum with a nonlinear observation model. The dynamics of the single pendulum are described by
\begin{equation}
\begin{bmatrix}
\dot{x}_1 \\ \dot{x}_2 
\end{bmatrix}
=
\begin{bmatrix}
x_2 \\
-\frac{g}{l}\sin(x_1)
\end{bmatrix},
\label{eq:nonlinear_pendulum}
\end{equation}
where $g=9.8$ denotes the gravitational acceleration and $l=1.2$ denotes the length of the pendulum~\cite{fbovi}. The initial state is $x_1(0)=0.5$ and $x_2(0)=0.5$. We use the forward Euler method to integrate Equation~\eqref{eq:nonlinear_pendulum} with time discretization interval $\Delta t=0.1$ and generate the reference trajectory for $50$ time steps. The resulting discrete-time dynamics are
\begin{equation}
\begin{bmatrix}
x_1(k) \\ x_2(k) 
\end{bmatrix}
=
\begin{bmatrix}
x_1(k-1)+x_2(k-1)\Delta t \\
x_2(k-1)-\frac{g}{l}\sin(x_1(k-1))\Delta t
\end{bmatrix},
\end{equation}
where $x_1(k)$ and $x_2(k)$ denote the values of the state components $x_1$ and $x_2$ at time step $k$, respectively. Using this reference trajectory, we generate the data according to the following nonlinear observation model
\begin{equation}
Y_k = \sqrt[3]{x_1(k)} + V_k, \quad V_k \sim \mathcal{N}(0, 0.01).
\label{eq:nonlinear_pendulum_obs}
\end{equation}

We assume that the observation model is known and seek to learn a stochastic dynamical model and the state of the pendulum. The dynamical model is parameterized as follows:
\begin{equation*}
\begin{bmatrix}
x_1(k) \\ x_2(k) 
\end{bmatrix}
=
\begin{bmatrix}
\theta_1 x_1(k-1)+x_2(k-1)\Delta t \\
\theta_1 x_2(k-1)-\theta_2\sin(x_1(k-1))
\end{bmatrix}
+
W_k, 
\quad
W_k \sim \mathcal{N}(0,0.01I_{2\times 2}),
\end{equation*}
where $\theta=(\theta_1,\theta_2)$ are unknown model parameters and $\Delta t=0.1$ is known. The priors for $\theta$ and the initial state are $\mathcal{N}(\mathbf{0},I_{2\times 2})$ and $\mathcal{N}([3,4.5]^T,4I_{2\times 2})$, respectively. The prior mean of the initial state is far from the true value, making the inference problem challenging.

We compare the results of FBOVI for the nonlinear observation model~\eqref{eq:nonlinear_pendulum_obs} with the results for the linear observation model
\begin{equation}
Y_k = x_1(k) + V_k, \quad V_k \sim \mathcal{N}(0,0.01).
\label{eq:linear_pendulum_obs}
\end{equation}
The reference state trajectory, dynamical model, parameter prior, and initial-state prior are kept the same across the tests with the two observation models. The results of FBOVI for the linear observation model~\eqref{eq:linear_pendulum_obs} were presented in our previous work~\cite{fbovi}.

\begin{figure}[t]
     \centering
    \begin{subfigure}[b]{0.5\linewidth}
        \centering
        \includegraphics[width=\textwidth]{figures/Lorentz/legends/highNoise/legend_data_fbovi.pdf}
    \end{subfigure}
     \vspace{0.4em}
     \includegraphics[width=1.0\textwidth]{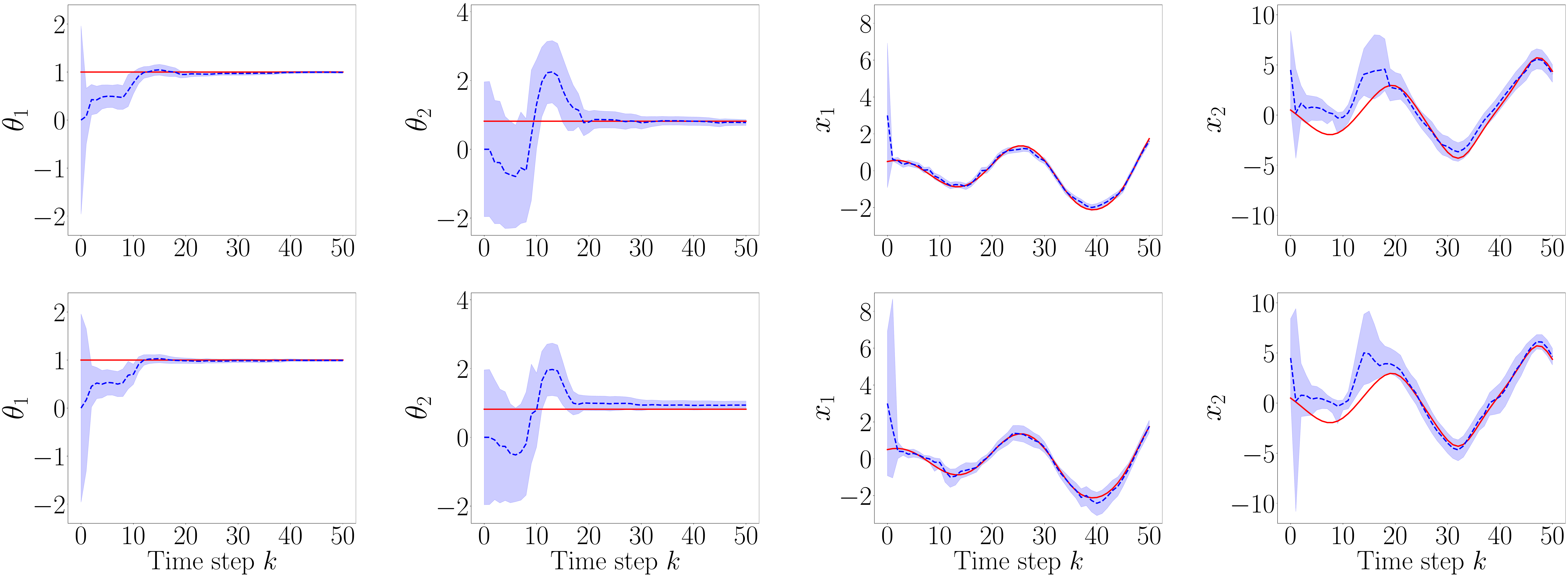}
         \caption{\textbf{Nonlinear single pendulum with linear and nonlinear observation models:}
         Top row, from left to right: approximate marginal posterior distributions of $\theta_1$, $\theta_2$, $x_1$, and $x_2$ obtained by FBOVI under the \textbf{linear} observation model~\eqref{eq:linear_pendulum_obs}. Bottom row, from left to right: approximate marginal posterior distributions of $\theta_1$, $\theta_2$, $x_1$, and $x_2$ obtained by FBOVI under the \textbf{nonlinear} observation model~\eqref{eq:nonlinear_pendulum_obs}. FBOVI provides satisfactory estimation performance for all parameters and states in both cases, although the estimation performance for $\theta_2$ under the linear observation model~\eqref{eq:linear_pendulum_obs} is slightly better than that under the nonlinear observation model~\eqref{eq:nonlinear_pendulum_obs}.}
         \label{fig:nonlinearObservations}
\end{figure}

Figure~\ref{fig:nonlinearObservations} depicts the marginal posterior distributions of the parameters and state components obtained by FBOVI at different time steps for the observation models~\eqref{eq:linear_pendulum_obs} and~\eqref{eq:nonlinear_pendulum_obs}. The results show that FBOVI successfully infers the parameters and state in both cases. The learning performance for $\theta_2$ in the case with the nonlinear observation model~\eqref{eq:nonlinear_pendulum_obs} is slightly worse than that in the case with the linear observation model~\eqref{eq:linear_pendulum_obs}. Apart from this difference, FBOVI provides accurate estimation of $\theta_1$ after step $20$ in both cases and rapidly starts tracking the true state values closely under both observation models.

\end{document}